\documentstyle[prd,eqsecnum,aps]{revtex}

\def\Nd{{N_{\rm d}}}
\def\eth{\epsilon_{\rm th}}
\def\sprop{{\cal S}}
\def\bprop{{\cal B}}
\def\pars{{\cal P}}
\def\drxn{{\bf n}}
\def\rvec{{\bf r}}
\def\Beff{B_{\rm eff}}
\def\Reff{R_{\rm eff}}

\def\km{\hbox{\enspace km}}
\def\s{\hbox{\enspace s}}
\def\erg{\hbox{\enspace erg}}

\def\MeV{\hbox{\enspace MeV}}

\def\gcc{\hbox{\enspace g}\;\hbox{cm}^{-3}}

\def\kpc{\hbox{\enspace kpc}}

\def\eg{{\it e.g.}}
\def\etal{{\it et~al.}}
\def\ie{{\it i.e.}}

\def\msol{M_\odot}

\def\eps{\epsilon}

\def\slr{\rm}
\def\araa{{\slr Ann. Rev. Astron. Ap.}}
\def\arnp{{\slr Ann. Rev. Nuc. Part. Phys.}}
\def\asap{{\slr Astr. Ap.}}

\def\apj{{\slr Astrophys.~J.}}
\def\apjl{{\slr Astrophys.~J. Lett.}}
\def\apjs{{\slr Astrophys.~J. Suppl. Ser.}}

\def\mnras{{\slr Mon. Not. R. Astron. Soc.}}
\def\nature{{\slr Nature (London)}}
\def\nyas{{\slr Ann. N. Y. Acad. Sci.}}
\def\prl{{\slr Phys. Rev. Lett.}}
\def\physrev{{\slr Phys. Rev.}}
\def\physrep{{\slr Phys. Rep.}}
\def\physlet{{\slr Phys. Lett.}}
\def\science{{\slr Science}}
\def\nuclear{{\slr Nucl. Phys.}}

\def\like{{\cal L}}
\def\Lmax{\like_{\rm max}}

\def\mnu{{m_{\bar \nu_e}}}

\def\tem{t^{\rm em}}
\def\tdet{t^{\rm det}}

\def\toff{t^{\rm off}}

\def\enu{E}
\def\epos{\epsilon}

\def\aem{A_c}

\def\nueb{{\bar \nu_e}}
\def\mwat{M_{\rm eff}}

\begin{document}
\draft


\title{Bayesian analysis of neutrinos observed from supernova 
SN~1987A\cite{THESIS}}

\author{Thomas J. Loredo}
\address{Department of Astronomy,
Space Sciences Building,
Cornell University,
Ithaca, New York 14853}

\author{Don Q. Lamb}
\address{Department of Astronomy \& Enrico Fermi Institute,\\
The University of Chicago,
5640 S. Ellis Ave.,
Chicago, Illinois 60637}

\date{14 July 2001}

\maketitle

\begin{abstract}
We present a Bayesian analysis of the energies and arrival
times of the neutrinos from supernova SN 1987A detected by the Kamiokande II,
IMB, and Baksan detectors, and find strong evidence for two
components in the neutrino signal:  a long time scale component from
thermal Kelvin-Helmholtz cooling of the nascent neutron star, and a brief
($\sim 1$ s), softer component similar to that expected from emission by
accreting material in the delayed supernova scenario.  In the context
of this model, we show that the data constrain the electron antineutrino
rest mass to be less than 5.7~eV with 95\% probability.
Our analysis takes advantage of significant advances that have occured
in the years since the detections in both our understanding of the
supernova mechanism and our ability to analyze sparse data.  This has
led to significant improvement over previous studies in two important
respects.  First, our comparison of the data with
parameterized models of the neutrino emission uses a consistent
and straightforward Bayesian statistical methodology.  This methodology helps
us distinguish the complementary tasks of parameter estimation and
model assessment, and fully accounts for the strong, nonlinear correlations
between inferred values of neutrino emission model parameters.
It also clarifies and improves the derivation of the likelihood function
(the probability for the data), improving on earlier derivations 
in two ways:  more consistent accounting for the energy-dependent
efficiencies of the detectors; and inclusion of the
empirically measured detector background spectra.  These improvements
lead to significant differences between our inferences and those found
in earlier studies.  Inclusion of detector background spectra
proves crucial for proper analysis of the Baksan data and for demonstrating
its consistency with data from other detectors.
Second, we compare the data with a much wider variety of neutrino
emission models than was explored previously, several of them inspired
by recent numerical calculations of collapse and explosion based on the delayed
supernova mechanism.  This allows us to compare predictions of both
the prompt and delayed mechanisms with the data, and insures that our
conclusions are robust.
We find that two-component models for the neutrino signal are
$\sim 100$ times more probable than single-component models.  Moreover,
single-component models imply a radius and binding energy for the nascent
neutron star significantly larger than those implied by even the stiffest
acceptable equations of state for neutron star matter.  In contrast,
the radius and binding energy implied by two-component models are in
agreement with predictions.  Taking this agreement with prior expectations
into account increases the odds in favor of two-component models by
more than an order of magnitude.
The inferred characteristics of the neutrino emission
are in spectacular agreement with the salient features of the theory
of stellar collapse and neutron star formation that had developed
over several decades in the absence of direct observational data.
We compare our work with previous work that used more conventional
``frequentist'' methods (including our own previous maximum likelihood
analysis).  We identify several methodological and technical weaknesses
in earlier analyses, and show how these are overcome in our
Bayesian analysis.

\end{abstract}

\section{Introduction}
\label{sec:intro}

The detection of neutrinos from supernova SN 1987A in the Large
Magellanic Cloud by the Kamiokande II (KII) \cite{KII-1,KII-2}
Irvine-Michigan-Brookhaven (IMB) \cite{IMB-1,IMB-2}
and Baksan \cite{Baksan-1,Baksan-2} detectors was a landmark
event in astrophysics.  Although only about two dozen
of the $\sim 10^{28}$ supernova
neutrinos that passed through the Earth were detected, they
provide us with our first glimpse of the collapsing core of a dying
star, and hence deserve careful scrutiny.

There is an extensive literature analyzing these epochal
detections, both qual\-i\-ta\-tive\-ly \cite{Bahcall-1,Sato-1,Arnett-87,%
Cowsik,Rosen,Sato-2,BL-87,Schramm-87,Burrows-88,Suzuki-88}
and quant\-i\-ta\-tive\-ly \cite{KST,BPPS,Adams,Krauss,Spergel-87,%
BS-88,ADW,SB-87,LML,Burrows-mass,LoSecco,LY-89,PS-89,JH-89,JH-89b,LL-89,Suz89}.
These previous studies
use a wide variety of methods, and although there is some agreement
among their conclusions, there are also important and troubling
differences.  Unfortunately, no criteria have been presented with
which one could evaluate and compare the various studies.  In addition,
there are technical deficiencies in many of the studies, including
inaccurate modeling of the detection process, and consideration of
unnecessarily restricted classes of models for the neutrino signal.
A consequence of these weaknesses is that the literature analyzing
the supernova neutrinos appears inconclusive or even contradictory.  Some
would argue that this is an inevitable consequence of the analysis
of a sparse data set.  We assert that it is a consequence only of
weaknesses in the analyses, and that probability theory is able to
precisely and consistently quantify the information in a data set, even
when the data set is small.  Indeed, it is in just such cases that a
careful quantification of our uncertainty is most necessary.

The years since the detection of the supernova neutrinos have seen
significant advances in our understanding of the supernova mechanism
and in our ability to analyze sparse data.  In 1987, the prompt
scenario for supernovae was favored, and almost all of the most
sophisticated analyses of the SN 1987A neutrino data used models based
on this scenario.  But in the intervening years, more careful
calculations have shown that the prompt mechanism probably fails to
create explosions, and that the delayed mechanism---relatively new in
1987---is more likely to be the cause of supernova explosions.  Through
the same decade there has been a parallel development in the
application of likelihood and Bayesian methods to the analysis of
inhomogeneous Poisson processes in astrophysics.  These theoretical and
analytical advances motivated us to undertake a new analysis of the
supernova neutrinos that significantly improves on previous analyses
both in its statistical methodology and in the variety of models
considered.

Our methodological improvements stem from 
consistent and straightforward application of
the principles of Bayesian inference.  The likelihood function---the
probability for the data given some hypothesis for their origin---plays
a key role in Bayesian inference, so aspects of our analysis bear
some similarity to earlier analyses based on likelihood functions
that used more conventional ``frequentist'' techniques, such as our
own earlier work \cite{LL-89}.  But there are crucial differences
both in the form of the likelihood function we use, and in the
manner in which we use it to make inferences.  

Our derivation
of the likelihood function reveals errors in previous attempts to
account for the energy dependence of the efficiencies of
the neutrino detectors; we show that these errors significantly corrupted
previous inferences.  In addition, our likelihood accounts for
the possibility that each event could have arisen from background sources
by using empirically measured detector background spectra.
Previous studies either ignored the detector background,
or tried to account for its effects by censoring the data.
We show that correct treatment of the background is crucial for proper
analysis of the Baksan data, and that it noticeably
affects the implications of the KII data.  Additionally, we include
the effects of dead time in the IMB detector, which has also been
ignored in most previous analyses.
Once the likelihood is available, Bayesian calculations
use it in a different manner than frequentist
likelihood analyses.    In particular, the
Bayesian methodology allows us to accurately summarize the implications
of the data for interesting subsets of the parameters needed to model
the data, in a way that fully accounts for the strong correlations
between inferred
values of neutrino emission model parameters.  These correlations
must be taken into account in order to fully compare the implications
of the data with the predictions of theory.  Also, Bayesian model comparison
implements an automatic ``Ockham's razor'' that takes into
account model complexity when comparing rival signal models; this
assures that complicated models are preferred only when there is
significant evidence in the data favoring them.

Complementing these methodological improvements are the improved
scientific insights gained by our use of
a much wider variety of neutrino emission
models than was explored previously.  Earlier studies explored either
a single model or an unnecessarily restricted class of models, almost
always presuming the prompt supernova mechanism to be true.  We
explore a variety of single-component models designed to mimic neutrino
emission from a cooling nascent neutron star (the principle detectable
component in the prompt scenario), and a variety of two-component
models that add to this cooling emission a component arising from
material that is heated upon passing through the stalled accretion shock
expected in delayed scenarios for supernova explosions.  We find that
all single component models lead to unacceptably large inferred
neutron star radii and binding energies.  We further show that the
data unambiguously prefer two component models, and that such models lead
to quite reasonable inferred radii and binding energies for the nascent
neutron star.
The wide variety of models we consider insures that our conclusions are
robust.

This paper is organized as follows.  We begin with a brief review of
Bayesian inference in Sec.~\ref{sec:bayes}.  We then devote two sections
to the derivation of the likelihood function.  Sec.~\ref{sec:dtxn}\
derives the probability for data from a neutrino detector, given some
parameterized form for the production rate of energetic charged leptons in the
detector; some details of the derivation appear in
Appendix~A.  Sec.~\ref{sec:lepton}\ describes how we derive the lepton
production rate from general models for the emission of neutrinos from
the supernova and their eventual interaction with earthbound detectors.

In Sec.~\ref{sec:models}\ we describe the wide variety of parameterized
models we have analyzed.  Included among these are phenomenological
models based on both the prompt and delayed mechanisms for
supernovae.  We present the best fit parameter values for each model
in Sec.~\ref{sec:results}, and we compare the models to one another
in light of the data, finding a definite preference for two-component
models.  The most tenable of the single-component models
is one with an exponentially decaying neutrinosphere temperature at
a constant radius; this model is also the one most extensively studied
in earlier analyses.
In Sec.~\ref{sec:exp}\ we analyze this single-component model in greater
detail.  We present
joint credible regions for the model parameters that display the strong
correlations between parameters, and that reveal an inconsistency
between the neutron star radius and binding energy implied by this model and
those predicted by current equations of state for neutron star matter.
In Sec.~\ref{sec:acn}\ we analyze the best two-component model in
greater detail.  We find the constraints implied by the data on parameters
describing both the cooling and accretion components of the emission, and
we demonstrate the consistency between the neutron star properties implied
by this model and those predicted by current equations of state.
In Sec.~\ref{sec:theory}\ we provide a brief review of theoretical
expectations for neutrino emission during and immediately after stellar
collapse, and compare these expectations with the observed characteristics
of the emission.

In the analyses presented in Sec.~\ref{sec:results}\ through
Sec.~\ref{sec:theory}, we assume that the rest mass
of the electron antineutrino, $\mnu$, is zero.  
In Sec.~\ref{sec:mass}\ we treat $\mnu$ as an additional parameter
to be inferred.  We find no significant evidence for a nonzero
mass, and we determine the upper limits implied by single-component
and two-component signal models.

Throughout the text we note technical differences between our work
and earlier work, particularly in regard to the form of the likelihood
function and the detector model.  In Sec.~\ref{sec:comp}\ we
elaborate on some of the weaknesses of earlier work, including
our own earlier frequentist likelihood analysis \cite{LL-89}.
We summarize our principle conclusions in Sec.~\ref{sec:conc}.

\section{Statistical Methodology}
\label{sec:bayes}

We carefully distinguish between the problems of
(1) estimating the value of parameters in a model for the neutrino signal, and
(2) assessing the adequacy of a particular parameterized model.
A major weakness of most previous analyses of the supernova signal is
the failure of investigators to distinguish between these complementary
statistical tasks, leading many to use model assessment methods
incorrectly to calculate ``confidence regions'' for parameters.

We address both parameter estimation and model assessment
problems with Bayesian methods.  In Bayesian inference, the viability of
each member of a set of rival
hypotheses, $\{H_i\}$, is assessed in the light of some observed data, $D$,
by calculating the probability for each
hypothesis, given the data and any background information, $I$, we may have
regarding the hypotheses and data.  Following a notation introduced
by Jeffreys \cite{Jeffreys}, we write such a probability as $p(H_i | D,I)$,
explicitly denoting the background information by the proposition, $I$,
to the right of the bar.
At the very least, the background information must specify the class of
alternative hypotheses being considered, and the relationship between
the hypotheses and the data (the statistical model).
In cases where the hypotheses of interest are labeled by the possible values
of a continuous parameter, $\theta$, the quantity $p(\theta | D,I)$ is
a probability {\it density}:  $p(\theta | D, I)d\theta$ is the
probability that the true value of the
parameter is in the interval $[\theta,\theta+d\theta]$, given the data
and the background information. We use the same symbol,
$p(\ldots)$, for densities and probabilities; the nature of the argument will
identify which use is intended.

Bayes's theorem gives $p(H_i | D,I)$ in terms of other probabilities,
\begin{equation}
p(H_i | D,I) = p(H_i | I)\; {p(D | H_i,I) \over p(D | I)}.\label{BT}
\end{equation}
The probabilities $p(H_i | I)$ for the hypotheses in the absence of $D$ are
called their prior probabilities, and the probabilities $p(H_i | D,I)$
including the information $D$ are called their posterior probabilities.
The quantity
$p(D | H_i,I)$ is called the sampling probability for $D$, or the likelihood
for $H_i$, and the quantity $p(D | I)$ is called the prior predictive
probability for $D$, or the (global) likelihood for the entire class of
hypotheses.

The rules of Bayesian inference lead one to use Bayes's theorem both
to estimate signal parameters and to assess a model as a whole by
comparing it to rival models.  But different types of calculations
are required to implement these two complementary tasks.
In this section we describe these applications of Bayes's theorem,
which we use freely throughout the remainder of this work; Bayesian
model comparison in particular has so far seen little use in physics,
motivating this brief pedagogical introduction.  We also briefly
describe the computational techniques we use to implement the calculations.
More complete derivations of the results in this section, with
simple examples and further references, are available in recent reviews
\cite{Gull-88,Brett-88,Brett-90,TJL-90,TJL-92,MacKay-92,L-99}.  The
{\em Bayesian Inference in the Physical Sciences} web site
\cite{BIPS} provides access to a variety of reviews and tutorials.

\subsection{Parameter estimation}

Many readers may be familiar with
the use of Bayes's theorem to estimate parameters
in a model.  Given some proposition, $M$,
specifying a model with parameters denoted collectively by $\theta$,
and a proposition, $D$, specifying data relevant to the model, one
calculates the posterior distribution
for the parameters, $p(\theta  | D,M)$, according to the continuous
version of equation~(\ref{BT}),
\begin{equation}
p(\theta  | D,M) = p(\theta | M)\; {p(D |\theta,M) \over p(D | M)}.
\label{BT-param}
\end{equation}
Of the factors in this equation, the likelihood function, $p(D |\theta,M)$,
is probably the most familiar.  It is the probability for the {\it data},
assuming the parameters have values given by $\theta$.  We often denote
the likelihood by the symbol $\like(\theta)$; this notation emphasizes that its
dependence on the parameters is what is important in Bayes's theorem,
but that it is not by itself a probability for the parameters.

The remaining terms in equation~(\ref{BT-param})\ are the prior for
$\theta$ and the prior predictive probability.  For the most part,
in this work we adopt uniform
(constant) priors for all parameters.  When the data are informative,
the posterior is robust to changes in the prior; we note those cases
where the data are uninformative as they arise.  The prior predictive,
$p(D | M)$, is independent of $\theta$ and merely plays the role of a
normalization constant whose value is given by integrating the product
of the prior and the likelihood:
\begin{equation}
p(D | M) = \int d\theta\; p(\theta | M)\; p(D |\theta,M).
  \label{gl-def}
\end{equation}
Thus the essential content of equation~(\ref{BT-param})\ may be summarized by
the statement that the posterior density is proportional to the
product of the prior and the likelihood.

Frequently a parameterized model will have more than one parameter, but
we will want to focus attention on a subset of the parameters.  For
example, at one point in this work we will want to focus on the
implications of the data
for the binding energy and radius of the neutron star formed by the
supernova, independent of the remaining parameters describing the
neutrino signal.  The
uninteresting parameters are known as {\it nuisance parameters}.
The posterior distribution for the parameters of interest
can be calculated by integrating out the nuisance parameters.
Explicitly, if model $M$ has two parameters, $\theta$ and $\phi$, and we
are interested only in $\theta$, then it is a simple consequence of the
sum and product rules of probability theory that,
\begin{equation}
p(\theta | D,M) = \int d\phi \; p(\theta,\phi | D,M).\label{marg}
\end{equation}
The procedure of integrating out nuisance parameters
is called {\it marginalization}, and $p(\theta | D,M)$ is called the
marginal posterior distribution for $\theta$.
In frequentist statistics there is no generally acceptable way to
eliminate nuisance parameters.  The ability to
marginalize parameters is thus an important advantage of the Bayesian
approach.

The Bayesian solution to the parameter estimation
problem is the full distribution, $p(\theta | D,M)$,
and not just a single point in parameter space.
Of course, it is often useful to summarize this distribution for textual,
graphical, or tabular display in terms of a ``best-fit'' value and
``error bars,'' indicating the location and width of the posterior.
Possible choices of
summarizing best-fit values are the posterior mode (the
value of $\theta$ with largest posterior density) or the posterior mean,
$\langle\theta\rangle = \int d\theta\; \theta\; p(\theta | D,M)$.
If the mode and mean are very different, the posterior distribution is
probably too complicated for its location to be adequately summarized by a
single number.  An allowed
range for a parameter with probability content $C$ is provided by a
{\it credible region},  $R$, defined so that
\begin{equation}
\int_R d\theta\; p(\theta | D,M) = C.\label{9}
\end{equation}
If $R$ is chosen so that
the posterior density inside $R$ is everywhere greater than that outside it,
then $R$ is a highest posterior density (HPD) credible region; all of
the credible regions we display in this work are HPD credible regions.
(Credible regions are not called ``confidence regions'' to distinguish them from
frequentist confidence regions, which are calculated in a very different
manner \cite{TJL-92}.)

In this work we present as a best-fit summary the posterior mode.
Since we are using flat priors, these estimates are identical to those
a frequentist maximum likelihood analysis would produce.  But Bayesian
and frequentist uses of the likelihood for finding allowed regions
differ (especially when nuisance parameters are present), so more
complete summaries (\eg, credible regions) will differ from their
frequentist counterparts.  To find the credible regions reported in
this work, we use {\em posterior sampling}---the use of Monte Carlo
methods to obtain a set of samples of parameter values from the full
joint posterior.  The ``cloud'' of such samples nicely summarizes the
full posterior; but more importantly, once the samples are available,
any marginal distribution can be easily estimated by simple
manipulations of the samples.  For example, samples from the marginal
distribution for any function of the parameters can be found simply by
evaluating the function on the samples.  A simple special case is when
we seek samples from the marginal distribution for a subset of the
parameters; they can be found simply by ignoring the nuisance parameter
coordinates of each sample from the full posterior.    We obtain the
samples using the rejection method \cite{NR}, and for plotting smooth
contours of one- and two-dimensional marginals we fit the cloud of
points to simple parameterized functions (exponentials of
polynomials).  Loredo \cite{L-99} provides further discussion of
posterior sampling and pointers to the literature.

\subsection{Model comparison}

In Bayesian inference, the success of a model is assessed by comparing
it to explicit alternative models.
To compare rival models, we again use Bayes's theorem.  This use of Bayes's
theorem is probably less familiar to most readers, though it is
analogous to use of Bayes's theorem for parameter estimation.  We begin
by specifying a set of competing models.  We use
the symbol $M_i$ to denote a proposition asserting that model $i$ describes
the data, and the symbol $I$ to denote a proposition asserting that one
of the models being considered describes the data ($I = ``M_1$ or $M_2$
or $\ldots$'').  Then we use Bayes's theorem to calculate
the probability for model $M_i$, assuming that one of the models
being considered describes the data:
\begin{equation}
p(M_i | D,I) = p(M_i | I)\; {p(D | M_i,I) \over p(D | I)}.\label{BT-mod}
\end{equation}
This is very much like equation~(\ref{BT-param}), with $M_i$ now playing the 
role of the parameter, and $I$ now playing the role of the model.
The term $p(M_i | I)$ is the prior probability for model $M_i$.
The proposition $(M_i,I)$ (``$M_i$ {\it and} $I$'') is true if and only if
model $M_i$ is true, that is,
it is equivalent to the proposition $M_i$ itself.
Thus $p(D | M_i,I) = p(D | M_i)$, the quantity calculated in equation~(\ref{gl-def}).  This quantity plays the uninteresting role of a normalization constant in
parameter estimation, but it plays a key role in model comparison:
it is the likelihood for model $M_i$ in equation~(\ref{BT-mod}).   
Equation~(\ref{gl-def}) reveals the likelihood for a model to
be equal to the {\em average} likelihood of its parameters (averaged
with respect to the prior for the parameters).  This is in stark
contrast to frequentist measures of model quality, which typically
maximize rather than average the likelihood for the parameters.  To
help distinguish the likelihood for a model's parameters from the
likelihood for the model as a whole, we use the term ``likelihood
function'' (a function of the parameters) for the former, and ``model
likelihood'' or ``average likelihood'' (a single real number) for the latter.

It is sometimes more convenient to work with ratios of model probabilities,
particularly when there is a special ``default'' model.  The ratio
of the probability for model $M_i$ to that for model $M_j$ is called
the odds in favor of $M_i$ over $M_j$.  We denote it by $O_{ij}$.  Using
Bayes's theorem, we can write the odds as
\begin{eqnarray}
O_{ij} &=& {p(M_i | D,I)\over p(M_j | D,I)}\nonumber\\
  &=&  {p(M_i | I) \over p(M_j | I)}
      {p(D | M_i,I) \over p(D | M_j,I)}\nonumber\\
  &\equiv& {p(M_i | I) \over p(M_j | I)}\; B_{ij},\label{Oij-def}
\end{eqnarray}
where the first factor is the prior odds ratio, and the second factor
is called the {\it Bayes factor}.  The Bayes factor is simply the ratio
of the likelihoods of the models.  Note that the normalization constant
in equation~(\ref{BT-mod}), $p(D | I)$, drops out of the odds ratio.
When the prior odds
does not strongly favor one model over another, the Bayes
factor can be interpreted just as one would interpret
an odds in betting; Table~\ref{table:B} summarizes the 
interpretation recommended in the extensive review of Bayes factors
by Kass and Raftery \cite{KR-95}.

An important aspect of Bayesian model comparison is
that the calculation of model likelihoods
implements an automatic and objective posterior ``Ockham's Razor,'' leading
one to prefer simpler models unless the data provide substantial evidence
in favor of a more complicated alternative, even when the rival models
are assigned {\it equal} prior probabilities.
In frequentist statistics, one commonly uses ratios of
maximum likelihoods to compare models.
However, more complicated models
almost always have higher likelihoods than simpler models, so
more complicated models are only accepted if the maximum likelihood
ratio in their favor exceeds some subjectively specified critical amount,
expressing a subjective {\it prior} preference for simplicity.  But Bayesian
methods compare averaged likelihoods, not maximum
likelihoods, and
tend to favor simpler models even when simple and complicated models
are assigned equal prior probabilities \cite{Gull-88,TJL-90,MacKay-92,JB-92}.

We can better understand the distinction between Bayesian and
frequentist model comparison and the nature of the Bayesian posterior
preference for simplicity by writing the model likelihood as the
product of the maximum parameter likelihood used in frequentist model
comparison, and an additional {\it Ockham factor}.
We thus implicitly define the Ockham factor $\Omega_\theta$
associated with the parameters $\theta$ of model $M$ by
writing $p(D | M) \equiv \Lmax\Omega_\theta$, where $\Lmax$
is the maximum value of the likelihood
function, $\like(\theta)\equiv p(D |\theta,M)$.  Recalling
equation~(\ref{gl-def})\ for the average likelihood, this implies
\begin{equation}
\Omega_\theta = {1 \over \Lmax}\; \int d\theta\; p(\theta | M)\;
  \like(\theta).\label{ockham}
\end{equation}
Assuming, as is generally the case, that the
prior varies slowly compared to the likelihood, the integral in this
equation is approximately equal to
$p(\hat\theta | M)\int d\theta \like(\theta)$, where $\hat\theta$ is the
maximum likelihood value of $\theta$.  If we write the integral of the
likelihood function as the maximum likelihood value times a characteristic
width of the likelihood, $\delta\theta$, we find that,
\begin{equation}
\Omega_\theta \approx p(\hat\theta | M)\; \delta\theta.\label{ockham-app}
\end{equation}
We can write the value of the prior at $\hat\theta$ as
$p(\hat\theta | M) \approx 1/\Delta\theta$, where $\Delta\theta$ is a
characteristic width of the prior (if the prior is flat over
some range of size $\Delta\theta$, the approximation is exact).
Then we find that
\begin{equation}
\Omega_\theta \approx \delta\theta/\Delta\theta,\label{ockham-app2}
\end{equation}
the ratio of the
posterior range for the parameter to its prior range.  This quantity will
be less than one, and in this manner the Ockham factor penalizes the
maximum likelihood.  This penalty generally grows with the number of
parameters, and in this way model likelihoods implement a posterior
preference for simpler models with fewer parameters, even when the
models are considered equally probable a priori.  In this way
Bayesian model comparison favors models that best predict the
data, not only for the best-fit parameters (which after all are
known only a posteriori), but taking into account uncertainty
in the parameters.

It is worth emphasizing that these Bayesian calculations provide
probabilities for {\em models} (or ratios of such probabilities), in
contrast to the ``false alarm'' probabilities provided by conventional
frequentist significance tests, which are probabilities for {\em data}
(i.e., probabilities for data more extreme than observed).  This
fundamental difference leads to different interpretations for the
probabilities these procedures report.  In frequentist statistics, it
is common to consider a departure from the null hypothesis at a 5\%
significance level to be barely significant.  In contrast, if a
Bayesian calculation gives the null hypothesis a probability of 5\%
(i.e., a Bayes factor of 19 against the null), this is considered quite
significant evidence against the null (see Table~\ref{table:B}).
Indeed, one often finds that a Bayesian analysis of data discrepant at
the 5\% significance level produces a Bayes factor of order unity---the
Bayesian calculation is confirming the conventional interpretation of
this significance level by providing a quantity with a more straightforward
and intuitive interpretation.  Sellke, Bayarri, and Berger \cite{SBB-99}
provide further discussion of this issue, with guidelines for a
Bayesian interpretation of significance tests.

The integrals needed to calculate average likelihoods for Bayes factors
are often challenging.  In this work, we often use an asymptotic
approximation to the Bayes factor applicable when comparing two nested
models (i.e., the simpler model corresponds to the more complicated one
when additional parameters are set at default values).  The
approximation is known as the Bayesian Information Criterion (BIC) or
the Schwarz criterion \cite{KW-95}.  The BIC uses a Gaussian
approximation for calculating average likelihoods, and an ``automatic''
prior with a width roughly corresponding to the width of the individual
data factors in the likelihood.  The result is that the log Bayes
factor can be approximated as
\begin{equation} \ln B_{21} \approx \ln \left[{\cal
L}_2(\hat\theta,\hat\phi) / {\cal L}_1(\hat\theta)\right]
   - {1\over 2} m_\phi \ln N,\label{BIC}
\end{equation} 
where model 2 is the more complicated model, with additional parameters
$\phi$, $m_\phi$ is the dimension of $\phi$, and $N$ is the number
of data.  When the approximate
results warrant interest in an accurate Bayes factor, we use
adaptive quadrature to calculate average likelihoods \cite{L-99}.

The key ingredient in Bayesian parameter estimation and model
comparison is the likelihood function.  We now turn to calculation of
the likelihood function based on the neutrino data.  This requires us
to model the production of neutrinos at the supernova, their
propagation to Earth, their interaction with terrestrial detectors, and
the detection of the energetic charged lepton produced upon such
interaction.  The last step of this modeling chain is the most
complicated one, and the place where the differences between our
likelihood function and those appearing in some earlier analyses are
greatest.  We therefore treat it first.

\section{Modeling Neutrino Detection}
\label{sec:dtxn}

Our task in this section is to calculate the probability for the data
produced by a neutrino detector, given the charged lepton production
rate throughout the detector.
Before beginning the calculation, we first introduce a number of notational
conventions that will streamline the derivation.  We also review some
basic results on inhomogeneous Poisson processes (Poisson processes with
varying event rates) that play an important role in the derivation.
We presume the reader is familiar with the basic setup of neutrino detectors
(see, e.g., \cite{Totsuka} for a detailed description of the Kamiokande II
detector).

The ``input'' to our calculation is specification of the charged lepton
production rate throughout a detector.  This rate has two components.
First, there is a background
component due to particles entering the detector from cosmic ray interactions
or radioactive decay in the surrounding rock.  We also
formally include other sources of false triggers (such as noise in the
detectors) in the background rate.
Second, there is the physically interesting
signal component due to astrophysical neutrinos.  We presume here that
both rates are given.  In practice, the background rate is inferred from
measurements, and the signal rate is the result of modeling, as we describe
in the following section.

The KII, IMB, and Baksan detectors
most efficiently detect neutrinos through capture of electron antineutrinos
on protons, resulting in the production of an energetic positron.
Thus throughout this work we will refer to the charged leptons produced
by the astrophysical neutrino signal
as positrons, even though many of our results
apply equally well to detection of energetic electrons.  The background
component may be due to positrons, electrons, or muons.  To simplify
the discussion, we will refer to the production of a charged lepton
of any type as an ``event.''  One must be careful to distinguish
occurrence of an event from detection of an event:  not every event
that occurs is detected.

We use
$B(\rvec,\drxn,\epos)$ to denote the differential background rate, so that
$B(\rvec,\drxn,\epos)dV d\drxn d\epos dt$ is the probability that a
background event will occur in an infinitesimal time interval $dt$ in
a volume $dV$ at position $\rvec$ in the detector, with a direction
in the solid angle $d\drxn$ about the unit vector
$\drxn$, and with an energy in the interval $[\epos,\epos+d\epos]$.
We presume the background rate is
constant in time over the duration of the observations.  It is not
constant in space, however, because background events due to sources
in the surrounding rock appear preferentially near the detector walls.

We use $R(\drxn,\epos,t)$ to denote the differential
signal rate: the rate of production of positrons in the detector
per unit time, energy, and steradian due to interactions with
neutrinos from the supernova.  Unlike the background
rate, it is time-dependent.  However, it is constant throughout the detector
volume since the detectors are optically thin to neutrinos.
When we need the signal rate per unit volume, it is thus simply given by
\begin{equation}
R(\rvec,\drxn,\epos,t) = {R(\drxn,\epos,t) \over V},\label{R-vol}
\end{equation}
where $V$ denotes the detector volume.
The signal rate will depend on some parameters, which we collectively
denote by $\pars$.  The number and type of parameters depends on the
model for the signal rate;
later sections describe the various models we consider.  We are
seeking the dependence of the likelihood on $\pars$ (and, implicitly,
on the choice of a parameterized signal model).

We will often need quantities such as the background rate for events
of any direction and at any position, but with energy in $d \epos$.
This requires integration over the other intervals.  For brevity, we
simply collapse the argument list to indicate the necessary integrations.
For example, $B(\epos) \equiv \int dV \int d\drxn\,
B(\rvec,\drxn,\epos),$ and an unadorned $B$ is the total background
rate per unit time.
We adopt similar conventions for the signal rate, so that
$R(\epos,t) \equiv \int d\drxn\, R(\drxn,\epos,t)$, and $R(t)$ is the
total signal rate per unit time at time $t$.

Our earlier work, and that of
others using likelihood functions, attempted to calculate the likelihood by
considering the data to be the inferred energies and arrival times of detected
positrons (i.e., the ``best fit'' values as reported by the detector
teams).  However, the actual data is not a set of time-tagged energy
values, but is instead a more complicated time series of pulse heights
in the thousands of photomultipliers surrounding each detector that
allows us to infer (with uncertainty) the properties of detected
positrons.  Although
this time series is not publicly available, a more accurate likelihood
calculation results if we imagine it were available and try to calculate
the probability for such a time series given the signal and background rates
and detailed knowledge of the detector.

Accordingly, we let $D$ denote all the available data, reported as a
time series specifying the state of the instrument at regular intervals
separated in time by $\delta t$.  The duration of $\delta t$ is unimportant,
so long as it is small enough that no more than one event is ever seen
in an interval.  We separate the data into two groups,
{\it detection data}, $d_i$, specifying all the data associated with
detected event number $i$; and {\it nondetection data}, $\bar d_j$,
specifying that no triggered event happened in time intervals indexed
by $j$.  We always use $i$ to index quantities associated with detected
events.  In particular, $t_i$ denotes
the time of event $i$.  Similarly, we always use $j$ to index quantities
associated with nondetections.
In particular, we use $\delta t_j$ to denote the time interval
$[t_j, t_j+\delta t]$ associated with $\bar d_j$.

We will presume that, given the signal and background rates,
the probability for a detection
in some interval $\delta t$ is independent of whether an event was
detected in other time intervals.  This implies that the likelihood
function is simply the product of independent probabilities for
the detections and nondetections,
\begin{equation}
\like(\pars) = \left[\prod_{i=1}^\Nd p(d_i | \pars,M)\right]\;
     \prod_{j} p(\bar d_j | \pars,M),\label{L-dnd}
\end{equation}
where $\Nd$ is the number of detected events and $j$ runs over all
intervals for which no event was detected.  As will become apparent, the
number of nondetection intervals has no bearing on the analysis; only their
total duration matters.  Here we use the symbol $M$ to denote all of
the modeling assumptions needed to calculate the required probabilities,
including specification of the signal model discussed in the next section.

We presume that, given the rates, the probability for an event occurring
in any specified infinitesimal
interval of time, volume, direction, and energy is
independent of whether or not an event occurred in some other interval.
This implies that the probability for $n$ events occurring in
an interval of finite size is given by the Poisson probability,
\begin{equation}
p_n = {\bar n^n \over n!} e^{-\bar n},\label{Poisson}
\end{equation}
where $\bar n$ is the expected number of events in the interval, found
by integrating the relevant differential rate over the interval.

Let us focus attention on a particular $\delta t$ interval, and let
$\sprop^0$ denote the proposition asserting that no signal events
occurred in the time interval.  The probability for $\sprop^0$ is given
by equation~(\ref{Poisson}), with $\bar n$ equal to the signal rate integrated
over $\delta t$:
\begin{eqnarray}
p(\sprop^0 | \pars, M)
   &=& \exp\left[-\int_{\delta t} dt\, R(t)\right]\nonumber\\
  &\approx& e^{-R(t)\delta t}.\label{Pois-0}
\end{eqnarray}
To get the last line, we have assumed that $\delta t$ is much smaller
than the timescale over which the rate varies, so that the integral
over $\delta t$ is well approximated by $R(t)\delta t$, with $t$ equal
to any time in $\delta t$.  Similar equations hold for the probability
for no background event; since the background rate is presumed constant,
there is no $t$ dependence and the $\delta t$ product form is exact.

Let us now focus attention on some specified time interval, and
let $\sprop(\rvec,\drxn,\epos)$ denote the proposition
asserting that a single signal event occurred in the $\delta t$
interval under consideration, and that it
had a position, direction, and energy in $dV d\drxn d\epos$
about the point $(\rvec,\drxn,\epos)$.  We
write the probability for this proposition as
\begin{equation}
p(\sprop(\rvec,\drxn,\epos)  | \pars, M) dV d\drxn d\epos,\label{P-S}
\end{equation}
so that $p(\sprop(\rvec,\drxn,\epos)  | \pars, M)$ is a probability
{\it density}.  This
proposition is the conjunction (logical ``and'') of two simpler
propositions:  (1) one signal event occurred in
$dV d\drxn d\epos \delta t$; and (2) no other signal event occurred in
$\delta t$ with a different position, direction, or energy.
The Poisson probability for the first of these propositions is
\begin{equation}
\left({R(\drxn,\epos,t) \over V} dV d\drxn d\epos \delta t \right)\,
   \exp\left[-{R(\drxn,\epos,t) \over V} dV d\drxn d\epos \delta t\right].
   \label{Pois-1-1}
\end{equation}
The Poisson probability for the second is
\begin{equation}
   \exp\left[-\left(R(t) -{R(\drxn,\epos,t) \over V} dV d\drxn d\epos\right)
   \delta t\right].\label{Pois-1-2}
\end{equation}
The probability (density) for $\sprop(\rvec,\drxn,\epos)$ is the product of 
these, divided by the differential $dV d\drxn d\epos$, giving
\begin{equation}
p(\sprop(\rvec,\drxn,\epos) | \pars, M)
  = {R(\drxn,\epos,t) \over V}  \delta t \;
         e^{-R(t)\delta t}.\label{Pois-1}
\end{equation}
We can write the probability for occurrence of a single, specific
background event similarly, substituting $B$ for $R$.

We now have all the ingredients we need to derive the form of
the likelihood function.  But before doing so for
realistic data, we will do so for ideal data produced by
a fictional detector that detects every
positron whose energy is above some threshold, $\eth$, and
that measures the locations, directions, and energies
of detected events with negligible uncertainty.
We will also presume there is no background rate in this detector.
This calculation will make clear the origin of the most important terms
in the more accurate likelihood function.

\subsection{Idealized likelihood}

We begin by calculating the probability for ideal nondetection data.
This is simply the Poisson probability for seeing no events when the expected
number of events is
\begin{equation}
\bar n = \delta t\int dV\int d \drxn \int d\epos  \,
     \Theta(\epos-\eth) {R(\drxn,\epos,t_j) \over V}.
     \label{ideal-nbar}
\end{equation}
Here $\Theta(x)$ is the unit step function, equal to 1 when its argument
is nonnegative, and 0 otherwise.  Thus $\Theta(\epos-\eth)$
is the efficiency for detecting
events of energy $\epos$, which is either 1 or 0 for this idealized
detector.  The efficiency assures that only
the {\it detectable} positron rate---that above the threshold---contributes to
$\bar n$.  With these definitions, the nondetection probability is
\begin{equation}
p(\bar d_j | M, I)
  = \exp\left[- \delta t \int dV\int d \drxn \int d\epos\,
         \Theta(\epos-\eth) {R(\drxn,\epos,t_j) \over V}\right]. 
       \label{pnd-ideal}
\end{equation}

To calculate the detection probability, we will presume that the nearly
ideal detection data specifies that one event occurred in $\delta t_i$
with energy $\epos_i$, direction $\drxn_i$, and position $\rvec_i$,
each measured with negligible uncertainties $\delta \epos$, $\delta
\drxn$, and $\delta V$.  The probability for such a datum is simply the
Poisson probability that one positron is produced in a time interval
$\delta t$ at $t_i$ with properties in the specified ranges, multiplied
by the probability that no other positron be produced in the same
interval but at another detectable energy, direction, or position.  We
derived such a probability above, although with infinitesimal ranges
(see equation (\ref{Pois-1})).  Thus we can write down the result,
\begin{equation}
p(d_i | M,I)
  = \left(\delta t\delta\epos\delta \drxn \delta V \,
          {R(\drxn_i,\epos_i,t_i)\over V}\right) \,
   \exp\left[-\delta t \int dV\int d \drxn \int d\epos\,
         \Theta(\epos-\eth) {R(\drxn,\epos,t_i) \over V}\right]. 
\label{pd-ideal}
\end{equation}

Assembling the detection and nondetection probabilities according
to equation~(\ref{L-dnd})\ gives the idealized likelihood function,
\begin{equation}
\like_{\rm ideal} = (\delta t\delta\epos\delta \drxn \delta V)^\Nd\,
   \exp\left[-\int_T dt\int dV\int d \drxn \int d\epos\,
           \Theta(\epos-\eth) {R(\drxn,\epos,t) \over V}\right]\;
       \prod_{i=1}^\Nd {R(\drxn_i,\epos_i,t_i) \over V}.  \label{L-ideal}
\end{equation}
The time integral in the exponent is over the entire duration of the data
and arose from combining the integrals in equation~(\ref{pnd-ideal})\ from all
the nondetection intervals with the integrals in the exponents of
the detection probabilities.  The exponent is thus the total expected
number of detectable positrons.  In general,
this is different from the (integer-valued)
number of positrons actually detected.  When the parameters of the
model specifying $R(\drxn,\epos,t)$ allow its amplitude to be freely
adjusted, one can show that the parameter values that maximize the
likelihood make the expected number of positrons equal the actually
detected number.

\subsection{Realistic Likelihood}

Realistic data differs from the idealized data in three important
respects.  First, the threshold for detection is not an energy
threshold, but is instead specified in terms of the number of
triggered photomultipliers.  In terms of positron energy, the
threshold is thus ``blurry,'' since the number of photomultipliers
triggered by a lepton of a particular energy cannot be precisely
predicted.
Second, the energies of detected leptons are inferred with
considerable uncertainty.  Finally, the KII and Baksan detectors
have nonnegligible background rates, so that triggers occasionally result even
when no energetic lepton has been produced by a neutrino.  We present
a detailed derivation of the likelihood, accounting for these
complications, in Appendix~A.  Though the calculation is somewhat lengthy,
its result is easy to understand in the light of the idealized calculation
described above.  The full likelihood function can be written,
\begin{eqnarray}
\like(\pars)
  &=& \exp\left[
       -\int_T dt \int d\drxn \int d\epos\,
           \bar\eta(\drxn,\epos) R(\drxn,\epos,t)\right] \nonumber \\
  &\quad\times& \prod_{i=1}^\Nd
   \left[B_i + \int d\drxn \int d\epos\,\like_i(\drxn,\epos)\,
             R(\drxn,\epos,t_i)\right].
  \label{Ltot}
\end{eqnarray}
Comparing this likelihood function with the
likelihood based on idealized data given by equation~(\ref{L-ideal}) reveals
three important differences, each associated with a new factor in the
likelihood.

First, the integral of the signal rate
appearing in the exponent (i.e., the effective rate) here has
the \emph{volume-averaged detection efficiency}, $\bar\eta(\drxn,\epos)$,
in place of $\Theta(\epos-\eth)$.  The sharp
energy threshold is replaced by a smooth threshold, due to the fact that
the detector trigger criteria are not simple functions of the actual
event energy.  There is a possible directional dependence in this factor.

Second, the product term has a weighted integral of the signal
rate in place of the signal rate evaluated at the direction, energy, and
time of the event.  The weighting function, $\like_i$, is the
\emph{event energy and direction likelihood}---the probability for seeing
the event data, presuming the positron that produced the data came
from direction $\drxn$ with energy $\epos$.
This integral accounts for uncertainty in the
inferred directions and energies of events.  

Finally, the {\it event background rate}, $B_i$, appears in the
product terms.  This quantity is just a weighted integral of the
background spectrum, the weighting function being $\like_i(\drxn,\epos)$.
It is the rate of background events resembling event $i$.  Recall that we are
ultimately interested in the functional dependence of $\like(\pars)$ on
$\pars$, determined by the dependence of the signal rate on $\pars$.
If, for a particular event, $B_i$ is much larger than the signal
rate (for any interesting choice of $\pars$), then that event's term in
the likelihood will remain nearly constant---the event is effectively
eliminated from consideration.  But the full likelihood function does
this ``background subtraction'' in a smooth way, reducing the weight
of information from potential background events according to the
relative probability that they are due to the background rather than
the signal.

We must add one further complication to the likelihood function.  Each of
the detectors has a fixed, known dead time, $\tau$, associated with
every detected event.  The likelihood
function corrected for dead time is found simply by
subtracting $[\Reff(t_i) + \Beff]\tau$ for each event from the
exponent in  equation~(\ref{Ltot}).  Since the $\Beff\tau$ parts of
these terms are constants (independent of the choice of model or
parameters for the signal), for simplicity we drop them from the likelihood. 

A further dead time correction is required for the IMB experiment.  This
experiment actually triggers on many more events than are reported as neutrino
events.  Characteristics of these events allow them to be justifiably
neglected as background events
(essentially, the experiment team eliminates events with a very
high $B_i$ from the reported data).  However, they each have dead time
associated with them, and they are numerous enough that this dead time
must be taken into account. In principle, we could subtract
$[R(t) + B]\tau$ for each such event from the exponent in
equation~(\ref{Ltot}).  In practice, the times of these events are not reported,
and they are numerous enough that it is adequate to simply multiply the
exponent by the live time fraction,
$f = 1-B_{\rm nr}\tau$, where $B_{\rm nr}$ is the rate of
background events that are not reported.  For the IMB detector,
$B_{\rm nr} = 2.7$ s$^{-1}$ and $\tau = 0.035$ s, so that
$f = 0.9055$.  For the KII and Baksan detectors $f=1$, since all events
are reported.
The likelihood function corrected for dead time is thus,
\begin{eqnarray}
\like(\pars)
  &=& \exp\left[
       -f\int_T dt \int d\drxn \int d\epos\,
             \bar\eta(\drxn,\epos) R(\drxn,\epos,t)\right] \nonumber \\
  &\quad&\times \prod_{i=1}^\Nd e^{\Reff(t_i)\tau}\,
   \left[B_i + \int d\drxn \int d\epos \,\like_i(\drxn,\epos)\,
           R(\drxn,\epos,t_i)\right].
  \label{Ldt}
\end{eqnarray}
This is the complete likelihood function based on data from a single
detector.  To combine the information from different detectors, we simply
calculate $\like(\pars)$ for each and multiply.

\subsection{Likelihood for isotropic signals}

The complete likelihood function is somewhat more general than what we
need.  As we note in the following section, the signal rate due to neutrinos
from SN1987A is essentially isotropic.  Thus we can perform some of
the volume integrals above, simplifying the likelihood function.  We
calculated the more general likelihood above both in order to illustrate some
of the complications hiding behind the isotropic form we are about to find,
and because it should prove useful in analyzing data from future supernova
neutrino observations, for which the anisotropic component of the signal
may not be negligible.  The complete likelihood function may also be
useful for analyzing other data, such as that produced by observing
solar neutrinos.

For an isotropic signal rate,
\begin{equation}
R(\drxn,\epos,t) = {R(\epos,t) \over 4\pi}.\label{R-iso}
\end{equation}
Inserting this into equation~(\ref{Ldt})\ allows us to write the likelihood for
isotropic signals as
\begin{eqnarray}
\like(\pars)
  &=& \exp\left[
       -f\int_T dt\int d\epos \,
             \bar\eta(\epos) R(\epos,t)\right] \\
  &\quad&\times \prod_{i=1}^\Nd e^{\Reff(t_i)\tau}\,
   \left[B_i + \int d\epos \,\like_i(\epos)\,
           R(\epos,t_i)\right],
  \label{Ldt-iso}
\end{eqnarray}
where the volume- and direction-averaged efficiency (hereafter simply
the {\it average efficiency}) is given by
\begin{equation}
\bar\eta(\epos) \equiv \int {dV \over V} \int {d\drxn \over 4\pi}
       \eta(\rvec,\drxn,\epos),\label{etab-iso}
\end{equation}
and the volume- and direction-averaged event likelihood function (hereafter
the {\it event energy likelihood}) is given by
\begin{equation}
\like_i(\epos) \equiv \int {dV \over V} \int {d\drxn \over 4\pi}
    \like_i(\rvec,\drxn,\epos).\label{Limarg}
\end{equation}
This is the likelihood function used in the calculations reported here.
It is simpler than equation~(\ref{Ldt})\ in the sense that
the experiment teams need only report the one-dimensional functions,
$\bar \eta(\epos)$ and $\like_i(\epos)$, rather than their more detailed
two-dimensional versions.  Similarly, the analyst needs to perform simpler
integrals for the analysis.  But it is important to realize that these
functions are intrinsically more complicated than they appear; the apparent
simplification
here simply reflects the fact that the experiment team can perform some
of the required integrals once and for all.

\subsection{The reported data}

We have derived the form of the likelihood function presuming that
the entire data set, in the form of a complicated time series, is available.
However, the final likelihood function depends only on some summaries
of this data.  The nondetection data are summarized in the efficiency
function.  The detection data are summarized in the form of
an event likelihood function for each detected event.  For making
inferences about isotropic signal models, all that is required is the
average efficiency, $\bar \eta(\epos)$, and the event energy likelihood
functions, $\like_i(\epos)$.  In addition, the data duration, $T$, the
equivalent water mass, $M$, the dead-time, $\tau$, and
live fraction, $f$, must be specified for each detector.  Finally, the
event-averaged background rate, $B_i$, must be specified for each event.

For our calculations, we use the reported detector efficiencies for
$\bar\eta(\epos)$.
In Figure 1 we plot the average efficiencies for the KII, IMB, and
Baksan detectors \cite{KII-2,IMB-2,Baksan-2,KII-bg}.  It is clear
that the three detectors sample the signal quite differently.
It is perhaps worth emphasizing that we are interpreting
these as the volume- and direction-averaged efficiencies for the detectors.
This implies that, in principle, these are {\it not} the efficiencies
one should use when analyzing signal models with an anisotropic
component (as would arise if there were a significant electron scattering
component).  But in practice, symmetries may make the
differences between the direction-averaged and direction-dependent
efficiencies negligibly small.  For example, electron scattering events
produced near the side of the detector closest to the source
are more likely to be
detected than those produced near the far side, since the latter will
produce Cerenkov photons preferentially directed out of the tank (thus
hitting few photomultipliers).  But the symmetry of the shapes of
detectors may result in near cancellation of the resulting
variations of the full efficiency upon integration over the detector
volume (this symmetry was broken for the IMB detector at
the time of the SN1987A observations, since power for a large number
of photomultipliers had failed).

Note that all of the reported average efficiency functions vanish
below some energy, $\epsilon_0$, that differs for each detector.
Formally, the efficiency probably
never identically vanishes (e.g., there is a small probability that
a low energy neutrino can trigger a large number of photomultipliers),
but it presumably becomes negligibly
small at the energy where the reported efficiency vanishes.

We also presume that the event energy likelihood
functions can be well-approximated by Gaussians,
\begin{equation}
\like_i(\epos) = C_i\,
     \exp\left[-{(\epos-\epos_i)^2 \over 2 \sigma_i^2}\right]
   \,\Theta(\epos-\epsilon_0),\label{Li-gauss}
\end{equation}
where $\epos_i$ is the reported ``best-fit'' energy for event $i$,
$\sigma_i$ is the reported uncertainty for the energy, and $C_i$ is a
normalization constant.  The $\Theta$ function appears for consistency
with $\bar\eta(\epos)$; it ensures that the event likelihood vanishes
at energies below the energy, $\epsilon_0$, where the reported
$\bar\eta(\epos)$ vanishes ($\epsilon_0$ is never closer to the peak
than two standard deviations).  The actual $\like_i$ function,
resulting from detailed fitting of the pattern of triggered
photomultipliers, is certainly not precisely a Gaussian.  But it must
be approximately Gaussian near its peak, since the leading order term
in the logarithm of $\like_i$ will be the second order, or Gaussian,
term.  The extent of the region over which this approximation is
adequate is impossible to ascertain without being provided the precise
likelihoods.  Since the detection teams have summarized their event
energy estimates with means and standard deviations, we have presumed
the approximation to be adequate to $\sim3$ standard deviations.

We note that normalization of $\like_i$ is simply a convention; $C_i$
can be changed to any value without affecting inferences, so long as
its value does not depend on the model parameters, $\pars$.  We choose
to normalize $\like_i$ with respect to $\epos$ (i.e., $C_i =
1/\sigma_i\sqrt{2\pi}$).  The only use we have made of this
normalization convention is in interpretation of $B_i$ in
equation~(\ref{Bi-def})\ as the rate of background events ``like''
event $i$.

Finally, we calculate $B_i$ for each event by integrating the product
of $\like_i(\epos)$ and an estimate of the background rate
spectrum $B(\epos)$.  The KII and Baksan teams have provided us
with measurements of $B(\epos)$ that we have used for this purpose;
the IMB experiment has a negligible background rate for events as
energetic as the reported events, so for the IMB events $B_i=0$.

In Table~\ref{table:det} we list the total background rates in the detectors,
as well as other detector characteristics
required for the likelihood calculation \cite{KII-1,KII-2,IMB-1,IMB-2,%
Baksan-1,Baksan-2,KII-bg,Chudakov}.
In Table~\ref{table:data} we list the
$\epos_i$, $\sigma_i$, and $B_i$ values for each event reported in
each detector.  For the KII and Baksan detectors, events are listed
that have not been included in other analyses.  Most earlier
analyses could not properly account for the background component,
and so had to exclude events suspected of being background events.
As already noted, the correct likelihood function weights events according
to the probability they come from the signal component, and so more
smoothly and consistently ``subtracts'' the background component from
the data.

The calculation of $B_i$ requires use of previously unpublished information,
and is based on some simplifying assumptions about the background rate.
Figure~2 shows the background rate measurements for
the KII and Baksan detectors that we use in the calculations.  Figure~2a shows
the KII empirical background rate spectrum \cite{KII-bg}, which is nonzero only
at low energies where the IMB efficiency is zero.  Figure~2b shows the
empirical background rate spectrum for the Baksan detector \cite{Chudakov};
it is significant even at high energies.
Most of the structure in the Baksan background rate spectrum can
be attributed to counting statistics, so the background spectrum we
actually use is the smooth curve in the figure, obtained by successively
performing a 3-point smoothing on the raw data points until a $\chi^2$
measure of the misfit between the data and the curve (a parabolic
interpolation of the smoothed data) is near its
expected value (two smoothings were used).
In fact, changing from the smooth curve to the raw
data has a negligible effect on our results, so the uncertainty in
the Baksan background spectrum need not be more carefully accounted for.
No 3-point smoothing of the KII spectrum could be tolerated, so we simply
interpolated between the measured values; again, the resulting background
uncertainty has a negligible effect on our results.
Note that both background spectra extend below the energies where
$\bar\eta(\epos)=0$ for each detector (c.f.\ Fig.~1).  The $B_i$ calculation
requires knowledge of $B(\rvec,\drxn,\epos)$ {\it before} ``filtering''
by the detection efficiency.  Thus it is best inferred by taking data
with no threshold criterion, resulting in background spectra extending
below the nominal instrumental cutoff.  Finally, a rigorous calculation
requires the background rate and event likelihoods as functions of
$\rvec$, $\drxn$, and $\epos$.  The available information is only a
function of $\epos$.  We have thus been forced to approximate
equation~(\ref{Bi-def})\ by
\begin{equation}
B_i = \int d\epos\; \like_i(\epos) B(\epos).\label{Bi-approx}
\end{equation}
This approximation ignores the position and direction information,
and thus could lead to over- or underestimation of $B_i$, depending
on the event location and direction, and the inhomogeneity and
anisotropy of the background.  Without detailed information about
the full event likelihoods and background rate, we cannot provide
a quantitative assessment of the quality of this approximation.
Nevertheless, it should be far superior to simple elimination of the
background events, which corresponds to the assumption of a very
high (formally infinite) $B_i$ value for the censored events.

\section{Positron Production Rate Model}
\label{sec:lepton}

In this section we describe how we model the lepton
production rate that was presumed given in the previous section.
As already noted, the detectors most efficiently detect neutrinos through 
capture of electron antineutrinos
on protons, resulting in the production of an energetic positron.
Thus we explicitly model only the emission of electron antineutrinos
by the supernova, and the production of positrons in the detector
(we later take into account the presence of neutrinos of other species
when inferring the total energy emitted by the supernova).
There are three steps in this modeling process.  First, we model the
electron antineutrino emission at the supernova.  Next, we model the
propagation of
this signal to Earth.  Finally, we model the interaction of these
neutrinos with neutrino detectors, leading to the production of energetic
leptons whose detection we have already modeled.

\subsection{Electron antineutrino emission model}

All of our signal models contain a component arising from the
cooling of the newly formed neutron star at the center of the supernova.
We refer to this part of the signal as the cooling component.
In addition to the cooling component, there may be a contribution to the
signal from hot, shocked accreting matter.  Such a contribution
arises in the delayed supernova mechanism.
We describe our models for these two components in turn.

\subsubsection{Cooling component}

Motivated by the results of numerical calculations of stellar
collapse \cite{BL-87,Burrows-88,JH-89,WW-86,BL-86,MWS-87,Bruenn87,%
Bruenn89a,Bruenn89b,MBHL,MB,W85,BW-85,Bethe90,Bruenn93a,BHF,%
Herant94,Janka93,JM-95,WMWW},
we assume that the newly formed neutron star emits electron
antineutrinos from a
neutrinosphere with a (possibly time-dependent) radius $R(t)$, and that
the instantaneous neutrino energy spectrum is well described by a thermal
Fermi-Dirac spectrum with time dependent temperature, $T(t)$, and
constant, nonnegative (usually zero) effective
``degeneracy parameter'', $\eta_\nu$ \cite{JH-89,JH-89b,Burrows-eta}.
The rate of emission of electron antineutrinos
with energies in the infinitesimal
range $[\enu,\enu+d\enu]$ is then $\dot N(\enu,\tem)d\enu$, with
\begin{equation}
\dot N(\enu,\tem) = \aem \enu^2 f[\enu,T(\tem)] r^2(\tem),\label{dn-emit}
\end{equation}
where $\aem$ is a constant with the value,
\begin{equation}
\aem = 4 \pi R^2 {g \pi c \over (hc)^3};\label{a-emit}
\end{equation}
$R = R(0)$ is the observed initial neutrinosphere radius;
$f$ is the Fermi-Dirac function,
\begin{equation}
f(\enu,T) = { 1 \over \exp\left({\enu\over T}-\eta_\nu \right) + 1};\label{FD}
\end{equation}
$\enu$ is the neutrino energy; $\tem$ is the emission time; and
$r(t) = R(t)/R(0)$.
The quantity, $g$, is the spin weight of the neutrino species in question;
$g=1$ for both massless and massive neutrinos \cite{BV-87}.  Here and
throughout this paper, temperature is measured in energy units.

We are presuming here that neutrinos are emitted isotropically.  Although
this is not expected to be rigorously true, current numerical
simulations indicate
the anisotropy of the emission resulting from the collapse of
a nonrotating star is not likely to be larger than of order 10\%.  The
effect of rotation on the neutrino emission (and on other features of
the collapse) remains an open question.

\subsubsection{Accretion component}

As accreted material flows through the stalled shock in the delayed
supernova mechanism, it is heated and produces $e^{\pm}$ pairs
\cite{Bethe90,Bruenn93a,BHF,Herant94,Janka93,JM-95,WMWW,Bethe93,Janka95,%
CBB-84}.
The accreted
material is neutron-rich (with neutron fraction $Y_n \approx 0.6$);
as a result, positron capture on neutrons produces electron
antineutrinos through the reaction $e^+ + n \rightarrow p + \nueb$.
Protons produced by this reaction (and those already in the flow)
can capture the thermal electrons to produce electron neutrinos
through the reaction $e^- + p \rightarrow n + \nu_e$.  These two
reactions proceed in local thermal equilibrium.  The resulting
electron antineutrino emission rate spectrum
per unit mass of emitting material is
\cite{Janka93,TS-75,Janka91,JM-89}
\begin{equation}
{\dot N(E) \over M_{\rm hot}}
  = A_a Y_n E^4 f(E,T_a),\label{acn-NdotM}
\end{equation}
where $M_{\rm hot}$ is the mass of hot accreted material emitting
the neutrinos, and $A_a$ is a constant with the value
\begin{equation}
A_a
 = {1 + 3 g_A \over 8}\, {\sigma_0 c \over m_n (m_e c^2)^2}\,
       {8\pi \over (hc)^3}.\label{Aacc-def}
\end{equation}
Here $g_A$ ($\approx 1.254$) is the coupling constant for axial
vector weak interactions, $m_n$ is the neutron rest mass, and
$\sigma_0$ ($= 1.7 \times 10^{-44}
\hbox{cm}^2$) is the standard weak interaction cross section.
This emission rate differs from equation~(\ref{dn-emit})\ primarily
through the factor $\sigma_0 E^2$ arising from the size and energy
dependence of the capture cross sections.  We always set $\eta_\nu=0$
for the accretion component.

To calculate the emitted spectrum,
we must multiply equation~(\ref{acn-NdotM})\ by the mass of hot material
emitting at any particular time, which we write as
\begin{equation}
M_{\rm hot}(\tem) = M_0 a(\tem),\label{Mhot-def}
\end{equation}
where $M_0$ is the maximum mass emitting during the event, and $a(\tem)$ is a
dimensionless function describing the temporal behavior of the
accretion emission, with $a(\tem) \le 1$.
We assume that the temperature of the emitting
material is constant in time, so that the electron antineutrino number
spectrum due to accretion can be written
\begin{equation}
\dot N(E,\tem) = A_a M_0 Y_n  E^4 f(E,T_a) a(\tem).\label{acn-Ndot}
\end{equation}

\subsection{Neutrino propagation}

If the distance to the neutron star is $D$, the neutrino number flux
per unit energy incident on detectors at the earth is
\begin{equation}
\Phi(\enu,\tdet) = {1 \over 4 \pi D^2} \dot N(\enu,\tem).\label{dn-inc}
\end{equation}
The times, $\tem$ and $\tdet$, are related by
\begin{equation}
\tdet = \tem + \Delta t(m_\nu,\enu) - \toff, \label{delay}
\end{equation}
where $\tdet \equiv 0$ for the first detected event, $\toff$ is
the (unknown) offset time between $\tdet = 0$ and the time of arrival of
the first neutrinos incident on the Earth, $m_\nu$ is the rest mass
of the electron antineutrino, and
\begin{equation}
\Delta t(m_\nu,\enu) = 2.57 \left({m_\nu \over {\rm eV}}\right)^2
\left({\enu \over {\rm MeV}}\right)^{-2} {D \over 50 \kpc} \quad \s.
\label{delta-t}
\end{equation}
A constant offset of $D/c$ has been dropped from equation~(\ref{delay}).

In our model, the flux of neutrinos at the earth as a function of detector time
is determined by specifying $T_c(t)$, $R$, and $r(t)$ for the cooling
component; $T_a(t)$, $M_0Y_n$, and $a(t)$ for the accretion component; and
$m_\nu$, and $\toff$.
If every detector had an accurate clock, we would need to specify only
a single $\toff$ parameter; it would represent the time between the detection
of the first neutrino detected by any detector and the unknown time
of arrival of the first neutrinos reaching the Earth.  However, accurate
absolute times are available only for those events detected by the IMB
detector.  Thus, a separate $\toff$ parameter must be considered for each
detector.  With the exception of Abbott, DeR\' ujula, and Walker \cite{ADW},
previous investigators have included at most only one such
parameter \cite{BL-87,Spergel-87,SB-87}.

\subsection{Charged lepton production}

Once emitted neutrinos reach the Earth, their detection involves two
distinct processes.  First, a neutrino must somehow produce
an energetic charged lepton in the detector.  Second, the Cer\v enkov
light produced in the detector by this charged lepton must be detected.  We
refer to these processes as the lepton production and detection processes,
respectively.  We have already discussed the detection
process in detail in the previous section; we thus conclude this section
by describing charged lepton production.  Often, these two
processes have not been
distinguished \cite{KST,BPPS,Adams,Krauss,Spergel-87,BS-88,ADW,SB-87,%
Burrows-mass,LoSecco,LY-89,PS-89,JH-89,JH-89b,Suz89}.

The dominant charged lepton production process is positron production
resulting from the absorption of electron antineutrinos ($\nueb$) on
free protons through the reaction,
\begin{equation}
\nueb + p \to e^+ + n.\label{rxn}
\end{equation}
All other processes have cross
sections at least an order of magnitude below the $\nueb-{\rm p}$ absorption
cross section \cite{KII-2,KST,BV-87}, and so we neglect them,
confining our analysis to this single species of neutrino.
The angular distribution of positrons produced by proton capture is nearly
isotropic \cite{Vogel}.
To a good approximation, we treat it as being isotropic,
allowing us to use the likelihood function for
isotropic rates described in the previous section.
The energy-dependent cross section for equation (10.12) has been calculated by,
for example, Tubbs and Schramm \cite{TS-75}.  It can be written as,
\begin{equation}
\sigma_{\nu p}(\enu) = 1.35 \sigma_0 \left({\enu \over m_e c^2}\right)^2
\kappa(\enu),\label{x-sxn}
\end{equation}
where $m_e$ is the electron rest mass, and
$\kappa(\enu)$ is a dimensionless function describing corrections to the
$\enu^2$ energy dependence.  This function is
\begin{equation}
\kappa(\enu) = \left(1 - {Q \over \enu}\right)
  \left[1 - {2Q \over \enu} + {Q^2 - m_e^2 \over 
     \enu^2}\right]^{1/2},\label{k-def}
\end{equation}
where $Q$ ($= 1.29$ MeV) is the neutron-proton mass deficit;
note that we have ignored small terms due to neutron recoil, and Coulomb
and radiative corrections \cite{BV-87,Vogel}.

If there are $N_{p}$ free protons in a detector, then its total cross
section is $N_{p} \sigma_{\nu p}$.  Using equation~(\ref{dn-inc})\ for
the incident neutrino flux, and considering first the cooling
component emission given by equation~(\ref{dn-emit}),
the capture rate per unit energy is
\begin{equation}
R^{\rm cap}(\enu,\tdet)
  = 1.35 \sigma_0 {\aem\over 4\pi D^2} N_{p} (m_e c^2)^2
    \left({\enu \over m_e c^2}\right)^4 f[\enu,T(\tem)] \kappa(\enu)
        r^2(\tem).\label{dn-cap1}
\end{equation}
To parameterize the amplitude we introduce the quantity,
\begin{equation}
\alpha \equiv {R \over 10 \km} \left({D \over 50 \kpc}\right)^{-1}
\sqrt g.\label{alpha-def}
\end{equation}
Other investigators have parameterized the amplitude in a more complicated
way.  The choice of $\alpha$, rather than the energy flux
$F$ \cite{BPPS,Spergel-87,SB-87,PS-89},
the total emitted number of electron antineutrinos $N$ \cite{BS-88,LML},
or the total neutrino luminosity $L$ \cite{Krauss}, permits straightforward
inferences about the neutrinosphere radius $R$.  The parameter
$\alpha$, or its equivalent, is as important as the remaining
parameters that describe the neutrino detection rate.  Unfortunately,
this parameter, or its equivalent, was fixed at its best-fit value in
some studies \cite{ADW,LY-89}, thereby artificially constraining the allowed
values of the remaining parameters.

Using $\alpha$, equation~(\ref{dn-cap1})\ can be written as
\begin{equation}
R^{\rm cap}(\enu,\tdet)
   = 1.22 \times 10^{-5} \alpha^2 \left({\mwat \over {\rm kton}}\right)
    \left(\enu \over {\rm MeV}\right)^4
    f(\enu, \tem) \kappa(\enu) r^2(\tem)
     \;{\rm MeV}^{-1}\,{\rm s}^{-1},\label{dn-cap}
\end{equation}
where $\mwat$ is the effective water mass of the detector.

We can calculate the capture rate for electron antineutrinos
from an accretion component in exactly the same manner as we did for
the neutrinos produced by cooling, starting with the spectrum
given by equation~(\ref{acn-Ndot}).  The resulting capture rate is
\begin{equation}
R^{\rm cap}(\enu,\tdet)
   = 2.14 \times 10^{-4} \mu \left({\mwat \over {\rm kton}}\right)
    \left(\enu \over {\rm MeV}\right)^6
    f(\enu, T_a) \kappa(\enu) a(\tem)
     \;{\rm MeV}^{-1}\,{\rm s}^{-1},\label{dn-cap-acn}
\end{equation}
where $\mu$ is a dimensionless parameter setting the amplitude
of the accretion emission given by
\begin{equation}
\mu = \left(M_0 \over M_\odot\right) \left(Y_n \over 0.6\right)
       \left(D \over 50 \kpc\right)^{-2}.\label{mu-def}
\end{equation}
The total capture rate in a model with such an accretion component is
simply the sum of the rates given in equations (\ref{dn-cap})\ and
(\ref{dn-cap-acn}).

Ignoring a small (angle dependent) term due to neutron recoil
\cite{BV-87,Vogel}, each captured electron
antineutrino produces a positron with energy $\epos = \enu - Q$.
The positron production rate per unit energy is thus the capture
rate evaluated at $\enu=\epos+Q$,
\begin{equation}
R(\epos,\tdet) = R^{\rm cap}(\epos+Q,\tdet).\label{dn-pos}
\end{equation}
This is the function needed to evaluate the likelihood function using
the formula developed in the previous section.

\section{Neutrino Emission Models}
\label{sec:models}

We have considered fourteen different models for electron antineutrino
emission from the supernova.  These fall into three groups.
First are seven single-component cooling models inspired by numerical
collapse calculations studying the  prompt supernova mechanism \cite{WW-86}.
These models have either constant or
monotonically decreasing neutrinosphere temperature, constant or
monotonically decreasing neutrinosphere radius, and a possibly nonzero
neutrino degeneracy parameter, $\eta_\nu$.  Next are five models
inspired by collapse calculations that produce delayed supernovae by means of
shocks that are revived by neutrino heating \cite{WW-86}.  These
models include both a cooling component and a component due to material
accreting through the stalled supernova shock.
Finally, we consider two {\it ad hoc} models with a distinctly different
structure that could be implied by the data:  temperatures
and fluxes that first increase and then decrease.  These models have
from three to six parameters describing the neutrino emission,
in addition to the required detector offset times.

We emphasize that our models are phenomenological, and are not meant
to reproduce in detail the behavior of any particular numerical calculation.
Given the sparseness of the data, excessive detail in the models seems
unwarranted.  Nevertheless, our analysis demonstrates that the data are
capable of distinguishing among the models we have studied, some of which
are considerably more structured than those studied previously.

\subsection{Cooling models}

{\em (a) Constant temperature.}
The simplest model we consider is emission from a constant
temperature, constant radius neutrinosphere over a time $t_{\rm burst}$,
after which emission ceases:
\begin{equation}
T(t) = \left\{ \begin{array}{ll}
  T_0 & {\rm for} \quad 0 < t < t_{\rm burst}, \\
   0 & {\rm otherwise};
  \end{array}\right.
  \label{T-const}
\end{equation}
\begin{equation}
r(t) = 1.\label{r-const}
\end{equation}
This is the simplest model that can fully characterize the data.  It
has a single energy scale that is determined by the energy distribution
of the events, a single time scale that is determined by their temporal
extent, and an amplitude, $\alpha$, that is determined by the number of
events seen.

{\em (b) Exponential dilution.}
Next we consider a model with constant neutrinosphere temperature,
but exponentially decreasing neutrinosphere radius:
\begin{equation}
T(t) = T_0\label{T-cntxn}
\end{equation}
\begin{equation}
r(t) = \exp(-t/2\tau).\label{r-cntxn}
\end{equation}
Here $\tau$ is the luminosity time constant.
As with the constant temperature model, this model has the smallest
number of parameters that can fully characterize the data.  However,
this model allows us to test the hypothesis that the flux
of the emitted neutrinos decreased in time.  Moreover, the flux
produced by this model
bears some similarity to that of some collapse calculations
in which the color temperature of the emitted neutrinos stays
roughly constant over timescales $\sim 10$ s, with the flux decreasing
due to dilution as the opacity in the layers below the neutrinosphere
gradually shifts from being absorption dominated to being scattering
dominated \cite{JH-89,JH-89b,Janka95,Janka91,Janka91b}.
In this case, $r(t)$ is more correctly interpreted as
a dilution factor than an actually decreasing physical radius; this
is why we term this model ``exponential dilution.''

{\em (c) Exponential cooling.}
The next model we consider is an exponential cooling model
described by the equations,
\begin{equation}
T(t) = T_0 \exp(-t/4\tau),\label{T-expc}
\end{equation}
\begin{equation}
r(t) = 1.\label{r-expc}
\end{equation}
Again, $\tau$ is the luminosity time constant.
As with the previous two models, this model has the smallest
number of parameters that can fully characterize the data.  However,
this model allows us to test the hypothesis that the characteristic
energy of the emitted neutrinos varied in time.  This model
exhibits the most basic characteristics of those numerical calculations
of the cooling of the neutron star
that show smoothly decaying neutrinosphere temperatures and a neutrinosphere
radius that falls to within $\approx 10$\% of its asymptotic value within about
0.5 sec \cite{BL-86}.

{\em (d) Exponential cooling and dilution.}
Our next model combines exponential dilution and exponential
temperature decay:
\begin{equation}
T(t) = T_0 \exp(-t/4\tau_T),\label{T-expcc}
\end{equation}
\begin{equation}
r(t) = \exp(-t/2\tau_r).\label{r-expcc}
\end{equation}
This model, with four parameters, allows us to test whether the data
provide evidence for evolution of both the characteristic energy
of the neutrinos and the radius of the neutrinosphere.

{\em (e) Displaced power-law cooling.}
For the exponential cooling model,
the cooling timescale, $- T / \dot T = 4\tau$, is constant
in time.  As a next level of complexity, we consider a
model with constant radius for which the cooling timescale
increases linearly in time, that is, we set
\begin{equation}
-{T \over \dot T} = 4\tau\left(1 + {t \over 4\gamma\tau}\right).\label{TTdot}
\end{equation}
Here $\gamma$ is the timescale on which the cooling
{\it rate} changes, in units of the initial cooling timescale $\tau$.
The temperature remains roughly constant for a time $4\gamma\tau$, and then
decreases like a power law afterward.
Such a model is capable of qualitatively describing the results
of several cooling calculations, including both those that show
neutrino emission with a temperature that decays monotonicly from
early times \cite{BL-86}, and those that show a roughly constant
temperature for times $\sim 10s$, followed by a monotonic decrease.
Also, such a growing  timescale might better account for
the three late events detected by KII.  Solving for
$T(t)$, the functions defining this model are,
\begin{equation}
T(t) = T_0 \left( 1 + {t \over 4 \gamma \tau} \right)^{-\gamma},\label{T-pl}
\end{equation}
\begin{equation}
r(t) = 1.\label{r-pl}
\end{equation}
This is the ``displaced power law'' cooling model of Bludman and
Schinder \cite{BS-88}.
It has one more parameter than the exponential cooling model, $\gamma$.
As $\gamma \rightarrow \infty$, this
model becomes simple exponential cooling.  We exclude values of
$\gamma$ less than $1/3$ as unphysical, because they imply an infinite
number of emitted neutrinos.

{\em (f) Nonzero degeneracy parameter.}
Monte Carlo calculations of neutrino radiation transport in the cooling
neutron star \cite{JH-89,Burrows-eta} indicate that the emitted
neutrino spectrum is nonthermal
and well modeled by a Fermi-Dirac distribution with positive neutrino
``degeneracy parameter,'' $\eta_\nu$.  Thus we consider an additional
model, the exponential cooling model described by equations (\ref{T-expc})\
and (\ref{r-expc}), but
with $\eta_\nu$ allowed to vary.  This fourth parameter allows us
to test whether there is evidence in the data for a nonthermal neutrino
spectrum.

{\em (g) Delayed exponential cooling.}
Finally, we consider emission at a constant temperature for a time
$t_{\rm dur}$, followed by exponential decay, with a constant neutrinosphere 
radius throughout:
\begin{equation}
T(t) = \left\{ \begin{array}{ll}
   T_{\rm max} & {\rm for\ } t < t_{\rm dur},\\
   T_{\rm max} \exp[-(t-t_{\rm dur})/4\tau]
    & {\rm for\ } t > t_{\rm dur};
  \end{array}\right.
\label{T-dexp}
\end{equation}
\begin{equation}
r(t) = 1.\label{r-dexp}
\end{equation}
This model has only one more parameter than the exponential cooling model,
the duration, $t_{\rm dur}$, of the constant temperature period.  It has
a ``plateau'' period that might account for enhanced emission at early
times without requiring an accretion component.

\subsection{Models with accretion and cooling components}

The above models were inspired by calculations studying the prompt supernova
mechanism, which produce neutrinosphere temperatures and neutrino luminosities
that decrease monotonically in time.  In contrast, in the delayed
scenario neutrino emission arises both from the cooling core, and
from material that is heated as it passes through the stalled
shock that will eventually produce the supernova explosion.
To see if there is significant evidence in the data for such behavior,
we considered five models that combine a cooling flux modeled with
one of the behaviors described above, with an accretion flux described
by one of two alternative models.

\subsubsection{Models with truncated accretion}

For four of our two-component models,
we model the accretion flux as that from accreted matter
with constant temperature, $T_a$,
with the amount of emitting matter proportional to
\begin{equation}
a(t) = {\exp\left[-\left(t \over \tau_a\right)^{10}\right] \over
          1 + {t \over 0.5s}}.
\label{a-trunc}
\end{equation}
The denominator is meant to mimic the properties of the accretion signal
observed in numerical calculations of the delayed scenario, in which
accretion is roughly constant for a few tenths of a second, and then
decreases like $t^{-1}$ until the supernova shock is revived and the
accretion ceases.
The form of the exponential factor is chosen to be
nearly constant for times less than $\tau_a$,
and then drop exponentially very quickly thereafter, thus
implementing a smooth truncation of the accretion.
We add to this accretion flux a variety of cooling fluxes, as follows.

{\em (h) Exponential cooling and truncated accretion.}
We will find the exponential cooling model to be the most interesting
single component model, so our first accretion model has a cooling flux
with an exponentially decreasing temperature, $T_c(t)$, at constant radius,
\begin{equation}
T_c(t) = T_{c,0} \exp(-t/4\tau_c),\label{T-c}
\end{equation}
\begin{equation}
r_c(t) = 1.\label{r-c}
\end{equation}
This model is thus a ``bridge'' between the single component models and
models with accretion.

{\em (i--k) Displaced power-law dilution/cooling and truncated accretion.}
A more accurate model for the cooling behavior observed in numerical
calculations of the delayed scenario is a displaced power law, with
the temperature or dilution factor roughly constant for a timescale
of order 10 s, and then falling.  Accordingly, we model the cooling
component with the following temperature and radius factor time dependences:
\begin{equation}
T_c(t) = {T_{c,0} \over (1 + {t \over \tau_c})^n},\label{T-dpl}
\end{equation}
\begin{equation}
r_c(t) = {1 \over (1 + {t \over \tau_c})^m}.\label{r-dpl}
\end{equation}
We consider three such models.  For model (\emph{i}), we set $n=1$ and
$m=0$.  For model (\emph{j}), we set $n=0$ and $m=1$.  For model
(\emph{k}) we set $n=1$ and $m=1$.  These models let us explore to what
extent the cooling component in two component models can be explained
by decreasing temperature or increasing dilution.

\subsubsection{Power-law accretion}

{\em (l) Exponential cooling and power-law accretion.}
In some recent calculations, the accretion rate decays smoothly, and
is roughly proportional to $t^{-1}$ during the first several tenths
of a second after collapse\cite{Janka-pc}.  To model emission
from these calculations, we add to an exponential cooling flux like that in
model ({\em c}) an accretion flux with temperature $T_a$ and
temporal behavior given by
\begin{equation}
a(t) = {1 \over \left(1 + {t\over\tau_a}\right)^\delta}.\label{a-pl}
\end{equation}
Thus the mass of emitting material is roughly constant over a
timescale $\tau_a$, after which it decreases like a power law with
index $-\delta$.  We fix $\delta$ at $1.5$.  This shallow value gives
temporal behavior roughly consistent with the $t^{-1}$ behavior observed
at early times in calculations, but avoids the logarithmic integral
divergence associated with a pure $t^{-1}$ power law.

\subsection{Other models}

{\em (m) Thermal rise and fall.}
All of the models described above have temperatures and fluxes that never rise.
Our final two models are single component models that depart from this
pattern.  The first has a linear temperature rise, followed by exponential
cooling, with the neutrinosphere radius constant throughout:
\begin{equation}
T(t) = \left\{ \begin{array}{ll}
   T_0 {t \over t_{\rm rise}} & {\rm for\ } t < t_{\rm rise},\\
   T_0 \exp[-(t-t_{\rm rise})/4\tau] & {\rm for\ } t > t_{\rm rise};
  \end{array}\right.
  \label{T-rf}
\end{equation}
\begin{equation}
r(t) = 1.\label{r-rf}
\end{equation}

{\em (n) Thermal rise and fall with contraction.}
The second has the same thermal evolution as the first, and a neutrinosphere
radius that contracts linearly during the period of rising temperature,
and remains constant thereafter:
\begin{equation}
T(t) = \left\{ \begin{array}{ll}
   T_0 {t \over t_{\rm rise}}
      & {\rm for \ } t < t_{\rm rise},\\
   T_0 \exp[-(t-t_{\rm rise})/4\tau] & {\rm for\ } t > t_{\rm rise};
  \end{array}\right.
   \label{T-rfc}
\end{equation}
\begin{equation}
r(t) = \left\{ \begin{array}{ll}
  1 + a\left(1-{t \over t_{rise}}\right)
     &{\rm for\ } t < t_{\rm rise},\\
  1 {\rm \quad for \quad} t > t_{\rm rise}.
  \end{array}\right.
  \label{r-rfc}
\end{equation}

In these models, the neutrino number flux can rise and sharply peak at some
time $\lesssim t_{\rm rise}$ with a temperature $\lesssim T_0$,
and fall slowly afterward, potentially accounting for the large number
of low energy events seen within the first second of the KII burst without
requiring an accretion component.

\section{Best-Fit Parameter Values and Model Comparison}
\label{sec:results}

In this section we briefly summarize some of the results of our analysis
of the models just described.  We present best-fit parameter values
for all the models.  We identify the exponential
cooling model as the most successful single-component model, and the
displaced power law cooling plus truncated accretion model as the most
successful two-component model; we consider these models further in
the following two sections.  We also discuss the consistency of the
Baksan data with the KII and IMB data, and the effect of proper treatment
of background on our inferences.

\subsection{Best-fit parameter values}

We list the best-fit values for the parameters of our
single component cooling models in Table~\ref{table:one}.
Also listed are the values of the neutron star binding energy
implied by the best fit parameters, calculated according to
\begin{equation}
{E_b \over 10^{53} \erg} = 3.39 \times 10^{-4} \alpha^2
  \left({D \over 50 \kpc}\right)^2
\int_0^\infty dt\, \left(T(t)\over {\rm MeV}\right)^4 r^2(t).\label{Eb-def}
\end{equation}
This expression assumes three flavors (six species) of neutrinos
and antineutrinos,
with each carrying away an equal part of the binding energy;
numerical calculations show this to be a reasonable
ap\-prox\-i\-ma\-tion
\cite{BL-87,Burrows-88,JH-89,WW-86,BL-86,MWS-87,Bruenn87}.
The tabulated values of $E_b$ and $R = 10\alpha (D/50 \kpc)$~km
were calculated assuming $D = 50 \kpc$, a value consistent with recent
measurements of the distance to SN 1987A based on observations of
its circumstellar ring \cite{Panagia,Gould}.

In Table~\ref{table:one}, four cooling models are not listed because they have
best-fit parameter values that make them identical to one of the listed models.
The model combining exponential cooling and exponential dilution has
a best-fit temperature timescale of $\tau_T = \infty$; this corresponds
to the pure exponential dilution model.  The remaining unlisted models
all have best-fit parameters that make them equivalent to the exponential
cooling model.  That is, all additional parameters have best-fit values of
zero.  These models are:  the exponential decay model with neutrino degeneracy
parameter, $\eta_\nu$; the delayed exponential decay model (equations
(\ref{T-dexp})\ and (\ref{r-dexp}));
and the linear temperature rise, exponential temperature decay model
(equations \ref{T-rf}\ and \ref{r-rf}).
Also, the best-fit values of the detector offset times for
models with neutrino fluxes and temperatures that never increase are
necessarily zero, and are not listed in Table~\ref{table:one}.

We present the best-fit values for the parameters of our two-component
models in Table~\ref{table:two}.  The radii listed are those associated
with the cooling component, so that $R = 10\alpha (D/50 \kpc)$~km, as
in Table~\ref{table:one}.  The binding energies are the sum of the
binding energy associated with the cooling component (given by equation
[\ref{Eb-def}]) and the energy of the neutrinos emitted by the
accretion component, calculated according to
\begin{equation}
{E_a \over 10^{53} \erg} = 4.14 \times 10^{-2} \mu
      \left({D \over 50 \kpc}\right)^2 
       \int_0^\infty dt \left(T(t)\over {\rm MeV}\right)^6 a(t).\label{Ea-def}
\end{equation}
The $E_a$ contribution is also listed separately, in parentheses.
Equation (\ref{Ea-def})\ assumes that equal energy is emitted in electron
neutrinos
and electron antineutrinos, and that negligible energy is emitted in
neutrinos of other flavors since thermal production of
mu and tau particles in the accreted matter is suppressed due to the large
masses of these leptons.  This suppression is not complete, so the actual
accretion energy may be slightly higher than $E_a$.

Since the neutrino flux and temperature never increase for any of the
two-component models, the best-fit offset times are necessarily zero,
and are not listed in Table~\ref{table:two}.

In Table~\ref{table:two}, we have set $\mu=0.5$ for all accretion
models.  As we will demonstrate in Sec.~\ref{sec:acn}, 
the likelihood function for
the two-component models varies rather weakly with $\mu$, and has a
very broad maximum at values of $\mu$ significantly larger than one.
The maximum likelihood values are significantly larger than expected
theoretically, and imply an amount of accreted material that would lead
to formation of a black hole on the timescale of $t_a$, which is
clearly incompatible with the detection of neutrinos at later times.
We thus set $\mu=0.5$ for these models, this being a characteristic
value in numerical calculations.  This value is not excluded by the
broad likelihood function; in essence, we are using prior information
to fix a parameter not usefully constrained by the data.

Two sets of best-fit parameters are presented in each table:  values
resulting from a joint analysis of all three data sets, and values
resulting from a joint analysis of only the KII and IMB data.  The
latter are included for comparison with previous studies that did not
include the Baksan data, and to give an indication of the consistency
of the Baksan data with the KII and IMB data; we comment further on
this later in this section.  Since we find all the data to be
consistent, all of our discussion of parameter values and model choice
will be based on results from the KII-IMB-Baksan analysis, except where
noted.

We defer comparison of the parameter values with theoretical
expectations until after the best models are identified and further
studied.

\subsection{Model comparison}

Tables~\ref{table:one} and \ref{table:two} also list the value of the
maximized likelihood function for each model.  The actual value of the
maximum likelihood is not directly meaningful; however, when models are
nested, the ratio of the maximum likelihoods of competing models can be
used to evaluate the BIC approximation to the Bayes factor, and it can
be used for a frequentist likelihood ratio significance test.  For
convenience, the likelihood values have been scaled to the value found
for the exponential cooling model.  Note that the BIC penalizes models
according to the number of their parameters, so that the (approximate)
Bayes factor can favor a complicated model only if its maximum
likelihood is larger than that of a simpler competitor.  Likelihoods
for calculations with and without the Baksan data have been scaled
separately; these two classes of calculations cannot be compared with
each other because they use different sets of data.

All of the models have scaled likelihoods of order unity or greater,
with the exception of the constant temperature and radius model, whose
scaled maximum likelihood is $\sim 10^{-5}$.  Further, models with
phases of constant or increasing luminosity all have best fit
parameters indicating that the duration of any such phase is short,
$\lesssim 1$~s.  Thus there is strong evidence in the data for a
neutrino luminosity that monotonically decreases throughout most of the
burst, and the constant temperature and radius model can be rejected.

The simplest of the remaining single-component cooling models are the
exponential dilution model and the exponential cooling at constant
radius model, each of which describe the neutrino emission with three
parameters.  The likelihood of the dilution model is slightly larger
than that of the cooling model.  Also, the model combining cooling and
dilution has a best-fit cooling timescale $\tau_T=\infty$, indicating a
preference for dilution over cooling.  However, this preference is
weak; the maximum likelihood for the dilution model is only $1.77$
times higher than that for the cooling model.  Thus although the data
indicate a neutrino flux that decreases significantly over timescales
$\sim 10$ s, they cannot conclusively distinguish dilution from cooling
as the cause for the flux decrease in a single-component model.  We
consider the exponential cooling model to be the more viable of these
models because the characteristic radius and luminosity timescale
associated with the dilution model are much more difficult to reconcile
with theoretical expectations than are the characteristics of the
cooling model.

The remaining two cooling models (displaced power law cooling, and
thermal rise and fall with contraction) have maximum likelihoods larger
than that of the exponential cooling model.  However, they are both
more complicated than this model, requiring four or more parameters (in
addition to the three offset times) to describe the neutrino emission.
The BIC penalty for additional parameters (see eqn.~\ref{BIC})
corresponds to a factor of $1/5.4$ per extra parameter for the
KII-IMB-Baksan fits, and $1/4.9$ per extra parameter for the KII-IMB
fits.  The approximate Bayes factors for the two more complicated
models are thus approximately unity or less.  In addition, more careful
accounting of our prior information about properties of the neutron
star formed by the supernova would likely decrease the Bayes factors
for the complicated models even further. This can be seen as follows.

The likelihood for each model is the prior-weighted average of the
likelihood function for its parameters.  The exponential cooling model
has best-fit parameter values that imply binding energies and radii
significantly in excess of those expected for a neutron star, even
presuming the stiffest acceptable equation of state and substantial
expansion due to the high temperature and lepton fraction of the
nascent neutron star.  (We assess this discrepancy more fully in
the following section.)  Its model likelihood will therefore be small,
since the prior probability in the vicinity of the maximum likelihood
peak will be negligible.  But the best-fit radii and binding energies
for the two remaining cooling models are significantly larger still.
We thus expect their model likelihoods to be smaller even than that
for the exponential cooling model, both because their prior
probabilities are spread out over more dimensions, and because the
prior in the vicinity of the mode for each model will be smaller than
that in the vicinity of the subspace of each model corresponding to
exponential cooling.  Essentially, the exponential cooling model is the
model among those single-component models with large maximum
likelihoods that has the most reasonable implications for the
parameters of the nascent neutron star.  We explore it more thoroughly
in the following section.

All of the accretion models have maximum likelihoods over 100 times
greater than that for the exponential cooling model.  The two-component
model with the highest maximum likelihood is the displaced power law
cooling and truncated accretion model.  We have used adaptive
quadrature methods to calculate the Bayes factor in favor of this model
over the single-component exponential cooling model; we find $B\approx
125$ (with $\mu$ fixed at 0.5 for the two-component model).  
This indicates strong evidence for an accretion
component.  This calculation used flat priors for the model parameters
over fairly broad ranges \cite{B-priors}.  One
might additionally consider the effect of our prior knowledge of the
nascent neutron star's possible size and binding energy on the Bayes
factor.  All of the two-component models that have a cooling component
with decreasing temperature have best-fit parameters implying neutron
star radii and binding energies much closer to expected values than any
single-component cooling model.  Accounting for this should more
strongly favor the two-component models.  This is borne out by
calculations.  We inserted a lognormal prior factor chosen to
qualitatively account for our expectations of the radius and binding
energy of the neutron star.  The (log) mean radius was set to 11~km,
and the (log) mean binding energy to $3\times10^{53}$~erg; the (log)
standard deviations were chosen corresponding to a $\pm4$\% variation
in radius and $\pm63$\% variation in binding energy, reflecting
uncertainties in equations of state of neutron stars of mass $\approx
1.4 M_\odot$ (see the discussion of Figure~5 in the following
section).  This prior increases the Bayes factor favoring the
two-component model to $\approx 2500$.  We conclude that there is
compelling evidence in the data for an accretion component in the
neutrino flux.

The two-component model with the highest maximum likelihood is the
displaced power law cooling and truncated accretion model.  We analyze
it in greater detail in Sec.~8.  Its likelihood is not significantly
greater than that of the model combining exponential cooling and
truncated accretion.  The latter model acts as a ``bridge'' between our
best cooling model and the models with accretion components.  But we
focus instead on the accretion model with displaced power-law cooling,
not only because its likelihood is larger, but also because it offers
us the opportunity to explore different cooling behavior, and because
displaced power-law cooling more closely resembles the cooling behavior
exhibited in recent supernova calculations.

A common, approximate frequentist significance test also indicates a
significant preference for two-component models.  Twice the logarithm
of the ratio of the maximum likelihoods of two nested models has an
asymptotic $\chi^2_\nu$ distribution, with $\nu$ equal to the
difference in the number of parameters of the models being fit to the
data \cite{Eadie}.  For example, the model combining displaced power
law cooling and truncated accretion has two more fitted parameters than
the exponential cooling model, and a likelihood 624 times greater.  The
chance of seeing an improvement this large or larger by chance if the
exponential cooling model is the true model is asymptotically given by
the tail area beyond $2\log(624) = 12.87$ in the $\chi^2_2$
distribution.  This probability is $Q=1.6\times10^{-3}$.  This
probability is approximate, in that it is based on an asymptotic
distribution.  Also, it ignores the size of the parameter space
searched and the extent to which the inferred parameter values agree or
disagree with expectations.  Nevertheless, it indicates significant
evidence for an accretion component, even from a frequentist
perspective.

Note that, in contrast to the single-component models, the accretion
models show a definite preference for a decrease in temperature of the
cooling component over an increase in dilution.  For example, the
truncated accretion model with displaced power law cooling has a
maximum likelihood over five times larger than that for the model with
displaced power law dilution.  Without more complete study of the
parameter dependence of the likelihood (i.e., rigorous calculation of
the model likelihood) it is not clear how strong this preference is.
We comment further on the characteristics of these models in Sec.~9.

\subsection{Consistency with Baksan data}

As noted above, Tables~\ref{table:one} and \ref{table:two} present
results both from joint analysis of the KII-IMB-Baksan data, and from
joint analysis of only the KII and IMB data.  Nearly all previous
analyses have ignored the Baksan data.  When these data were first
reported, there was a discrepancy between the time of the pulse
observed at Baksan and that reported by IMB, the Baksan data having
been detected approximately 30 s after the pulse observed by IMB
\cite{Baksan-1} (the KII detector has an absolute time uncertainty of
$\pm 1$ m and thus could not settle the issue).  But within a month of
the supernova, the Baksan group discovered a subtle, cumulative error
in their clock, rendering their absolute timescale uncertain over $-54$
to $+2$ s, and eliminating the discrepancy \cite{Baksan-2}.
Nevertheless, the Baksan data has been largely ignored, perhaps because
no methodology existed that could consistently account for the
relatively large background rate in the Baksan detector.  An exception
is the work of Piran and Spergel\cite{PS-89}.  But though they find
exponential cooling models for which the KII, IMB, and Baksan data are
consistent, they had to presume all Baksan events were signal events,
leading to acceptable models with unnecessarily large neutrino fluxes.

Our analysis easily accounts for strong,
energy-dependent backgrounds, and demonstrates that the Baksan data
are fully consistent with the KII and IMB data.  This is partly apparent
in the Tables, where the deviations between KII-IMB-Baksan estimates
and KII-IMB estimates appear relatively small.  As will become apparent
in the following sections, these deviations are indeed small compared
to the uncertainties in the parameter values.  More formally, we can
quantitatively assess the consistency simply by setting the offset
time for the Baksan detector to be large (negative or positive), so
that the data are considered to be entirely due to background, and
comparing the likelihood of such a case to the likelihood when the Baksan
events are allowed to be coincident with the supernova signal.  The
likelihood associated with the hypothesis that the Baksan data is
entirely background will just be the likelihood listed in the KII-IMB
column in the Tables, multiplied by a constant factor arising from
the Baksan data.  This factor is $1.5\times 10^{-5}$, and once introduced
allows comparison across columns of the table.  For example,
for the exponential cooling model
a model attributing the Baksan data entirely to background has a
maximum likelihood $1.5\times 10^{-5}$ smaller than the likelihood of
a model attributing part of the Baksan data to the supernova signal.
These results leave little question about the presence
of a supernova neutrino signal in the Baksan data consistent with that
detected by the KII and IMB experiments.

\subsection{Effect of background}

Proper treatment of background spectra plays a key role in settling the
issue of the consistency of the Baksan data with the other supernova
neutrino data.  The KII detector also has a significant background
rate.  To assess the effect that our inclusion of the KII background
rate has on our results, we performed the following calculation,
designed to mimic how other investigators dealt with the KII
background.  We analyzed the KII and IMB data jointly, but we set all
KII event background rates, $B_i$, equal to zero.  Duplicating the
efforts of others who attempted to account for background by
introducing an artificial energy threshold and censoring the data, we
also made the KII detection efficiency vanish for energies below $7.5$
MeV, and we omitted event 6 and events 13--16 from the KII data.
Analysis of the exponential cooling model then gives the following
best-fit values:  $\alpha=4.31$, $T_0 = 3.66$ MeV, and $\tau = 4.50$ s,
implying a binding energy of $E_b = 5.1\times 10^{53}$ erg and a
neutrinosphere radius of 43.1 km.  Comparing these results to the
KII-IMB results in Table~\ref{table:one} reveals little change in
$\alpha$ or $T_0$, but a more substantial change (over 15\%) in
$\tau$.  This is because there is a  nonnegligible probability that KII
events 10--12 are due to background.  It is not likely that {\it all}
of these events are background events, but it is likely that at least
one of them is a background event.  The analysis incorporating
background information accounts for this, and thus prefers a shorter
neutrino signal.  The small change in the inferred temperature is also
easily understood.  That found with background is somewhat higher
because the KII background spectrum peaks at low energy, which relaxes
the constraint imposed on the model neutrino spectrum by the low-energy
KII events.

It is interesting to note that the best-fit duration for the constant
temperature model is 10.43~s, thus excluding event no.~12 from the
neutrino signal.  This timescale is roughly five times more likely
than the 12.44~s timescale that would include this event.  This is
because there is a reasonable probability that event no.~12 is a
background event.  Previous analyses that ignored the background spectrum
would assign our best-fit constant temperature
model {\it zero} likelihood.

Finally, we note that our results are insensitive
to the removal of events 13--16 from the KII data because the
likelihood function finds it overwhelmingly likely that these events
are background events.  For example, the best-fit $\tau$ for the
exponential cooling model inferred from an analysis of the KII and IMB
data ignoring these late events is less than 4\% smaller than that
found including them; this is the parameter most affected by their
inclusion.

Table~\ref{table:bg} gives the probability that each KII and Baksan event is a
background event for the best-fit exponential cooling model and for
the best-fit displaced power law cooling plus truncated accretion
model.  These are obtained simply by dividing $B_i$ by the sum of
$B_i$ and the predicted signal rate for events like event $i$,
\begin{equation}
\int d\epsilon\, \like_i(\epsilon) R(\epsilon,t_i).\label{sig-rt-i}
\end{equation}
This is just the ratio of the background part of event $i$'s contribution
to the likelihood to its total contribution.  The formal (model dependent)
probability that each event is a background event requires integration
over the model parameters; the tabulated values are thus merely indicative.
Most striking is how the brief, low temperature component of
the accretion model and the resulting higher temperature for the
cooling component reduces the background probabilities for
KII events 1--6 and Baksan events 1--3 to roughly half the values
implied by the exponential cooling model.

\section{The Exponential Cooling Model}
\label{sec:exp}

We now explore more fully the implications of the data in the context
of the exponential cooling model.  First, we determine the allowed
ranges for the parameters of this model.  Then we examine the implications
of this model for the radius and binding energy of the neutron star
presumably created by the stellar collapse.  Then
we graphically demonstrate how the best-fit model accounts for
the data.  We defer most discussion of the comparison of these inferences
with theory to Sec.~9.  A frequentist assessment of the goodness-of-fit
of the best-fit model appears in Appendix~B.

\subsection{Credible regions}

A few previous investigators noted that their best-fit values for the
radius and binding energy of the neutron star were somewhat
higher than those predicted by current equations of state
\cite{Spergel-87,BS-88}.  Our best-fit parameter values imply
a radius and binding energy significantly greater than those found
by previous investigators, indicating an even more serious discrepancy.
It is therefore important to determine, not only the best-fit parameter
values, but the entire region in parameter space allowed by the
data.

The exponential cooling model has three physical parameters, $\alpha$, $T_0$,
and $\tau$.  Additionally, there is an unknown offset time, $\toff$,
for each detector.  In an analysis of the KII-IMB-Baksan data, there are
thus six parameters.  We summarize the full, six-dimensional joint
posterior by presenting marginal credible regions
for various interesting subsets of the parameters.  
Here and elsewhere we use ``68\%'' and ``95\%'' to denote
the probability content of credible regions formally including
68.3\% and 95.4\% of the posterior probability; our calculations are
based on Monte Carlo sampling and are accurate to $\approx 1$\%.

Figure 3 shows the one-dimensional marginal distributions for each of
the six parameters.  Each of the six curves shown in the Figure summarizes
the implications of the data for one of the parameters without regard
to the values of the other parameters.  In particular, one should not
quote credible regions from these marginal distributions jointly,
since correlations between inferred values of the parameters are ignored
in these plots.

Figures 3a--3c show marginal distributions for the three cooling model
parameters, $\alpha$, $T_0$, and $\tau$, with dashed lines indicating
the value of the marginal distribution bounding 68\% and 95\% credible
regions.  The modes and 95\% credible regions for these marginal distributions
are as follows: $\log_{10}\alpha = 0.64^{+0.26}_{-0.32}$,
$T_0 = 3.58^{+0.99}_{-0.70}$ MeV, $\tau = 5.29^{+3.45}_{-2.38}$ s.
Here and elsewhere we plot distributions as a function of $\log_{10}\alpha$
rather than of $\alpha$ itself.  The distribution as a function of $\alpha$
is broad and very skew; working in terms of $\log_{10}\alpha$ simplifies
the appearance of the posterior, particularly later when we show joint
credible regions.  Note that the modes of the marginal distributions are
at somewhat different locations than is the mode of the joint distribution.
This is simply a consequence of the integration involved in the marginal
distribution:  there is a greater volume of parameter space with high
probability at the mode of the marginal than at the joint mode, due to
asymmetry in the full distribution.  The
changes in location are small compared to the size of the credible
regions, however.

Figure 3d shows the marginal distributions for the three offset times,
with the location of the endpoint of a 95\% credible region for each
offset time noted by a dot.  The lower boundary of these credible regions
is at zero for all three offset times.  The credible regions are as
follows:  $t^{\rm off}_{\rm KII} = 0.09^{+1.10}_{-0.09}$ s,
$t^{\rm off}_{\rm IMB} = 0.00^{+1.01}$ s,
$t^{\rm off}_{\rm Bak} =0.00^{+3.28}$ s.

The most interesting parameters are the three parameters, $\alpha$, $T_0$,
and $\tau$, describing the cooling model.  Figure 4 shows three
two-dimensional marginal distributions that reveal how strongly the
inferred values of these parameters are correlated.  The dots show the
coordinates of 500 samples from the marginal distributions to illustrate
our use of posterior sampling to find the marginals; the contours show
68\% (dashed) and 95\% (solid) joint credible regions.  The inferred
values of $\alpha$ and $T_0$ are particularly strongly correlated (note
that the vertical coordinate is logarithmic, so that $\alpha$ and $T_0$
exhibit a semilogarithmic rather than a linear correlation).
This is because the expected number of neutrinos increases strongly
and nonlinearly with $T_0$:  the incident number grows with the standard
thermal dependence of $T_0^3$, but the $E^2$ dependence of the
capture cross section and the energy dependence of the detection efficiency
make the detectable number grow more quickly than $T_0^5$.  To keep
the expected number, which is also proportional to $\alpha^2$, near the
observed number, $\alpha$ must therefore decrease strongly with $T_0$,
as shown in the plot.

Each choice of $\alpha$, $T_0$, and $\tau$ implies a radius and binding
energy for the nascent neutron star.  The joint probability
distribution for the model parameters thus implies a joint distribution
for $R$ and $E_b$.  Figure 5 shows the 68\% and 95\% joint credible
regions of the marginal posterior for $\log R$ and $\log E_b$.  Also
shown are $E_b$ vs.\ $R$ curves for a representative set of equations
of state from the compendium of ``classic'' neutron star models
compiled by Arnett and Bowers \cite{AB-77}\ (dashed curves labeled
P$(\Lambda)$, BJV, and PS(tensor) following Arnett and Bowers) and
state-of-the art models calculated by Akmal, Pandharipande, and
Ravenhall \cite{APR-98} (solid curves labeled APR).  The classic models
span the softest and hardest models that have been seriously considered
in the past; the two APR curves are believed to bound the truth.  For
these models, the observed radius $R$ was calculated from the proper
radius $R_p$ according to $R = R_p (1-2GM_G/R_pc^2)^{-1/2}$, where
$M_G$ is the gravitational mass of the neutron star.  There is a
significant discrepancy between the data and all but the stiffest (and
currently disfavored) equations of state.

A number of effects might work in the direction to reduce the discrepancy.
One must first keep in mind that the neutrinosphere radius (the quantity
we actually infer) is in general distinct from the radius of the nascent
neutron star.  However, the Kelvin-Helmholtz
cooling calculations of Burrows and Lattimer \cite{BL-87}\ show that
the neutrinosphere falls to within 10\% of the radius of
the neutron star within $\approx 0.5$ s, and that the neutron star
radius changes by only 10\% as it cools after this time, even though significant
neutrino emission continues for $\sim 10$ s.  The work of Gudmundsson
and Buchler \cite{GB-80} elucidates this somewhat curious behavior.
In their study of the effects of lepton fraction on neutron star structure,
they found that neutron stars with masses of order $1.3$ M$_\odot$
or greater shrink by less than 30\% as their lepton fraction $Y_l$
decreases from $Y_l=0.45$.  There is significant rearrangement of mass, but
in a manner that keeps the overall radius roughly constant.  This behavior
is a consequence of the fact that
the leptons in the neutron star are relativistic,
while the nuclei are nonrelativistic and by themselves exhibit a very
stiff equation of state.  The loss of leptons from the star stiffens
the dense regions of the star where nuclear effects dominate the equation
of state, but softens those regions where Coulomb effects are important
(i.e., the inner crust).  Thus as $Y_l$ decreases, the core expands and
the crust shrinks.  The overall result is that high mass neutron stars
(which have large cores)
expand as $Y_l$ decreases, but low mass neutron stars shrink.
By coincidence, for masses near $1.3$~M$_\odot$ the two effects nearly
cancel, and the radius of the star suffers little change as the lepton
fraction decreases.  Thus the relatively large lepton fraction of
the nascent neutron star cannot account for its large inferred radius.

Another effect that might reduce the discrepancy is rotation.  If the
star is born as a fast rotator, its observed radius might be larger
than expected from nonrotating models.   Indeed, Cook et
al.\ \cite{Cook-94} find that rotation can increase the equatorial
radius of a cold neutron star by $\approx 40$\%.  However, they find
that the effect is strong only for angular velocities very near the
breakup velocity.  Further, we would have to be observing the neutron
star along its spin axis to see the full enhancement.

Thus, even allowing for the high temperature and lepton fraction of
the nascent neutron star and the effects of rotation,
there is a significant discrepancy between the inferred neutron star
radius and the predictions of current equations of state,
especially for realistic equations
of state which would require $\sim 50$\% expansion just to
reach the boundary of the 95\% credible region.

\subsection{Quality of fit}

Figure 6 provides an informal, graphical display of
how the best-fit exponential cooling
model compares with the observed data.  The Figure shows contours
of the detectable event rate, $\bar\eta(\epos)R(\epos,t)$ for the three
detectors.  Integrals of this rate give the expected number of events
in the region of integration.  The plotted contours bound regions that include
68\% (dashed curve) and 95\% (solid curve) of the total number
of detected events predicted by the best-fit model.
Also shown are the energies and times of the detected events.  The ``ridge''
at low energies in the KII plot in Figure 6a is due to the detector
background, as is the ragged structure in the Baksan plot in Figure 6c
(c.f.\ Figure 2).  The striking contrast between the shapes of
the KII, IMB, and Baksan contours illustrates how differently
the efficiency functions of the three detectors filter the neutrino signal.
Roughly two thirds of the events lie within the 68\% contours for all three
detectors and all of the events except for KII event 11 lie within the 95\%
contours, indicating broad compatibility of the model with the data.

A two-dimensional generalization of the Kolmogorov-Smirnov test, a
frequentist test of goodness of fit, can be used to attempt to quantify
the graphical comparison we present in Figure 6; several
earlier investigations employed such tests.
We present the results of such tests in Appendix~B along with a critique
of them.  Such tests are rather weak.  They verify the adequacy of
the exponential cooling model, but they fail to display the level
of improvement offered by this model over the constant temperature model,
or by accretion models over this model, to the degree it is displayed by
an explicit model comparison using the likelihood function (Bayesian
or frequentist).

From a purely statistical point of view, the exponential cooling model
appears adequate to account for the data when viewed in isolation from
reasonable alternative models.  However, its implications
for the parameters of the nascent neutron star conflict strongly with
prior expectations, and argue against acceptance of this model.

\filbreak
\section{Displaced Power-Law Cooling \\ and Truncated Accretion Model}
\label{sec:acn}

As noted in Sec.~\ref{sec:results}, models with an accretion component
not only have much larger maximum likelihoods than single component
models, but they also lead to inferred neutron star parameters much
closer to those expected based on theoretical and observational
knowledge of neutron stars.  Thus, these models have much
higher probabilities than single component models.  Here we explore
more fully the implications of the data for the best accretion model:
that combining displaced power law cooling with truncated accretion
(hereafter referred to simply as the cooling plus accretion model).  As
we did with the exponential cooling model, we first present credible
regions for model parameters, and then discuss how well the best-fit
model accounts for the data.  Comparison of our inferences with
theoretical expectations appears in the following section.

\subsection{Credible regions}

The cooling plus accretion model has nine parameters:  three describing
the displaced power law cooling component $(\alpha,T_{c,0},\tau_c)$,
three describing the accretion component $(\mu,T_a,\tau_a)$, and three
detector offset times.  The sparsity of the data, combined with the
complicated structure of the emitted rate and spectrum, result in
a posterior distribution that is significantly more complicated
than the unimodal posterior found for the exponential cooling model.
This is illustrated in Figure 7, which presents simple summaries of our
inferences for the three accretion parameters.

Figure 7a shows the
profile likelihood for the accreted mass parameter, $\mu$.
The profile likelihood, $\like_p(\mu)$,
is found by calculating, for each $\mu$, the maximum
value of the likelihood (maximized over all the remaining parameters).
The plotted value has been normalized so that it gives directly the
maximum likelihood ratio between a model with specified $\mu$ and
the exponential cooling model.
A profile likelihood can provide an approximate marginal distribution.
In particular, for posteriors that are multidimensional Gaussians
(with arbitrary amounts of correlation), normalized profile
posteriors are identical to the corresponding marginal distributions.
More generally, the approximation can range from excellent to very
poor, depending on how strongly the characteristic scale of variation
of the posterior varies with the parameters.  While we have not
quantified how accurately these profile posteriors approximate
the corresponding marginals, our investigations of the behavior of
the likelihood as a function of the maximized parameters indicate that
these curves adequately display the regions of parameter space where
most of the posterior probability lies.

Figure 7a shows that the likelihood varies rather weakly with $\mu$, with
values over the entire range we searched, from $\mu=0.1$ to $\mu=3$,
having profile likelihoods that vary by less than a factor of 8
(roughly the range of variation across two standard deviations of a Gaussian
distribution).  As already noted, we focus attention on models with
$\mu=0.5$ as being representative of those found in supernova calculations
based on the delayed scenario.  The point on the curve corresponding to
this model is indicated by a dot.

Figure 7b displays the $(T_a,\tau_a)$ dependence of the posterior for
the $\mu=0.5$ model.  For each value of $(T_a,\tau_a)$, we maximized the
posterior with respect to the $\alpha$ and $T_{c,0}$ cooling parameters.
The cooling timescale $\tau_c$ was fixed at its best-fit value of
14.7 s for this calculation because maximization with respect to this
parameter proved problematical away from the peaks (extreme values were
preferred); the most probable $\tau_c$ values in the vicinity of the peaks
are near this best-fit value.  This Figure clearly reveals the complicated
structure of the posterior.  Three local modes are apparent.  One is at
very small values of $\tau_a$ corresponding to accretion components
that account only for the first event in each detector.  Another
is near $\tau_a = 0.1$ s, giving a duration just sufficient to include
the second KII event.  The global
mode at $T_a = 2.00$~MeV and $\tau_a = 0.74$~s has a peak density
about twenty times greater than that at $\tau_a\approx 0.1$~s and thus
contains most of the posterior probability; the 0.74~s duration includes
the first six KII events.  The posterior density falls very
steeply with increasing temperature, setting a firm upper limit
on $T_a$ of $\approx 2.5$~MeV for the most probable values of
$\tau_a$ ($> 0.2$ s).  It falls less steeply with decreasing temperature,
but $T_a < 1.5$ MeV is strongly excluded.
There is an additional very small mode, not shown, at $\tau_a \approx 12$~s,
due to the late, soft KII events, nos.~10--12.

The complicated structure of the posterior has prevented us from
calculating rigorous marginal credible regions for the parameters of
this model using the rejection method described earlier.  In the remainder
of this section, we present inferences conditioned on $\mu=0.5$ and
on the resulting best-fit values of $T_a$ and $\tau_a$, listed
in Table~\ref{table:two}.  More rigorous calculations (for example, using
Markov chains instead of the rejection method; see
\cite{L-99}) should result in somewhat
broader credible regions than those we will show here, as a result of
averaging over other values of the accretion parameters.  But since
$T_a$ and $\tau_a$ are fairly well determined for the global mode,
and since their best-fit values do not change greatly with $\mu$, we do not
believe more rigorous credible regions would be substantially larger than
those displayed here.

Figure 8 displays marginal distributions for the three parameters of
the cooling component and for the three offset times, conditioned on
the best-fit accretion temperature and timescale for $\mu=0.5$.
It is instructive to compare these inferences with those displayed
in Figure 3, based on the exponential cooling model.  The inferred
value of $\alpha$ when an accretion component is present is substantially
smaller, because a significant number of the earliest, softest events
is attributed to the accretion component.
The temperature of the cooling
component is significantly higher than that in a single-component model
because the constraint placed on the temperature by those early, soft
events has been relaxed.
Inferences for the cooling timescale must be more cautiously compared,
since the cooling components of these models have different temporal
behavior.  In particular, for the exponential
cooling model, $\tau$ was the {\it luminosity} timescale, so that
$4\tau$ is the temperature timescale.  In the two-component model
studied here, $\tau_c$ is the temperature timescale.  Its inferred
value is somewhat smaller than $4\tau$ for the exponential cooling
model, but the rate of cooling is significantly less in this model
(with its displaced power-law cooling component) than in the exponential
cooling model.  The timescales are thus comparable.
Finally, the offset times are better constrained in the
accretion model, in order to keep the
early events of all three detectors coincident with the brief accretion
component.  The modes and 95\% credible regions for these marginals
are as follows:  $\log_{10}\alpha = 0.31\pm0.41$,
$T_{c,0} = 4.23^{+1.58}_{-1.07}$ MeV, $\tau_c = 14.5^{+18.5}_{-6.7}$ s,
$t^{\rm off}_{\rm KII} = 0.00^{+0.23}$ s,
$t^{\rm off}_{\rm IMB} = 0.00^{+0.80}$ s,
$t^{\rm off}_{\rm Bak} = 0.00^{+1.01}$ s.

Figure 9 displays two-dimensional marginal distributions
for the cooling component parameters (again, conditional on the
parameters for the accretion component).  These illustrate the
correlations between the inferred values of the parameters, which
show the same qualitative behavior as that displayed in Figure~4 for
the exponential cooling model.

Figure 10 shows the implications of this model for the radius and
binding energy of the nascent neutron star.  These results are
conditional on the best-fit $T_a$ and $t_a$ for $\mu=0.5$, so
by assumption there is an accretion contribution to the binding
energy given by equation~(\ref{Ea-def}); this contribution is
$E_a = 6.1\times10^{52}$ erg.  Added to this is an
uncertain contribution due to the cooling component; the figure
shows the joint distribution for logarithms of the total binding
energy, $E_b$,
and the observed radius, $R = 10\alpha (D/50 \kpc)$ km.  Also shown
are the same representative $E_b$ vs.\ $R$ curves shown in
Figure~5.  Clearly, the radius and binding energy implied by
this two-component model are easily compatible with
the values predicted by all viable equations of state.

\subsection{Quality of fit}

Figure 11 graphically illustrates how the best-fit accretion model
(with $\mu=0.5$) compares with the observed data.  Comparison with
Figure~6 (for the exponential cooling model) reveals how this model can
so substantially increase the probability for the data.  The brief, low
temperature accretion component accounts for the early, soft KII
events, nos.~2--6.  This relaxes the constraint these events placed on
the temperature of the cooling component, allowing it to be higher.
The higher temperature cooling component better accounts for the
remaining early KII events (that have significantly higher energies
than events 2-6), and also better accounts for the energetic events
seen in IMB and Baksan.  Results of a two-dimensional
Kolmogorov-Smirnov (KS) test further demonstrating the adequacy of the
best-fit model appear in Appendix~B.

\section{Comparison With Theory}
\label{sec:theory}

Here we review the basic predictions
of supernova theory for the characteristics of the neutrino emission,
and then compare these with the characteristics inferred above.

\subsection{Core collapse and bounce}

Several reviews describe the collapse of a massive ($\gtrsim 10
\msol$) star, such as the progenitor of SN 1987A, and the subsequent
birth of a neutron star
\cite{BL-87,Burrows-88,JH-89,WW-86,BL-86,MWS-87,Bruenn87,HM-nus,%
Baym-SN,Coop-nus,Burrows-review,ABKW-89}.  Here we summarize the
basic features of the supernova event and the resulting neutrino
signal, following closely the descriptions of Woosley and Weaver
\cite{WW-86} and of Arnett, Bahcall, Kirshner, and
Woosley \cite{ABKW-89}.

Once the massive progenitor of the supernova begins fusing oxygen, its
neutrino luminosity exceeds its photon luminosity.  Neutrinos thus play
a dominant role in the evolution of the star well before the drama of
stellar collapse begins, though the neutrino luminosity is far below
the limit of detectability.

Nuclear burning proceeds in the progenitor core until an iron core is
produced with mass $M_c \sim 1.26 \msol$, radius $R_c \sim$ few $\times
10^3 \km$, central density $\rho_c \sim 10^{10} \gcc$, and central
temperature $T_c \sim 1$ MeV.  The pressure in the iron core is
dominated by the degeneracy pressure of relativistic electrons ($\mu_e
\sim 10 \MeV$), so the core resembles a degenerate dwarf star with an
equation of state with effective adiabatic index, $\Gamma = (\partial
\ln P / \partial \ln \rho)_s$, near that of an ideal, relativistic
electron gas, and therefore only slightly above the critical value
$\Gamma = 4/3$ at which gravitational collapse will occur.

Since iron is at the peak of the nuclear binding energy curve, at this
point the progenitor has exhausted its supply of thermonuclear fuel.
The core contracts and heats, causing photodissociation of the iron
nuclei through the reaction $\gamma + ^{56}{\rm Fe} \rightarrow
13\alpha +4{\rm n}$.  This reaction is endothermic, requiring $\approx
124 \MeV$ per dissociation, which depletes the kinetic energy of the
electrons, reducing their pressure support of the star.  Additionally,
electron captures on nuclei in the core occur through reactions of the
form,
\begin{equation}
{\rm e}^- + (Z,A) \rightarrow (Z-1,A) + \nu_{\rm e}.\label{ecap}
\end{equation}
Initially, the neutrinos produced by these reactions leave the core,
carrying away most of the kinetic energy of the captured electrons,
and further reducing the electron pressure support.

Through the combined effect of these two processes, the effective
adiabatic index in the core falls below $\Gamma=4/3$ and dynamical
collapse ensues.  The inner $\approx 0.8\msol$ of the core remains
partially pressure supported (\ie, the infall velocity remains
subsonic) and collapses homologously with velocity proportional to
radius.  Outside the inner core, the infall velocity is supersonic, and
is approximately the free fall velocity.

The neutrinos produced by electron capture have energies typical of the
electrons that produced them, $E_\nu \lesssim 10\MeV$.  Their
wavelengths ($\lambda_\nu \sim 20$ fm) are thus long compared with
nuclear sizes, so they scatter coherently from nuclei, with a cross
section proportional to the number of nucleons squared
\cite{Freedman}.  The mean free path for elastic scattering of a
neutrino of energy $E$ consequently becomes much smaller than the radius
of the core \cite{LP-76}.  Initially, the neutrino diffusion
timescale, $R^2/c l_\nu$ (where $l_\nu$ is the neutrino mean free
path), is shorter than the dynamical timescale,
and the neutrinos leave the collapsing core, carrying away entropy.
But once the density exceeds $\approx 3\times 10^{11}\gcc$, the
diffusion timescale exceeds the dynamical timescale, and the neutrinos
are trapped in the collapsing material.  Thus soon after collapse
begins, the lepton fraction of the core is frozen at $Y_l \approx
0.35$, and the collapse proceeds adiabatically.  The degenerate
electrons and electron neutrinos in the core store the gravitational
energy of the collapse.

On the dynamical timescale of a few milliseconds, the density reaches
$\rho_c \sim 10^{14}\gcc$, at which point degenerate, nonrelativistic
nucleons become the dominant source of pressure in the inner core.  The
resultant stiffening of the equation of state abruptly halts the
collapse of the inner core.  Pressure waves propagating outward
coalesce into a shock $\approx 0.2 \msol$ beyond the edge of the inner
core.  The shock begins moving outward and dissociating the outer core
material, so that the post-shock material consists mostly of neutrons
and protons.  In this environment of free nucleons, the electron
capture rate rises and the neutrino cross section decreases, causing
electron neutrinos to pile up behind the shock.

Several milliseconds later, the shock reaches a density of $\rho \sim
10^{11}\gcc$, where the optical depth outward is of order unity, and
the electron neutrinos behind the shock are released in a dynamical
timescale of a few milliseconds.  This is the first significant
neutrino signal produced during collapse.  The electron neutrinos
released during this ``breakout'' phase have a spectrum like that of
the degenerate electrons that produced them ($\mu_\nu \approx 40$
MeV).  A small number of thermally produced pairs of electron
neutrinos and antineutrinos and other neutrino flavors are also emitted.

As the shock propagates through the outer core, it weakens due to
neutrino emission and photodisintegration of heavy nuclei.  The
temperature of the shocked material is so high that destruction of iron
down to free nucleons occurs, releasing an energy $E_{\rm photo} \sim 1.5 \times
10^{51}$ erg for each $0.1 \msol$ of matter that is
photodisintegrated.  The shock cannot endure such losses for long, and
will die unless it quickly reaches the edge of the outer core, where
the density is low and the heat capacity is high.  Here the shock
temperature falls, becoming less effective in producing neutrinos and
too low to disintegrate iron.

In order for the shock to survive the energy losses due to
photodissociation of the outer core, the total mass of the core must be small
\cite{WMWW,CBB-84,BW-82,Hill82a,Hill82b,HNW-84,BL-83,BL-85,%
Kahana-87}.
Thus an iron core as
large as $2 \msol$ almost certainly cannot produce a
supernova by a hydrodynamical bounce,
but cores smaller than $\approx 1.35 \msol$ might.  Even in this case,
however, a successful explosion is problematical and depends on the
equation of state used and details of the hydrodynamic and neutrino
transport codes employed.  Thus Hillebrandt et al.\ \cite{HNW-84}\ get a strong
explosion for an 8.8-$\msol$ star, Wilson et al.\ \cite{WMWW}\ get a weak
one, and Burrows and Lattimer \cite{BL-85}\ get none at all.  While
Hillebrandt \cite{Hill82a}\ obtains a marginal hydrodynamic explosion for a
10-$\msol$ star, Wilson et al., Bruenn, and Burrows and
Lattimer do not \cite{WMWW,BL-85,Bruenn85};
and while Baron et al.\ \cite{Baron-85a,Baron-85b}, using a ``softer''
nuclear equation of state than hitherto accepted, obtain prompt
explosions for 12-$\msol$ and 15-$\msol$ stars, Wilson et al.\ \cite{WMWW},
using a more standard equation of state, do not.

\subsection{Prompt explosion}

If a prompt explosion occurs, the shock moves outward rapidly upon
reaching the edge of the outer core, ejecting the mantle and envelope
of the star.  Electron neutrinos gradually diffuse out of the inner
core on a diffusion timescale $\sim 2 \s$.  Electron captures at
first replenish them, and through this diffusion process the lepton
fraction of the core begins to decrease.  These electron neutrinos are
created with energies of the order of the Fermi energy of the captured
electrons ($\sim 40\MeV$), but do not leave the core until they have
down-scattered to energies for which the scattering mean free path is of
order the size of the core ($\sim 5\MeV$).  As a result of this
``deleptonization'' phase, the outer core is {\it heated}, leading to
thermal production of neutrinos of all flavors in the outer core
through electron-positron annihilation.  For a few
tenths of a second after shock breakout,
production of electron neutrinos in the optically thin region just behind
the shock dominates that of other species.  But soon the neutrino
emitting region becomes optically thick and pairs of
all flavors of neutrinos are produced in roughly equal numbers.

These thermal pairs cool the core, which has now reached its final
radius $\approx 10\km$.  The integrated energy of these thermal
neutrinos is very nearly equal to the full binding energy of the
collapsed core, $E_B \sim 10^{53}\erg$.

To summarize, in the prompt explosion picture, the neutrino signal is
expected to consist of two principle phases.  First, there is a brief,
intense burst of electron neutrinos from shock breakout, with a
degenerate spectrum of high energy.  Though intense, this burst is so
brief that very little of the binding energy and lepton number of the
collapsed core is carried away by it.  Following this burst is a much
weaker signal of thermally produced neutrinos of all flavors with lower
energies, but lasting for a much longer time, $\sim 10$s.  The separate
timescales for shock breakout ($\sim 0.2$ s) and Kelvin-Helmholtz
cooling ($\approx 2$~s) may be discernible in the neutrino signal.
The integrated energy of the
later, thermal signal equals the binding energy of the neutron star,
and the integrated number of neutrinos in this phase of emission
exceeds the number of leptons originally contained in the collapsing
core by an order of magnitude.  The signal in water C\` erenkov
detectors, whose cross sections for interaction with $\nu_e$ (and
neutrinos of other flavors) is an order of magnitude lower than that
for absorption of $\nueb$, is expected to be dominated by thermally
produced $\nueb$'s.

\subsection{Delayed explosion}

If the prompt explosion fails, as all recent numerical simulations
find, the deposition behind the shock of a small amount of
energy by neutrinos streaming out of the core may produce a
delayed explosion \cite{W85,BW-85,Bethe90,Bruenn93a,BHF,%
Herant94,Janka93,JM-95,WMWW,RJ00,Mezz01,Lieb01,Janka01}.

Following the failure of the shock, a nearly stationary
``neutrinosphere'' develops at about 40 km, where the density $\rho
\sim 10^{11}$ g cm$^{-3}$ and the neutrino emission temperature $T_\nu
\sim 5$ MeV.  The stalled shock
lies at $\sim 100-300$ km, well beyond the neutrinosphere, where
the post-shock temperature ($\sim 1.5$ MeV) and density ($\sim 10^8$ gm
cm$^{-3}$) are much smaller.  Capture of a small fraction ($\lesssim 5$\%)
of the $\sim 10^{53}$ erg s$^{-1}$ neutrino luminosity by neutrons
and protons and, later, by scattering off electron-positron pairs
behind the shock heats the matter, and eventually revives the shock.
After $\sim 0.1 \msol$ or more of matter has accreted onto the core over a
period of $\sim 1$ sec, the outward motion of the shock resumes, ejecting
the mantle and envelope of the star.  During this accretion phase, the
hot material behind the shock copiously emits electron neutrinos and
antineutrinos; the production of neutrinos of other flavors is
suppressed because they can be produced only in neutral current interactions.
The amount of material finally accreted is
uncertain, but may be as much as $0.5 \msol$ in order to
leave behind a neutron star with a mass near the typical observed
value of 1.4 $\msol$.

To summarize, in the delayed explosion picture, the
neutrino signal is expected to consist of three principal components.  Two
of these, the emission of electron neutrinos at breakout, and the
diffusion of neutrinos out of the inner core, which heats the outer core
and produces neutrino-antineutrino pairs of all flavors, are identical
to those of the prompt explosion picture.  In addition, there is a
third component, lasting $\sim 1$ sec, during which the flow of accreting
matter through the stalled shock at $\approx 100$-200~km produces
electron neutrino-antineutrino pairs, possibly with a luminosity $L_a
\sim 10^{53}$~erg~s$^{-1}$ and temperature $T_a\approx 3$ to 5~MeV.

\subsection{Comparison with neutrinos from SN 1987A}

The inferred values for the neutrino cooling timescale and
characteristic cooling temperature, both for single component
cooling models and for the cooling component of models including
an accretion component, are in
remarkable agreement with that expected in the above scenario, which
had developed in the absence of direct observations
\cite{BL-87,Burrows-88,JH-89,WW-86,BL-86,%
MWS-87,Bruenn87,Burrows-review}.
The $\nueb$ energy, $3.15 \; T_0 \approx 15 \; {\rm MeV}$,
is typical of that expected for the neutral current diffusion of
$\nueb$ out of the hot outer core.  Finally, the cooling time scale
$\tau \approx 4$~s is of the order of the expected
timescale for deleptonization of the inner core.

However, for single component models, the inferred values of $R$, $E_b$,
and the total number of $\nueb$ are all well above theoretical expectations.
This is most clearly displayed by the $(R,E_b)$ credible regions
for the exponential cooling model plotted in Figure~5.  All other single
component cooling models we explored had best-fit $(R,E_b)$ values
even more excessive than those found with the exponential cooling
model.

When an accretion component is added to the signal, not only does the
fit substantially improve, but the inferred values for $R$, $E_b$,
and the number of thermally emitted $\nueb$ are all in agreement
with theoretical expectations.  Figure~10 displays the agreement
between inferred and expected $R$ and $E_b$ values.
The implied number of $\nueb$ from the cooling component
($\approx 3\times10^{57}$) is
comparable to that expected from $\nu_e$ diffusing out of the inner
core and heating the outer core by neutral current scattering and
absorption.  Approximately $5\times10^{57}$ additional $\nueb$
are emitted by the accreted material.

The inferred timescale of the accretion component ($\approx 0.74$ s) is
in agreement with the timescales $\sim 0.5$-1~s observed in numerical
calculations\cite{W85,BW-85,Bethe90,Bruenn93a,BHF,%
Herant94,Janka93,JM-95,WMWW,RJ00,Mezz01,Lieb01,Janka01}.  The best-fit 
temperature of the
accretion component is 2 MeV, and there is a sharp upper limit of
$\approx 2.5$ MeV.  The temperatures observed in current numerical
calculations are 1 to 3 MeV higher, but are highly uncertain.  Finally,
the data prefer large values for the amount of accreted material ($> 1
M_\odot$).  However, this preference is not of great statistical
significance, and models with $0.1$ to $0.8 M_\odot$ of accreted
material all make the data substantially more probable than
single-component models.

Surprisingly, then, these relatively sparse data are able to discern between
models with and without an accretion component, due to broad spectral
and temporal features in the data.  However,
the data have proved too sparse to discern some interesting details
about the spectral evolution of the neutrino signal.  Detailed
studies of the transport of neutrinos through the core during the
deleptonization and cooling phases show that the emitted spectrum
is significantly nonthermal \cite{JH-89,Burrows-eta}.  The strong
energy dependence of neutrino scattering cross sections ($\propto
\eps^{-2}$) leads to a spectrum that is well modeled by a Fermi-Dirac
spectrum with positive chemical potential $\mu_\nu$ (or effective
degeneracy parameter, $\eta_\nu = \mu/T$), with $\eta_\nu \approx 2$
to 4.  But the data are too sparse
to provide a significant measure of these transport effects:  when the
$\eta_\nu$ parameter is added to the exponential cooling model, its
best-fit value is zero, and its 95\% credible region extends to
$\eta_\nu \approx 5$.  Similar conclusions were reported earlier by
Hillebrandt \etal\ \cite{HHJM}.
In addition, there is some ambiguity among the calculations regarding
the evolution of the spectrum of the cooling component.  Calculations
that treat the neutrino transport in a limited way by considering
only a ``luminosity temperature'' for the neutrinos necessarily find
a neutrino temperature that decreases in time as the luminosity decreases
\cite{BL-86}.
More sophisticated calculations seem to indicate that the neutrino
temperature stays roughly constant over $\sim 10$ s (perhaps even
rising slightly during the first few tenths of a second
\cite{Burrows-review}), with the luminosity
decreasing as the opacity just below the neutrinosphere becomes
more and more scattering-dominated, leading to dilution of the
neutrino spectrum \cite{JH-89,JH-89b,Janka95,Janka91,Janka91b}.
In our study of single-component models, the
data were not able to discern between cooling and dilution, although
a slight preference for dilution appeared.  In our study of models
including an accretion component, the initial temperature was larger
than in single component models, with the result that models with
a decreasing temperature for the cooling component are
preferred over models with pure dilution.  This preference is not
decisively strong, however.

\section{Inferring the Electron Antineutrino Mass}
\label{sec:mass}

The calculations of the preceding sections all presume that the
electron antineutrino rest mass, $\mnu$, is zero.  We derived the
likelihood function allowing for nonzero $\mnu$, so it is
straightforward to test this assumption.  For several of the models we
considered, the likelihood is maximized with $\mnu=0$, indicating no
evidence for a nonzero rest mass in the supernova neutrino data.  This
is true for the exponential cooling model that was the focus of
Sec.~\ref{sec:exp}.  For others, the likelihood is maximized for small
values of $\mnu$ (a few eV), but the likelihood is increased only
slightly above its $\mnu=0$ value, indicating no significant evidence
for a nonzero mass.  This is true for the cooling plus accretion model
that was the focus of Sec.~\ref{sec:acn}; Table~\ref{table:m} provides the
best-fit parameters for this model found using the KII-IMB-Baksan
data.  The ${\cal L}/{\cal L}_0$ entry gives the ratio of the maximum
likelihood to that found with $\mnu=0$.  These results are
representative of models with nonzero best-fit $\mnu$:  best-fit masses
of a few eV; best-fit detector offset times $\sim 0.1$~s; negligible
changes in other parameters; and insignificant improvement in the
maximum likelihood.

Presuming that there is nevertheless a small nonzero rest mass, we can
calculate marginal posterior distributions for $\mnu$ for any model of
interest to obtain constraints on the mass.  Figure~12 shows such
marginal distributions for the exponential cooling model (dashed curve)
and the cooling plus accretion model (solid curve; here $\mu$, $T_a$,
and $\tau_a$ were fixed as in Sec.~\ref{sec:acn}).  The dots indicate
the upper bounds of 95\% credible regions and are at $\mnu = 8.9$~eV
for the exponential cooling model and $\mnu = 5.7$~eV for the cooling
plus accretion model.  It is interesting to note that these upper
limits for $\mnu$ are substantially better than the laboratory limits
that were available at the time of the supernova neutrino detections
(and comparable to current limits).  Formally, a complete summary of
the implications of the data for $\mnu$ would additionally marginalize
over the choice of signal model, essentially producing a weighted
average of the individual marginals shown in the Figure (this is called
Bayesian model averaging \cite{W-97,HMRV-00}).  But the cooling plus
accretion model is so much more probable than single-component models
that model averaging would essentially reproduce the solid curve, which
we thus consider to summarize our results for $\mnu$.

Note that the marginal posterior based on the exponential cooling model
peaks at positive $\mnu$, even though the joint posterior based on that
model peaks at $\mnu=0$.  Also, for the cooling plus accretion model,
the ratio of the peak of the marginal to its value at $\mnu=0$ is
greater than the likelihood ratio of 2.3 listed in
Table~\ref{table:m}.  These differences between the joint distributions
and their marginals are further examples of the phenomenon discussed in
Sec.~\ref{sec:exp} (see the discussion of Figure~3).  There is somewhat
more allowed volume in the parameter space for slightly positive values
of $\mnu$, and the integration yielding the marginal for $\mnu$
accounts for this, increasing the marginal density for $\mnu$ in that
region.  Such effects are common, and provide an illustration of the
difference between using profile likelihoods and true marginal
distributions.

\section{Comparison With Methodologies of Previous Studies}
\label{sec:comp}

We have reached substantially different conclusions than previous studies
of the supernova neutrinos.  One of the major improvements of this work
is our more thorough exploration of the space of alternative signal models,
and thus it may not seem surprising that we might discover a signal
component missed by others.
However, this alone does not account for the differences between our
results and those of others.  For example, the exponential cooling model
has been studied by several investigators, yet the best-fit radius
we find is 70\% larger than that found by Spergel et al.\ \cite{Spergel-87}
based on a likelihood analysis, and our best-fit binding energy is
over 40\% larger than that found both by these investigators and
by Bludman and Schinder \cite{BS-88}, who also used a
likelihood function.  

Previous analyses of the neutrino data are extremely diverse, using
a wide variety of statistics and methods.  A detailed comparison
of all these methods with the present analysis would be lengthy.
We here choose instead to emphasize two points of departure between
our analysis and earlier ones that appear to us to offer the most
important lessons for analysis of data like the SN~1987A neutrino
data.

\subsection{The form of the likelihood}

It is clear that there are important differences between our likelihood
function and those used by others, since our best-fit parameter values
(equivalent to maximum likelihood estimates) are significantly
different from those found earlier.  Comparing our likelihood function,
equation~(\ref{Ldt-iso}), with those used by other investigators,
several differences are apparent.  Most obvious, perhaps, is the
presence of the background term that allows us to correctly incorporate
information about the energy-dependent background rates of detectors.
We have already noted, in Sec.~6, how important such terms are for
incorporating the Baksan data.  But we also noted that their effect on
inferences using only the KII and IMB data, although noticeable, is not
significant compared to the uncertainties in inferred parameter
values.  Thus these terms do not explain the differences between our
results and those of others.

Another difference is the presence of terms to correct for deadtime.
But for the most part, these terms affect only the overall amplitude of
the effective signal in IMB (decreasing it by roughly 10\%), and thus also
do not account for the significant differences.

The remaining difference is the absence of a factor of $\eta(\epsilon)$
from inside each event integral in the likelihood function.  That is,
all previous studies replaced the integral in the product
term of equation~(\ref{Ldt-iso})\ with a term proportional to
\begin{equation}
\int d\epsilon\, \bar\eta(\epsilon) \like_i(\epsilon)
          R(\epsilon,t_i).\label{bad-Li}
\end{equation}
We have verified that inclusion of such an additional, incorrect
$\bar\eta$ factor indeed results in best-fit exponential cooling
parameter estimates very close to those found in earlier studies.  This
factor reduces the low energy contribution to the integral, so that
somewhat larger temperatures are needed to make the likelihoods of the
events reasonably large.  The expected number of detectable neutrinos
varies very strongly with $T$ (more strongly than $T^5$, due to the
$E^2$ dependence of the capture cross section and the strong energy
dependence of $\bar\eta(\epsilon)$), so the value of the amplitude
parameter $\alpha$ (i.e., of the neutron star radius) found in the fit
is strongly affected by the presence or absence of the $\bar\eta$
factor, as is the binding energy, which scales like $\alpha^2 T^4$.

The detection probability is already built into the $\like_i$ function;
insertion of an additional $\bar\eta(\epsilon)$ factor represents an
attempt to take into account a selection already accounted for in
$\like_i(\epsilon)$.  This is perhaps most easily seen by considering a
simple situation in which detection occurs only if the number of
photomultiplier (PMT) ``hits'' exceeds a threshold value, $n_{\rm th}$,
and the detection data for event $i$ is simply the number of
photomultipliers hit, $n_i$.  Suppose also that the probability for $n$
hits is a Poisson distribution with a mean that is an increasing
function of the event energy.  The detection efficiency is the
probability for hitting more than $n_{\rm th}$ PMTs, given the event
energy.  It would be calculated by summing Poisson probabilities for $n
> n_{\rm th}$.  The event likelihood for event $i$ is the Poisson
probability for seeing exactly $n_i$ hits.  It is {\it not} the product
of this probability and the efficiency; this product has no meaningful
interpretation.  We could multiply it by the product of detection, {\it
given} that $n_i$ PMTs were hit (since the Poisson factor already takes
that into account).  But since $n_i$ must have been larger than $n_{\rm
th}$ for the event to have been detected, this extra conditional
probability is equal to unity.  The inclusion of an $\bar \eta$
factor in the detection likelihoods is thus incorrect.

It is worth noting that straightforward application of the rules of
probability theory led us to the correct likelihood in a more-or-less
automated way, once we set out to calculate the probability for the data
from first principles, and not merely write it down based on our
intuition.  The derivation is Bayesian in that
we freely assigned probability distributions to the energies (and
directions and positions) of detected events, despite the fact that
these quantities cannot be considered to be ``random variables.''

\subsection{Distinguishing parameter estimation \\ from model assessment}

As noted in Sec.~II, frequentist and Bayesian statistics both
divide questions about parameterized models into two classes.  First is
the class of {\it estimation} questions that assess the implications of
assuming the truth of a particular model, usually by estimating values
or allowed ranges for the model parameters.  Second is the class of
{\it model assessment} questions that assess the viability of a model.
We have outlined Bayesian methods for treating these questions in
Sec.~\ref{sec:bayes}.  A clear discussion of the application of
frequentist methods for estimation and model assessment to problems in
the physical sciences is available in the text by Eadie,
\etal\ \cite{Eadie}.

Frequentist procedures used for estimation are fundamentally different
from those used for model assessment.  Unfortunately, nearly every
previously published statistical analysis of these data has incorrectly
used model assessment procedures to address estimation problems.  In
particular, a number of studies used goodness-of-fit (GOF) procedures
to specify ``confidence'' regions, based either on statistics of the
Kolmogorov-Smirnov type \cite{Bahcall-1,Spergel-87,SB-87}, a likelihood
statistic \cite{Krauss}, or an ad hoc ``$\chi^2$'' type statistic
\cite{Burrows-mass}.  In these studies, the boundary of the calculated
``confidence region'' was determined by finding parameters for which
the significance level of a GOF test is equal to the desired confidence
level (i.e., significance levels were confused with confidence
levels).  Such misapplication of GOF procedures to parameter estimation
problems is commonplace in astrophysics; we have been guilty of it
ourselves in the past.  Loredo and Wasserman discuss the problem in
detail in the context of the analysis of gamma ray burst data (see
Appendix~A of \cite{LW3}).  Using a simple example based on inferring
the mean of a Gaussian distribution, they show that use of a $\chi^2$
GOF test to determine ``confidence'' regions in the manner of earlier
studies not only fails to reproduce the familiar ``$\sigma/\sqrt{N}$''
68.3\% confidence region, but produces an erroneous region whose
average size is larger than the correct region, with an error that {\it
grows} as the amount of data increases.

It is interesting to speculate about why such a basic mistake is so
frequently made.  One reason is that, for the
familiar case of Gaussian statistics, the same function---the $\chi^2$
statistic---is used both to define the statistic used in a GOF test
(the minimum value of $\chi^2$), and the interval-valued statistic used
for a confidence region (the parameter range for which $\chi^2$ is
within some critical value, $\Delta\chi^2$, of its minimum value).
This may have motivated those investigators who attempted to use the KS GOF
statistic to define confidence regions, although we know of no
statistical literature suggesting that this statistic is useful for
estimation problems.  More fundamentally, the confusion may arise
because there are several qualitatively different probabilities in
frequentist statistics.  Covering probabilities for confidence regions,
Type I error probabilities, Type II error probabilities---all of these
are quantities that span $[0,1]$ that scientists can use to assess the
reasonableness of hypotheses.  But none of them are probabilities {\it
for hypotheses}, so it is easy for nonexperts to confuse which is most
closely related to the question they are asking.  This confusion is
exacerbated by the fact that all frequentist probabilities must
condition on a particular point hypothesis, even those that refer to an
entire class of hypotheses.  For some problems (particularly for
confidence region calculations), the hope is that the final result is
independent of the particular hypothesis used.  But this is seldom true
in real problems, so that one hypothesis must inevitably be chosen to
represent a class of hypotheses (\eg, approximate confidence regions
are found using calculations conditioning on the best-fit hypothesis).

This confusion cannot arise in the Bayesian approach.  One always
calculates probabilities for hypotheses, so there is never ambiguity
over what kind of hypothesis a probability is associated with:  one
must explicitly state it in order even to start the calculation.  If
one seeks a measure of how plausible it is for a parameter to lie in
some region, one simply calculates the probability that it is in that
region (parameter estimation).  If instead one wishes to assess an
entire model, one calculates the probability for that model as a whole
(model comparison).  The formalism forces one to distinguish between
these options.

\section{Conclusions}
\label{sec:conc}

Using the tools of Bayesian inference, we have performed an analysis of
the neutrinos from SN~1987A that differs significantly from previous
analyses, both in its methodology and in its results.

Methodologically, the key ingredient in our analysis is the likelihood
function, and our likelihood function differs from those used in
previous studies in several important respects.  It more consistently
accounts for the energy-dependent efficiencies of neutrino detectors,
it incorporates detailed information about the background spectra of
the detectors, and it accounts for dead time.  Our methodology allows
us to carefully quantify the uncertainty in our inferences in a way
that fully displays the strong correlations between inferred parameter
values.  Also, we have studied a much wider variety of neutrino
emission models than were studied previously.  The Bayesian approach
lets us use the likelihood function to calculate probabilities for
rival models that account for parameter uncertainty and implement an
automatic penalty for model complexity.  These features of our approach
insure that our conclusions are robust with respect to model 
uncertainty.

Our calculations indicate that the neutrino data strongly favor signal
models that have two components:  a long timescale component due to
Kelvin-Helmholtz cooling of the nascent neutron star, and a brief
($\lesssim 1$~s), softer component due to emission from material accreting
through a stalled supernova shock, as expected in the delayed scenario
for supernova explosions.  Such models make the data significantly more
probable than single-component cooling models motivated by the prompt
scenario for supernova explosions.  In addition, the radius and binding
energy of the nascent neutron star implied by single-component models
deviates significantly from the values predicted by current neutron
star models, whereas those implied by models with an accretion
component are in complete agreement with the predictions.  As a result,
two-component models are hundreds to thousands of times more probable
than single-component models.  The neutrino data thus provide the first
direct observational evidence in favor of the delayed scenario over the
prompt scenario.  Furthermore, the inferred characteristics of the
signal are in spectacular agreement with the salient features of the
theory of gravitational collapse and neutron star structure,
particularly when correlations between parameters are fully taken into
account in the comparison of theory with observation.  In addition to
studying the implications of the neutrino data for the formation of the
nascent neutron star, we have also used the data to find
model-dependent upper limits on the rest mass of the electron
antineutrino that are competitive with laboratory limits.

The detection of neutrinos from supernova SN 1987A initiated a new era
in astrophysics, the era of extrasolar neutrino astronomy.  Years
later, the supernova neutrinos detected by the Kamiokande-II, IMB, and
Baksan detectors are still offering us important lessons, not only
about the physics of supernovae and neutrinos, but also about the
potential of Bayesian methods for improving the analysis of complicated
astrophysical data.

\acknowledgments

We would like to express our thanks to the many colleagues who have
provided us with technical assistance and encouragement with respect to
the work reported here.  We thank Yoji Totsuka, Al Mann, Soo-Bong Kim,
Weiping Zhang, and the Kamiokande-II collaboration for providing
unpublished information about the background rate and spectrum of the
KII detector and for many valuable conversations.  We similarly thank
Jim Matthiu and Jack van der Velde for providing unpublished
information about the Irvine-Michigan-Brookhaven detector, and
Alexandar Chudakov for providing unpublished information about the
Baksan detector.  We are very grateful to H.-Thomas Janka for bringing
us up to date on the results of supernova calculations based on the
delayed explosion scenario, for guiding our development of accretion
models, and for his enthusiasm for our work.  We thank Arya Akmal for
providing detailed information about recent neutron star structure
calculations in electronic form.

We would also like to thank the institutions whose resources have
helped make this work possible.   During some of this work, TJL was
supported by Farr, McCormick, and Harper fellowships from the
University of Chicago, by a NASA Graduate Student Researchers Program
grant (NGT-50189), and by a NASA Compton Gamma Ray Observatory
Fellowship (NAG 5-1758).  Much of our code development, and many
initial calculations, were performed with the resources of the National
Center for Supercomputing Applications at the University of Illinois at
Urbana-Champaign.  Additional support for this work was provided by
NASA under grants NAG 5-2762, NAG 5-2868, NAG 5-3097, NAG5-3427, and by
the NSF under grants AST 91-19475 and AST 93-15375.

\appendix
\section{Derivation of the Likelihood Function}

We present here a derivation of the
full likelihood function for the supernova neutrino data,
equation~(\ref{Ltot}) in the main text.
The calculation is straightforward and the result is easy to understand,
as explained in Sec.~III.
However, we make some effort here to go through it in detail, both
to reveal several errors that were made in previous studies, and to
demonstrate how straightforward the calculation of such likelihoods is from a
Bayesian perspective.  Loredo and Wasserman
\cite{LW-95}\ used similar methods
to derive likelihood functions for Bayesian analyses of gamma-ray burst
data.

As with the derivation of the idealized likelihood in Sec.~III,
we first consider the probability for nondetections.  To do this,
we will use a standard ``trick'' from probability theory that frequently
arises in Bayesian calculations.  When we cannot directly calculate
$p(A | C)$, we introduce an exhaustive, exclusive set of auxiliary
propositions, $\{B_i\}$ (one and
only one of the $B_i$ must be true), such that we can calculate
$p(A | B_i,C)$.  Then we can find the the desired probability from
\begin{eqnarray}
p(A | C) & = & \sum_i p(A,B_i | C) \nonumber \\
  & = & \sum_i p(A | B_i,C)\; p(B_i | C), \label{trick}
\end{eqnarray}
provided we can calculate or specify $p(B_i | C)$.  If the $B_i$ form
a continuum, the sum becomes an integral.
This trick is sometimes referred to as ``extending the conversation.''

To apply this trick to calculate $p(\bar d_j | \pars,M)$, we
begin by noting that there are
many situations that can result in a nondetection.  If neither a signal
nor background event occurs, no detection will be reported.  But even if
one or more signal or background event occurs, it is possible no event
will be reported, because of the instrument threshold.  If we
let $\sprop^m$ denote the proposition that $m$ signal events occurred in
the time interval under consideration, and $\bprop^n$ denote the proposition
that $n$ background events occurred, then we can write the nondetection
probability as
\begin{equation}
p(\bar d_j |\pars,M) =
  \sum_{m=0}^\infty \sum_{n=0}^\infty
   p(\bar d_j,\sprop^m,\bprop^n |\pars,M). \label{pnd-all}
\end{equation}
Each term will involve poisson probabilities for $m$ signal events,
proportional to $(R\delta t)^m$, and $n$ background events,
proportional to $(B\delta t)^n$.  Since the
$\delta t$ intervals are small (in the sense that $R\delta t\ll1$ and
$B\delta t\ll 1$), we can neglect possibilities involving more than one
event occurring in $\delta t$.  This leaves three possibilities.
\begin{equation}
p(\bar d_j |\pars,M) \approx p(\bar d_j,\sprop^0,\bprop^0 |\pars,M)
   + p(\bar d_j,\sprop^1,\bprop^0 |\pars,M)
   + p(\bar d_j,\sprop^0,\bprop^1 |\pars,M). \label{pnd3}
\end{equation}

To calculate the first term, we first apply the product rule, writing
\begin{equation}
p(\bar d_j,\sprop^0,\bprop^0 |\pars,M)
  = p(\bar d_j |\sprop^0,\bprop^0,M)\,
     p(\sprop^0 |\pars,M)\,
     p(\bprop^0 |\pars,M). \label{pnd-1-1}
\end{equation}
Here we have dropped $\pars$ from the right of the bar in the
first probability, since it is irrelevant to $\bar d_j$ once we specify
that no events have occurred.  Also, we factored
the joint probability of
$(\sprop^0,\bprop^0)$ as the product of their independent probabilities to get
the last two factors.
The first factor---the probability for reporting no detection if neither a
signal nor a background event occurs---is simply equal to 1.  The second
and third factors are simply given by the Poisson probability for no
event, given the expected number in $\delta t$
(c.f.\ equation~(\ref{Pois-0})).  Thus
\begin{equation}
p(\bar d_j,\sprop^0,\bprop^0 |\pars,M)
  = e^{-[B + R(t)]\delta t}. \label{pnd-1}
\end{equation}

To calculate the second term in equation~(\ref{pnd3}),
we extend the conversation, resolving $\sprop^1$
into a continuum of $\sprop(\rvec,\drxn,\epos)$ propositions.  This gives
\begin{eqnarray}
p(\bar d_j,\sprop^1,\bprop^0 |\pars,M)
  & = & \int d\epos\int dV \int d\drxn \;
     p(\bar d_j,\sprop(\rvec,\drxn,\epos),\bprop^0 |\pars,M) \nonumber \\
  & = & \int d\epos\int dV \int d\drxn \;
     p(\bar d_j |\sprop(\rvec,\drxn,\epos),\bprop^0,M)\,
     p(\sprop(\rvec,\epos,\drxn) |\pars,M)\,
     p(\bprop^0 |\pars,M). \label{pnd-2-1}
\end{eqnarray}
The first factor in the integrand is the probability that a signal event
occurring at a specified position, with a specified energy and direction,
will lead to a nondetection.  We presume that the experiment team can
calculate this probability by detailed modeling of the detector (perhaps
including results of calibration measurements).  It is simply the
probability that the specified event will produce triggers that do not
satisfy the detection criterion.  We write this probability as
\begin{equation}
p(\bar d_j |\sprop^1(\rvec,\drxn,\epos),\bprop^0,\pars,M)
  = 1 - \eta(\rvec,\drxn,\epos),\label{eta-def}
\end{equation}
where $\eta(\rvec,\drxn,\epos)$ is the detection efficiency for events
with the specified position, energy and direction; we call this the {\it
full} detection efficiency.

The second factor in the integrand of
equation~(\ref{pnd-2-1})\ is the probability
for detecting the specified signal event, and no other, in $\delta t_j$.
It is simply given by the Poisson distribution:
\begin{equation}
p(\sprop^1(\rvec,\drxn,\epos) |\pars,M)
  = {R(\drxn,\epos,t_j) \over V} e^{- R(t_j)\delta t}.\label{pnd-2-2}
\end{equation}

The third factor in the integrand of equation~(\ref{pnd-2-1})\ is the
probability for no background events that we needed for the first term in the
nondetection probability, equal to $\exp(-B\delta t)$.
We thus have all the factors needed to calculate equation~(\ref{pnd-2-1}).
Since only the full efficiency factor depends
on $\rvec$, we can pull the signal rate through the volume
integral, writing
\begin{eqnarray}
p(\bar d_j,\sprop^1,\bprop^0 |\pars,M)
  & = & \delta t e^{-[B + R(t_j)]\delta t} \int d\drxn \int d\epos\;
      R(\drxn,\epos,t_j)[1-\bar\eta(\drxn,\epos)] \nonumber \\
  & = & e^{-[B + R_\epos(t_j)]\delta t}
     \left[ R(t_j)\delta t -
             \delta t \int d\drxn\int d\epos\;
                   R(\drxn,\epos,t_j)\bar\eta(\drxn,\epos)\right].
     \label{pnd-2-3}
\end{eqnarray}
Here we have defined the {\it volume-averaged} detection efficiency
according to
\begin{equation}
\bar\eta(\drxn,\epos) \equiv \int {dV \over V} \;
       \eta(\rvec,\drxn,\epos).\label{etab-aniso}
\end{equation}
We can write equation~(\ref{pnd-2-3})\ more succinctly by introducing an 
effective (detectable) signal rate,
\begin{equation}
\Reff(t) \equiv \int d\drxn \int d\epos\; \bar \eta(\drxn,\epos)\,
     R(\drxn,\epos,t).\label{Reff-def}
\end{equation}
Using this, equation~(\ref{pnd-2-3})\ becomes
\begin{equation}
p(\bar d_j,\sprop^1,\bprop^0 |\pars,M)
  = e^{-[B + R_\epos(t_j)]\delta t}\, \delta t
       [R(t_j) - \Reff(t_j)].\label{pnd-2-3-eff}
\end{equation}

The last probability we need in order to calculate the nondetection
probability---the last term in equation~(\ref{pnd3})---is very similar
to the one we
have just calculated.  We can get it simply by switching the roles of
background and signal, taking into account the fact that the background
rate may depend on position and direction.  This gives
\begin{equation}
p(\bar d_j,\sprop^0,\bprop^1 |\pars,M)
  = e^{-[B + R(t_j)]\delta t}\,
     \delta t( B - \Beff ),
   \label{pnd-3}
\end{equation}
where the effective background rate is given by
\begin{equation}
\Beff = \int d\epos\int dV \int d\drxn\;
  \eta(\rvec,\drxn,\epos) B(\rvec,\drxn,\epos).\label{Bbar-def}
\end{equation}
We cannot use $\bar \eta(\drxn,\epos)$ here because $B(\rvec,\drxn,\epos)$
is a function of position in the detector (e.g., due to radioactivity
in the rock surrounding the detector).
We have presumed here that the full efficiency for detecting a background
event with specified position, direction, and energy is the same as that
for detecting a signal event with the same properties.  That is, we are
assuming that the detector does not distinguish background and signal
events by some other property.

Assembling all of the ingredients, we can now write down the full
nondetection probability:
\begin{equation}
p(\bar d_j | \pars,M) = e^{-[B + R(t_j)]\delta t}
  \left(1 + \delta t[R(t_j) + B] -
     \delta t[\Reff(t_j) + \Beff]\right).\label{pnd}
\end{equation}
Since we will need the product of many such probabilities,
its logarithm is easier to work with.  Taking advantage of the fact
that $R\delta t\ll 1$ and $B\delta t\ll 1$, and using
$\log(1+x)\approx x$ for small $x$, we find
\begin{equation}
\log[p(\bar d_j | \pars,M)]
  \approx - \delta t[\Reff(t_j) + \Beff].\label{log-pnd}
\end{equation}
The product of all the nondetection probabilities will thus be an
exponential with sums of the effective rates over all nondetection
intervals.  This sum is just the integral of the effective rates over
the nondetection intervals, so the product of nondetection probabilities
can be written
\begin{equation}
\prod_j p(\bar d_j | \pars,M)
  = \exp\left[-\Beff T_{\rm nd} -
       \int_{T_{\rm nd}} dt\, \Reff(t)\right],\label{pnd-prod}
\end{equation}
where $\int_{T_{\rm nd}} dt$ denotes integration of the (disjoint)
intervals of time without detections.

Now we turn to the detection probabilities.  A reported event can be
either a signal or a background event, so we have
\begin{equation}
p(d_i | \pars, M) = p(d_i,\sprop^1,\bprop^0 | \pars,M)
   + p(d_i,\sprop^0,\bprop^1 | \pars,M).\label{pd2}
\end{equation}
As with the nondetection probability, we ignore possibilities that are
higher than first order in $\delta t$.

We can calculate the first term by introducing
$\sprop(\rvec,\drxn,\epos)$ and applying the product rule, just as we
did in equation~(\ref{pnd-2-1}).  The result is
\begin{equation}
p(d_i,\sprop^1,\bprop^0 | \pars,M)
  = \delta t \int d\epos\int dV\int d\drxn \;
    \like_i(\rvec,\drxn,\epos)
     {R(\drxn,\epos,t_i) \over V}
     \exp\left(-[R(t_i)+B]\delta t\right).\label{pd-1}
\end{equation}
Here we have defined the {\it individual event likelihood function}
according to
\begin{equation}
\like_i(\rvec,\drxn,\epos) \equiv
  p(d_i | \sprop(\rvec,\drxn,\epos),M).\label{likei-def}
\end{equation}
This is just the probability for observing the detection data, presuming
the location, direction, and energy of the lepton producing the data have
the specified values.  It is the likelihood function we would use to
infer the properties of a particular detected event.
Detailed knowledge of the detector should allow
experimenters to calculate this function for each detected event (by
fitting the PMT data).  Since $\like_i(\rvec,\drxn,\epos)$
is a probability for $d_i$, it need not be normalized when
integrated over $(\rvec,\drxn,\epos)$.  However, $\like_i$ can be
multiplied by any constant without affecting our inferences (since the
constant will drop out in Bayes's theorem), and we will find it
convenient to adopt the convention that the reported individual
likelihood functions include a constant that makes them normalized when
integrated over $(\rvec,\drxn,\epos)$.

The second term in equation~(\ref{pd2})\ can be calculated in exactly the
same way, switching the roles of the signal and background rates.
Combining this term with equation~(\ref{pd-1}) gives us the 
detection probability,
\begin{equation}
p(d_i | \pars, M) =
\delta t e^{-[R(t_i)+B]\delta t}
  \int d\epos \int dV \int d\drxn\;  \like_i(\rvec,\drxn,\epos)\,
     \left[{R(\drxn,\epos,t_i)\over V} + 
            B(\rvec,\drxn,\epos)\right].\label{pd-full}
\end{equation}
We can take advantage of the homogeneity of the signal rate to replace the
signal-dependent integral with
\begin{equation}
\int d\epos \int d\drxn\,
 \like_i(\drxn,\epos)\, R(\drxn,\epos,t_i),\label{sig-part}
\end{equation}
where the volume-averaged event likelihood function is given by
$\like_i(\drxn,\epos) = \int dV \like_i(\rvec,\drxn,\epos)/V$.
We retain the simple likelihood notation for this and other averaged
likelihoods, because this is in fact the likelihood for the direction
and energy:
\begin{eqnarray}
\like_i(\drxn,\epos) &\equiv& p(d_i | \sprop(\drxn,\epos), M) \nonumber\\
  &=& \int dV p(d_i, \rvec  | \sprop(\drxn,\epos), M) \nonumber\\
  &=& \int dV p(\rvec | M) \like_i(\rvec,\drxn,\epos).
\label{avg-Li}
\end{eqnarray}
Taking the prior density for the event position to be uniform throughout
the tank reveals $\like_i(\drxn,\epos)$ to be the volume-averaged
event likelihood, as claimed.

To further simplify the appearance of our equations, we introduce
the event-averaged background rate, $B_i$, according to
\begin{equation}
B_i \equiv \int d\epos \int dV \int d\drxn \,
      \like_i(\rvec,\drxn,\epos)\,B(\rvec,\drxn,\epos). \label{Bi-def}
\end{equation}
With our convention of normalizing $\like_i$, this can be interpreted as
the rate of background events ``like'' event number $i$ in the sense of
having positions, directions, and energies consistent with the data
for that event.  These
definitions let us write the detection probability as
\begin{equation}
p(d_i | \pars, M) =
\delta t \exp\left(-[R(t_i)+B]\delta t\right)
  \left[B_i\delta t + \int d\epos \int d\drxn\,\like_i(\drxn,\epos)\,
      R(\drxn,\epos,t_i)\right].
   \label{pd-simp}
\end{equation}

Combining the detection and nondetection probabilities gives us the
full likelihood function, 
\begin{eqnarray}
\like(\pars)
  &=& (\delta t)^\Nd \exp\left[-\Beff T -
       \int_T dt\int d\epos \int d\drxn\,
           \bar\eta(\drxn,\epos) R(\drxn,\epos,t)\right] \nonumber \\
  &\quad\times \prod_{i=1}^\Nd
   \left[B_i + \int d\epos \int d\drxn\,\like_i(\drxn,\epos)\,
             R(\drxn,\epos,t_i)\right].
  \label{Ltot-B}
\end{eqnarray}
Here we have combined the exponentials in the detection factors
appearing in equation~(\ref{pd-simp})\ with the exponents in
the nondetection probabilities to give an integral over the
{\it entire} duration
of the data.  In doing so, we have neglected the difference between the
full and effective rates in the $N$ detection intervals; but this
difference is very small provided that $\delta t\ll T$, and one
can easily demonstrate that it has a negligible effect on inferences.

One last simplification can be made.  Since scaling by a
parameter-independent factor does not affect our inferences, we can drop
the $(\delta t)^N$ factor and the $\Beff T$ exponent from the likelihood.
This leads to equation~(\ref{Ltot}), the full likelihood used in the main
text.

\section{Two-Dimensional Kolmogorov-Smirnov Tests}

In Table~\ref{table:ks} we present the results of two-dimensional
Kolmogorov-Smirnov (KS) goodness-of-fit tests applied to the constant
temperature/radius model, the exponential cooling model, and the model
combining displaced power-law cooling and truncated accretion, each
with parameters fixed at their best-fit values.  We used the version of
the test devised by Fasano and Franceschini \cite{FF-87}.  This test
compares the fraction of the expected rate in four quadrants about the
point $(t_i,\epsilon_i)$ associated with each event with the fraction
of the number of detected events in that quadrant.  The largest
difference between the observed and expected values is the KS
statistic, $D$.  The model is rejected if $D$ is too large, the typical
critical value being that associated with a 95\% false rejection rate.
This test ignores the uncertainty in $\epsilon$ for each event, and the
quoted significance values are approximate (they are based on an
approximate expression for the distribution for $D$).  We performed the
test separately for each detector (using the best-fit parameters from a
joint fit), and then combined the test results using standard methods
to find the significance associated with the joint fit \cite{Eadie}.
These results indicate moderate incompatibility of the data with the
constant temperature model, and compatibility with the other models.

Bayesian inference does not include such a thing as an alternative-free
goodness-of-fit test; we provide these tests for those readers who find
them useful.  KS tests have several limitations that must be kept in
mind when interpreting their results.  First, the one-dimensional and
two-dimensional KS tests are sensitive only to the shape of a
distribution, not its amplitude.  The test may be straightforwardly
extended to include the amplitude, but the resulting test then becomes
insensitive to the shape of the distribution for the supernova neutrino
data because Poisson fluctuations in the number of events detected,
rather than the positions of the events in the time-energy plane,
dominate $D$.  Second, the two-dimensional test lacks the
distribution-free property that makes the one-dimensional test
attractive.  In fact, there are different generalizations of the test
to two dimensions, each with different sensitivity to the parent
distribution \cite{FF-87,Peacock}.  Thus, the test should ideally be
calibrated with extensive Monte Carlo calculations.  Finally, the
reliance of the test on the cumulative distribution of events, rather
than the differential distribution, can make it insensitive to local
structure present in the model (e.g., it can accept a model even if
there are data in regions of zero probability).

On a more subjective level, our extensive experience with application
of this test to these data has led us to be skeptical of its value.  We
have found it to be quite insensitive, accepting models that seem
clearly unacceptable on other grounds (either to the trained eye or
based on tests with likelihood functions).  Some evidence of this
behavior is obvious here:  the best-fit cooling and accretion models
have comparable values of $P(>D_{\rm obs})$, despite the fact that the
latter model makes the data over 600 times more probable than the
former.  Finally, we note that some earlier studies attempted to assess
joint fits by applying a single KS test to a fictitious ``sum''
detector whose expected rate is the sum of the rates of the considered
detectors, and whose data is the collected data of the detectors
\cite{Spergel-87}.  This procedure corrupts the test, as it ignores
information about which events to associate with which expected rate.
We have found that some models that are accepted with a KS test based
on such a ``sum'' detector can be rejected by a combination of tests
applied to the individual detectors, and vice versa.  This is because
no detected event represents a sample of the summed detector rates,
leading to erroneous results when the test is performed with the
``sum'' detector.


\newpage

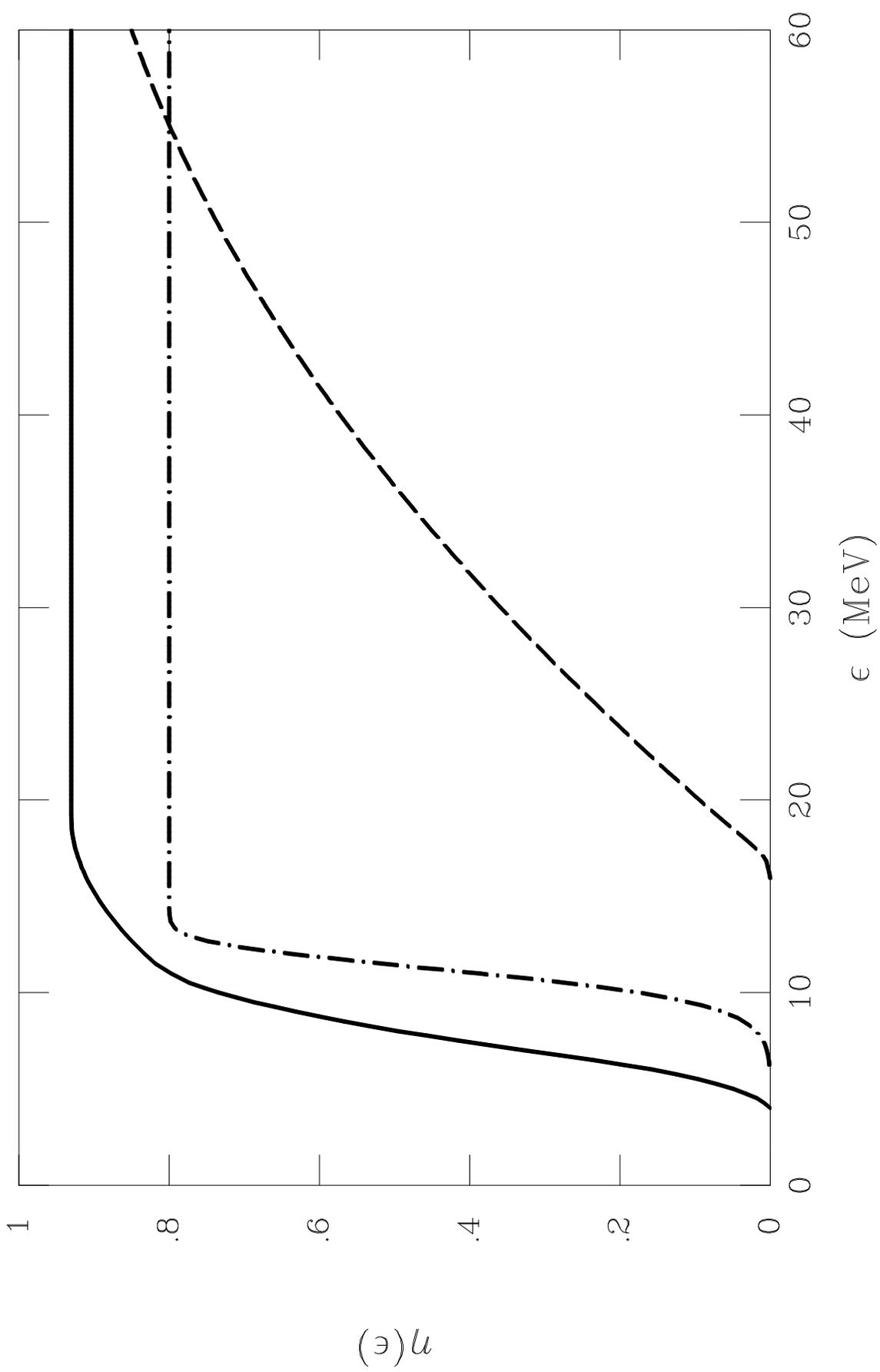
\begin{figure}
\caption{Average
efficiency functions, $\bar\eta(\epsilon)$, for the
KII detector (solid curve), IMB detector (dashed curve) and the Baksan detector
(dot-dashed curve).}
\label{fig-eta}
\end{figure}

\begin{figure}
\caption{{\it (a)} Background spectrum, $B(\epsilon)$,
for the KII detector; measured values are shown as points with error bars,
the interpolated function is shown as a solid curve.
{\it (b)} As in {\it (a)}, for the Baksan detector.}
\label{fig-B}
\end{figure}

\begin{figure}
\caption{One-dimensional marginal distributions for parameters of the
exponential cooling model.}
\label{fig-exp1d}
\end{figure}

\begin{figure}
\caption{Two-dimensional marginal distributions for the parameters of the
exponential cooling model.  Contours indicate the boundaries of
68\% (dashed) and 95\% (solid) credible regions. Points indicated
the coordinates of 500 samples from the distributions.}
\label{fig-exp2d}
\end{figure}

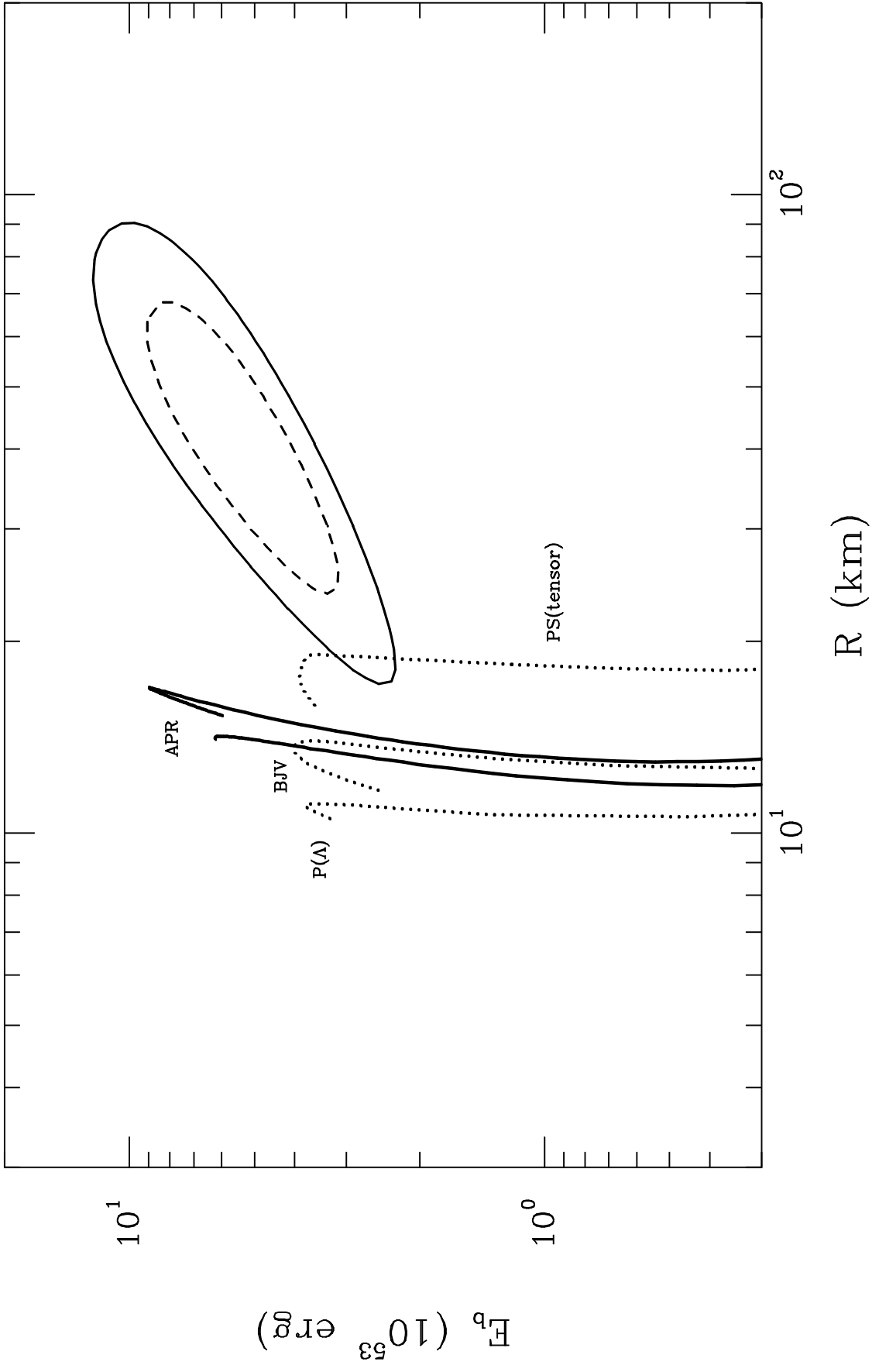
\begin{figure}
\caption{Joint marginal distribution for the logarithm of the
observed radius, $R$, and the logarithm of the
binding energy, $E_b$, of the nascent neutron star, based on the
exponential cooling model.  
Contours indicate the boundaries of
68\% (dashed) and 95\% (solid) credible regions.  $(R_{\rm obs},E_b)$-curves 
for neutron star models based
on a representative set of equations of state appear as solid lines.}
\label{fig-expER}
\end{figure}

\begin{figure}
\caption{Contour plots of predicted detection rates in each detector
for the best-fit exponential cooling model.  Contours enclose 68\%
(dashed curve) and 95\% (solid curve) of the total predicted number
of detectable neutrinos.  Points indicate the inferred energies and
arrival times of the detected events.}
\label{fig-expet}
\end{figure}

\begin{figure}
\caption{Summaries of the posterior distribution for parameters describing
the accretion component of the displaced power law cooling and truncated
accretion model.  ({\it a})  The profile likelihood for the dimensionless
accretion mass, $\mu$.  ({\it b})  The posterior for $T_a$ and $\tau_a$,
for $\mu=0.5$ and $\tau_c=14.7$ s, maximized with respect to
$\alpha$ and $T_{c,0}$.}
\label{fig-acn}
\end{figure}

\begin{figure}
\caption{One-dimensional marginal distributions for parameters of the
cooling component of the best accretion model.}
\label{fig-acn1d}
\end{figure}

\begin{figure}
\caption{Two-dimensional marginal distributions for the parameters of the
cooling component of the best accretion model.
Contours indicate the boundaries of
68\% (dashed) and 95\% (solid) credible regions.  Points indicate the
coordinates of 500 samples from the distributions.}
\label{fig-acn2d}
\end{figure}

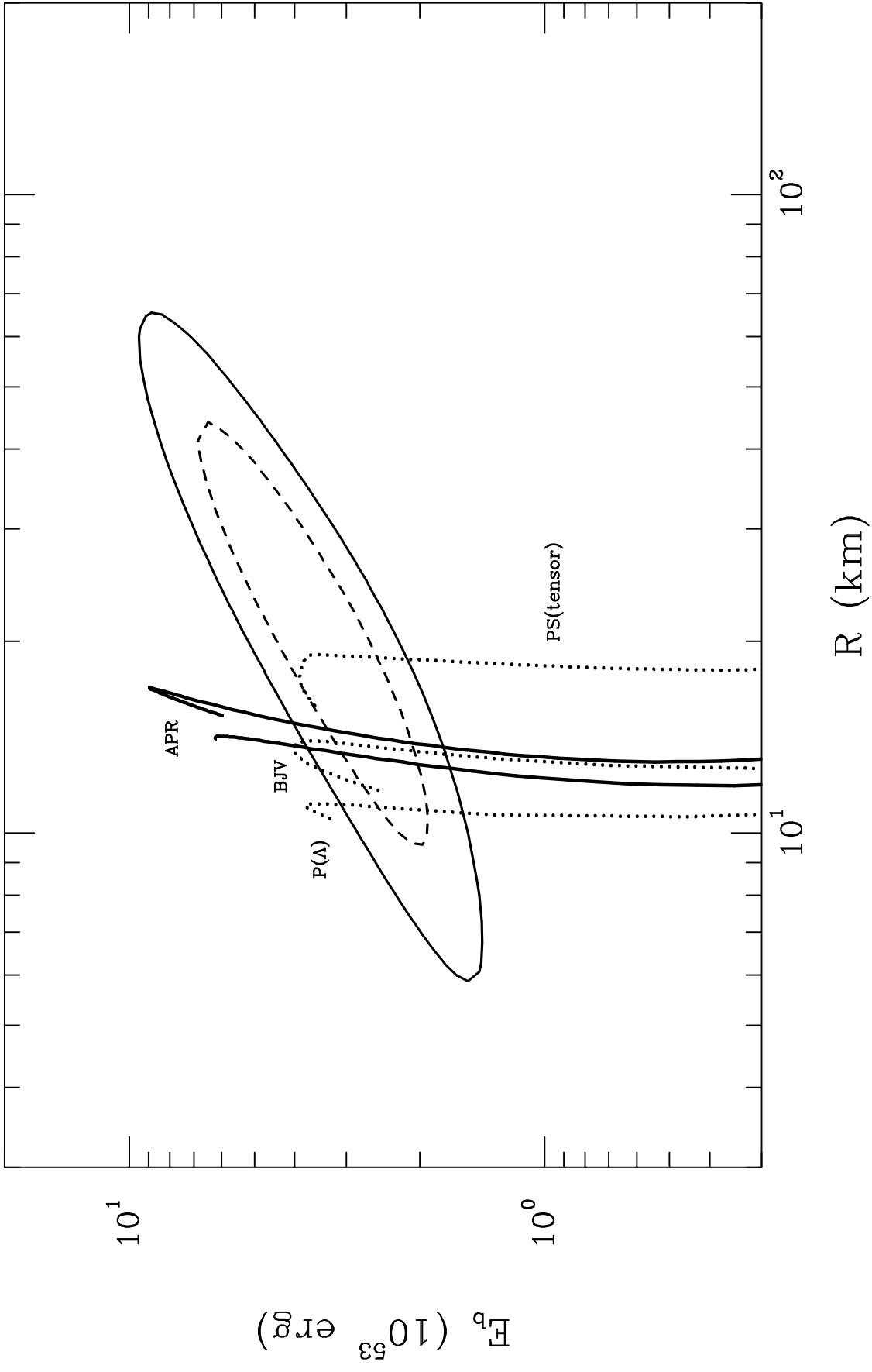
\begin{figure}
\caption{Joint marginal distribution for the logarithm of the
observed radius, $R$, and the logarithm of the
binding energy, $E_b$, of the nascent neutron star, based on the
displaced power law cooling and truncated accretion model, with accretion
parameters fixed at their best-fit values.
Contours indicate the boundaries of
68\% (dashed) and 95\% (solid) credible regions.  $(R_{\rm obs},E_b)$-curves 
for neutron star models based
on a representative set of equations of state appear as solid lines.}
\label{fig-acnER}
\end{figure}

\begin{figure}
\caption{Contour plots of predicted detection rates in each detector
for the best-fit accretion model.  Contours enclose 68\%
(dashed curve) and 95\% (solid curve) of the total predicted number
of detectable neutrinos.  Points indicate the inferred energies and
arrival times of the detected events.}
\label{fig-acnet}
\end{figure}

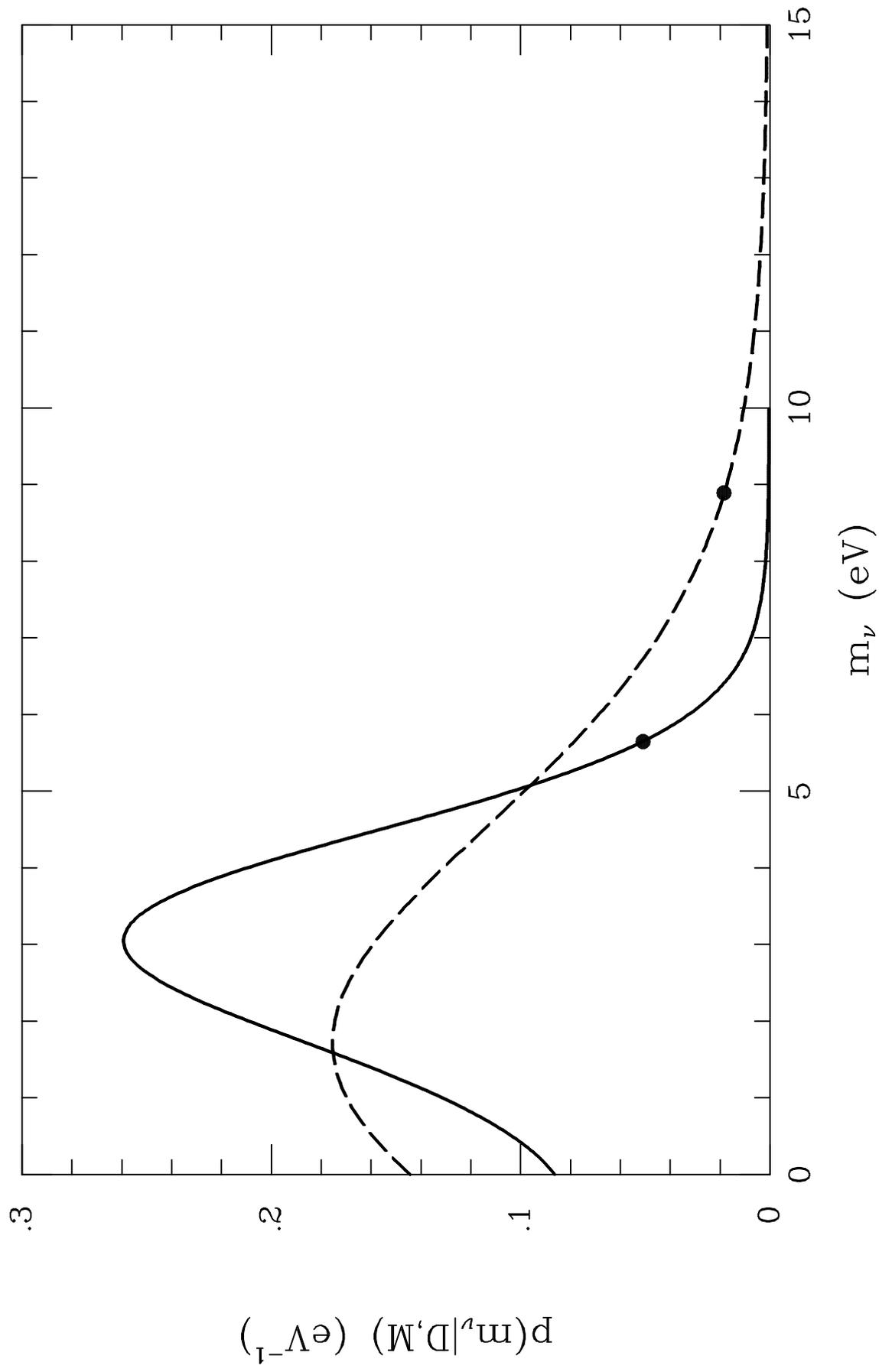
\begin{figure}
\caption{Marginal distributions for the electron antineutrino
rest mass $\mnu$, using the exponential cooling model (dashed) and
the cooling plus accretion model (solid).  Dots indicate the upper
bounds of 95\% credible regions.}
\label{fig-m}
\end{figure}

\newpage

\narrowtext
\begin{table}
\caption{Interpretation of Bayes Factors.\label{table:B}}
\begin{tabular}{ccl}
$\ln(B_{ij})$  &  $B_{ij}$  &  Evidence against $H_j$ \\
\tableline
0 to 1 & 1 to 3 & Not worth more than a bare mention\\
1 to 3 & 3 to 20 & Positive\\
3 to 5 & 20 to 150 & Strong\\
$>5$ & $>150$ & Very Strong\\
\end{tabular}
\end{table}

\mediumtext
\begin{table}
\caption{Detector characteristics used in evaluating the likelihood function.
\label{table:det}}
\begin{tabular}{lccc}
Characteristic &  Kamiokande II & IMB & Baksan\\
\tableline
Effective H$_2$0 Mass, $M_{\rm eff}$ (kton) & 2.14 & 6.8 & 0.28 \\
Background Rate, $R_{\rm bg}$ (events ${\rm s}^{-1}$) &
    $0.187$ & $0$ & 0.0345 \\
Dead Time, $\tau$ (s) &     0 & 0.035 & 0 \\
Live Fraction, $f$ &     1 & 0.9055 & 1 \\
\end{tabular}
\end{table}

\narrowtext
\begin{table}
\caption{Detected event data used in evaluating the likelihood function.
\label{table:data}}
\begin{tabular}{rdddc}
    & $t_i$ &  $\epsilon_i$ & $\sigma_i$  & $B_i$ \\
Event  & (s)  &  (MeV) &  (MeV)  & (s$^{-1}$) \\
\tableline
\multicolumn{5}{c}{Kamiokande II}\\
1 & $\equiv$ 0.0 & 20.0 & 2.9 & $1.6\times10^{-5}$ \\
2 & 0.107          & 13.5 & 3.2 & $1.9\times 10^{-3}$ \\
3 & 0.303          & 7.5 & 2.0 & $2.9\times10^{-2}$ \\
4 & 0.324 & 9.2 & 2.7 & $1.2\times10^{-2}$ \\
5 & 0.507 & 12.8 & 2.9  & $2.1\times10^{-3}$ \\
6\rlap{\tablenotemark[1]} & 0.686 & 6.3 & 1.7 & $3.7\times10^{-2}$ \\
7 & 1.541 & 35.4 & 8.0 & $4.5\times10^{-5}$ \\
8 & 1.728 & 21.0 & 4.2 & $8.2\times10^{-5}$ \\
9 & 1.915 & 19.8 & 3.2 & $1.5\times10^{-5}$ \\
10 & 9.219 & 8.6 & 2.7 & $1.5\times10^{-2}$ \\
11 & 10.433 & 13.0 & 2.6 & $1.9\times10^{-3}$ \\
12 & 12.439 & 8.9 & 1.9 & $1.6\times10^{-2}$ \\
13\rlap{\tablenotemark[1]} & 17.641 & 6.5 & 1.6\tablenotemark[2] %
    & $3.8\times10^{-2}$ \\
14\rlap{\tablenotemark[1]} & 20.257 & 5.4 & 1.4\tablenotemark[2] %
      & $2.9\times10^{-2}$ \\
15\rlap{\tablenotemark[1]} & 21.355 & 4.6 & 1.3\tablenotemark[2] %
      & $2.8\times10^{-2}$ \\
16\rlap{\tablenotemark[1]} & 23.814 & 6.5 & 1.6\tablenotemark[2] %
      & $3.8\times10^{-2}$ \\
\tableline
\multicolumn{5}{c}{IMB}\\
1  &$\equiv$ 0.0 & 38 & 7 & 0 \\
2  & 0.412 & 37 & 7 & 0 \\
3  & 0.650 & 28 & 6 & 0 \\
4  & 1.141 & 39 & 7 & 0 \\
5  & 1.562 & 36 & 9 & 0 \\
\noalign{\smallskip}
6  & 2.684 & 36 & 6 & 0 \\
7  & 5.010 & 19 & 5 & 0 \\
8  & 5.582 & 22 & 5 & 0 \\
\tableline
\multicolumn{5}{c}{Baksan}\\
1  &$\equiv$ 0.0 & 12.0 & 2.4 & $8.4\times10^{-4}$ \\
2  & 0.435 & 17.9 & 3.6 & $1.3\times10^{-3}$ \\
3  & 1.710 & 23.5 & 4.7 & $1.2\times10^{-3}$ \\
4  & 7.687 & 17.6 & 3.5 & $1.3\times10^{-3}$ \\
5  & 9.099 & 20.3 & 4.1 & $1.3\times10^{-3}$ \\
\end{tabular}
\tablenotetext[1]{Omitted as a background event
by other investigators.}
\tablenotetext[2]{Calculated using a linear
fit of $\sigma_i$ vs.\ $\epsilon_i$ for earlier events.}
\end{table}

\narrowtext
\begin{table}
\caption{Properties of best-fit single component cooling models.
\label{table:one}}
\begin{tabular}{lcc}
Quantity &  KII--IMB--Baksan & KII--IMB  \\
\multicolumn{3}{c}{Constant Temperature}\\
$\alpha$ & 3.20 & 2.43  \\
$T_0$ (MeV) & 3.30 & 3.51  \\
$t_{\rm burst}$ (s) & 10.43 & 10.43  \\
${\cal L}$ & \multicolumn{1}{c}{$2.4 \times 10^{-5}$} &  %
         \multicolumn{1}{c}{$2.5 \times 10^{-5}$}  \\
\noalign{\bigskip}
$N_{\rm det}$ (KII) & 16.6 + 5.6\tablenotemark[1]
          & 13.6 + 5.6\tablenotemark[1]  \\
$N_{\rm det}$ (IMB) & 4.3 & 4.1 \\
$N_{\rm det}$ (Baksan) & 1.8 + 1.0\tablenotemark[1] & -- \\
$R$ (km) & 32.0  & 24.3 \\
$E_b$ ($10^{53}$ erg) & 4.30 & 3.19 \\
\tableline
%
\multicolumn{3}{c}{Exponential Dilution}\\
$\alpha$ & 6.69 & 5.63 \\
$T_0$ (MeV) & 3.43 & 3.61  \\
$\tau$ (s) & 1.75 & 1.61 \\
${\cal L}$ & 1.77 & 1.66 \\
\noalign{\bigskip}
$N_{\rm det}$ (KII) & 15.1 + 5.6\tablenotemark[1]
         & 13.0 + 5.6\tablenotemark[1]  \\
$N_{\rm det}$ (IMB) & 4.0 & 4.0 \\
$N_{\rm det}$ (Baksan) & 1.6 + 1.0\tablenotemark[1] & --  \\
$R$ (km) & 66.9  & 56.3 \\
$E_b$ ($10^{53}$ erg) & 3.68 & 2.93 \\
\tableline
%
\multicolumn{3}{c}{Exponential Cooling}\\
$\alpha$ & 4.02 & 3.42  \\
$T_0$ (MeV) & 3.81 & 3.98 \\
$\tau$ (s) & 4.37 & 3.97 \\
${\cal L}$ & \llap{$\equiv$}1.0 & \llap{$\equiv$}1.0  \\
\noalign{\bigskip}
$N_{\rm det}$ (KII) & 16.9 + 5.6\tablenotemark[1]
           & 14.4 + 5.6\tablenotemark[1]  \\
$N_{\rm det}$ (IMB) & 4.0 & 3.9 \\
$N_{\rm det}$ (Baksan) & 1.8 + 1.0\tablenotemark[1] & --  \\
$R$ (km) & 40.2  & 34.2 \\
$E_b$ ($10^{53}$ erg) & 5.02 & 3.96 \\
\tableline
%
\multicolumn{3}{c}{Displaced Power Law Cooling}\\
\noalign{\smallskip}
$\alpha$ & 4.72 & 4.05 \\
$T_0$ (MeV) & 4.02 & 4.17 \\
$\tau$ (s) & 1.30 & 1.24 \\
$\gamma$ & 0.34\tablenotemark[2] & 0.34\tablenotemark[2]  \\
${\cal L}$ & 7.8 & 4.5 \\
\noalign{\bigskip}
$N_{\rm det}$ (KII) & $18.2 + 5.6$\tablenotemark[1]
         & $15.9 + 5.6$\tablenotemark[1]  \\
$N_{\rm det}$ (IMB) & 3.8 & 3.7 \\
$N_{\rm det}$ (Baksan) & $1.9 + 1.0$\tablenotemark[1] & -- \\
$R$ (km) & 47.2 & 40.5 \\
$E_b$ ($10^{53}$ erg) & 10.2 & 8.33 \\
\tableline
%
\multicolumn{3}{c}{Thermal Rise and Fall}\\
\multicolumn{3}{c}{With Contraction}\\
\noalign{\smallskip}
$\alpha$ & 2.44 & 2.20 \\
$T_0$ (MeV) & 4.01 & 4.16 \\
$t_{rise}$ (s) & 1.32 & 1.30 \\
$\tau$ (s) & 5.49 & 4.82 \\
$a$ & 15.4 & 13.3 \\
$t_{\rm off}$ (KII) (s) & 0.71 & 0.70 \\
$t_{\rm off}$ (IMB) (s) & 1.09 & 1.06 \\
$t_{\rm off}$ (Baksan) (s) & 0.74 & -- \\
${\cal L}$ & 5.5 & 1.4 \\
\noalign{\bigskip}
$N_{\rm det}$ (KII) & $16.9 + 5.6$\tablenotemark[1]
        & $14.4 + 5.6$\tablenotemark[1]  \\
$N_{\rm det}$ (IMB) & 4.2 & 4.0 \\
$N_{\rm det}$ (Baksan) & $1.8 + 1.0$\tablenotemark[1] & --  \\
$R$ (km) & 24.4 & 22.0 \\
$E_b$ ($10^{53}$ erg) & 28.3 & 19.6 \\
\end{tabular}
\tablenotetext[1]{Expected numbers of signal and background
events are listed separately.}
\tablenotetext[2]{Best-fit values are the lowest values permitted in
the fit.}
\end{table}

\narrowtext
\begin{table}
\caption{Properties of best-fit two-component cooling plus accretion models.
\label{table:two}}
\begin{tabular}{lcc}
Quantity &  KII--IMB--Baksan & KII--IMB  \\
\tableline
\multicolumn{3}{c}{Exponential Cooling}\\
\multicolumn{3}{c}{and Truncated Accretion}\\
\noalign{\smallskip}
$\alpha$ & 1.71 & 1.48 \\
$T_{c,0}$ (MeV) & 4.56 & 4.83  \\
$\tau_c$ (s) & 5.15 & 4.39  \\
$T_{a,0}$ (MeV) & 2.02 & 1.96  \\
$\tau_a$ (s) & 0.74 & 0.76  \\
$\mu$ & \llap{$\equiv$}0.5 & \llap{$\equiv$}0.5  \\
${\cal L}$ & 577 & 101   \\
\noalign{\bigskip}
$N_{\rm det}$ (KII) & $15.8 + 5.6$\tablenotemark[1]
        & $13.4 + 5.6$\tablenotemark[1]  \\
$N_{\rm det}$ (IMB) & 4.5 & 4.3 \\
$N_{\rm det}$ (Baksan) & $1.7 + 1.0$\tablenotemark[1] & -- \\
$R$ (km) & 17.1 & 14.8 \\
$E_b$ ($10^{53}$ erg)\tablenotemark[2] & 2.84 (0.63) & 2.31 (0.54) \\
\tableline
%
\multicolumn{3}{c}{Displaced Power Law Cooling}\\
\multicolumn{3}{c}{and Truncated Accretion}\\
\noalign{\smallskip}
$\alpha$ & 1.80 & 1.58 \\
$T_{c,0}$ (MeV) & 4.64 & 4.89  \\
$\tau_c$ (s) & 14.7 & 12.5  \\
$T_{a,0}$ (MeV) & 2.00 & 1.94  \\
$\tau_a$ (s) & 0.74 & 0.76  \\
$\mu$ & \llap{$\equiv$}0.5 & \llap{$\equiv$}0.5  \\
${\cal L}$ & 624 & 118   \\
\noalign{\bigskip}
$N_{\rm det}$ (KII) & $15.9 + 5.6$\tablenotemark[1]
       & $13.6 + 5.6$\tablenotemark[1]  \\
$N_{\rm det}$ (IMB) & 4.5 & 4.3 \\
$N_{\rm det}$ (Baksan) & $1.7 + 1.0$\tablenotemark[1] & -- \\
$R$ (km) & 18.0 & 15.8 \\
$E_b$ ($10^{53}$ erg)\tablenotemark[2] & 3.08 (0.61) & 2.53 (0.51) \\
\tableline
%
\multicolumn{3}{c}{Displaced Power Law Dilution}\\
\multicolumn{3}{c}{and Truncated Accretion}\\
\noalign{\smallskip}
$\alpha$ & 5.75 & 4.79 \\
$T_{c,0}$ (MeV) & 3.73 & 3.94  \\
$\tau_c$ (s) & 1.31 & 1.20  \\
$T_{a,0}$ (MeV) & 1.88 & 1.82  \\
$\tau_a$ (s) & 0.73 & 0.76  \\
$\mu$ & \llap{$\equiv$}0.5 & \llap{$\equiv$}0.5  \\
${\cal L}$ & 138 & 32   \\
\noalign{\bigskip}
$N_{\rm det}$ (KII) & $15.4 + 5.6$\tablenotemark[1]
         & $13.1 + 5.6$\tablenotemark[1]  \\
$N_{\rm det}$ (IMB) & 4.34 & 4.2 \\
$N_{\rm det}$ (Baksan) & $1.6 + 1.0$\tablenotemark[1] & -- \\
$R$ (km) & 57.5 & 47.9 \\
$E_b$ ($10^{53}$ erg)\tablenotemark[2] & 3.26 (0.40) & 2.61 (0.35) \\
\tableline
%
\multicolumn{3}{c}{Displaced Power Law Dilution/Cooling}\\
\multicolumn{3}{c}{and Truncated Accretion}\\
\noalign{\smallskip}
$\alpha$ & 1.99 & 1.76 \\
$T_{c,0}$ (MeV) & 4.47 & 4.72  \\
$\tau_c$ (s) & 20.1 & 16.9  \\
$T_{a,0}$ (MeV) & 2.00 & 1.94  \\
$\tau_a$ (s) & 0.74 & 0.76  \\
$\mu$ & \llap{$\equiv$}0.5 & \llap{$\equiv$}0.5  \\
${\cal L}$ & 399 & 81   \\
\noalign{\bigskip}
$N_{\rm det}$ (KII) & $15.6 + 5.6$\tablenotemark[1]
          & $13.2 + 5.6$\tablenotemark[1]  \\
$N_{\rm det}$ (IMB) & 4.5 & 4.3 \\
$N_{\rm det}$ (Baksan) & $1.7 + 1.0$\tablenotemark[1] & -- \\
$R$ (km) & 19.9 & 17.6 \\
$E_b$ ($10^{53}$ erg)\tablenotemark[2] & 2.77 (0.61) & 2.27 (0.51) \\
\tableline
%
\multicolumn{3}{c}{Exponential Cooling}\\
\multicolumn{3}{c}{and Displaced Power-Law Accretion}\\
\noalign{\smallskip}
$\alpha$ & 2.33 & 2.01 \\
$T_{c,0}$ (MeV) & 4.10 & 4.32 \\
$\tau_c$ (s) & 5.43 & 4.74  \\
$T_{a,0}$ (MeV) & 2.45 & 2.40 \\
$\tau_a$ (s) & 0.05\rlap{\tablenotemark[3]} &
      0.05\rlap{\tablenotemark[3]}  \\
$\mu$ & \llap{$\equiv$}0.5  & \llap{$\equiv$}0.5  \\
${\cal L}$ & 384  & 32  \\
\noalign{\bigskip}
$N_{\rm det}$ (KII) & $16.5 + 5.6$\tablenotemark[1]
         & $14.2 + 5.6$\tablenotemark[1]  \\
$N_{\rm det}$ (IMB) & 4.1 & 3.9 \\
$N_{\rm det}$ (Baksan) & $1.8 + 1.0$\tablenotemark[1] & --  \\
$R$ (km) & 23.3 & 20.1 \\
$E_b$ ($10^{53}$ erg)\tablenotemark[2] & 3.27 (0.44) & 2.64 (0.40)  \\
\end{tabular}
\tablenotetext[1]{Expected numbers of signal and background
events are listed separately.}
\tablenotetext[2]{Total $E_b$ is given, with part due to accretion
in parentheses.}
\tablenotetext[3]{Minimum value allowed in fit.}
\end{table}

\narrowtext
\begin{table}
\caption{Background probabilities for KII and Baksan events for the best-fit
Exponential Cooling model and the best-fit Displaced Power Law Cooling and
Truncated Accretion model.
\label{table:bg}}
\begin{tabular}{rcc}
Event & Cooling & Accretion \\
\tableline
\multicolumn{3}{l}{Kamiokande II}\\
1 & $5.8\times10^{-5}$ & $2.4\times10^{-5}$\\
2 & $6.3\times10^{-3}$ & $1.9\times10^{-3}$\\
3 & 0.16 & $4.7\times10^{-2}$\\
4 & $5.4\times10^{-2}$ & $1.7\times10^{-2}$\\
5 & $7.6\times10^{-3}$ & $3.2\times10^{-3}$\\
6 & 0.25 & 0.15\\
7 & $1.2\times10^{-3}$ & $1.5\times10^{-3}$\\
8 & $5.7\times10^{-4}$ & $1.0\times10^{-3}$\\
9 & $9.9\times10^{-5}$ & $1.9\times10^{-4}$\\
10 & 0.33 & 0.49\\
11 & 0.11 & 0.12\\
12 & 0.54 & 0.60\\
13 & 0.92 & 0.89\\
14 & 0.97 & 0.94\\
15 & 0.97 & 0.93\\
16 & 0.99 & 0.94\\
\tableline
\multicolumn{3}{l}{Baksan}\\
1 & $2.1\times10^{-2}$ & $4.9\times10^{-3}$\\
2 & $3.6\times10^{-2}$ & $1.9\times10^{-2}$\\
3 & $7.4\times10^{-2}$ & 0.12\\
4 & 0.30 & 0.35 \\
5 & 0.55 & 0.52\\
\end{tabular}
\end{table}

\narrowtext
\begin{table}
\caption{Properties of best-fit two-component model with nonzero neutrino mass.
\label{table:m}}
\begin{tabular}{lc}
Quantity &  KII--IMB--Baksan  \\
\tableline
%
\multicolumn{2}{c}{Displaced Power Law Cooling}\\
\multicolumn{2}{c}{and Truncated Accretion}\\
\noalign{\smallskip}
$\alpha$ & 1.78  \\
$T_{c,0}$ (MeV) & 4.65  \\
$\tau_c$ (s) & 14.7  \\
$T_{a,0}$ (MeV) & 2.04  \\
$\tau_a$ (s) & 0.56  \\
$\mu$ & \llap{$\equiv$}0.5  \\
$\mnu$ (eV) & 3.02  \\
$t_{\rm off}$ (KII) (ms) & 0.07 \\
$t_{\rm off}$ (IMB) (ms) & 0.04 \\
$t_{\rm off}$ (Baksan) (ms) & 0.13 \\
${\cal L}/{\cal L}_0$ & 2.3  \\
\noalign{\bigskip}
$N_{\rm det}$ (KII) & $15.7 + 5.6$\tablenotemark[1] \\
$N_{\rm det}$ (IMB) & 4.5 \\
$N_{\rm det}$ (Baksan) & $1.7 + 1.0$\tablenotemark[1] \\
$R$ (km) & 17.8 \\
$E_b$ ($10^{53}$ erg)\tablenotemark[2] & 3.04 (0.56) \\
\end{tabular}
\tablenotetext[1]{Expected numbers of signal and background
events are listed separately.}
\tablenotetext[2]{Total $E_b$ is given, with part due to accretion
in parentheses.}
\end{table}

\narrowtext
\begin{table}
\caption{Two-dimensional Kolmogorov-Smirnov test results for the best-fit
Constant Temperature model, the best-fit
Exponential Cooling model and the best-fit Displaced Power Law Cooling and
Truncated Accretion model.
\label{table:ks}}
\begin{tabular}{lcccc}
Quantity &  KII & IMB & Baksan & Joint  \\
\tableline
\multicolumn{5}{c}{Constant Temperature}\\
$D_{\rm obs}$ & 0.38 & 0.45 & 0.52 & -- \\
$P(>D_{\rm obs})$ & $3.2\times 10^{-2}$ & $6.9\times 10^{-2}$ & $8.4\times %
      10^{-2}$ & $8.7\times 10^{-3}$ \\
\tableline
\multicolumn{5}{c}{Exponential Cooling}\\
$D_{\rm obs}$ & 0.31 & 0.28 & 0.38 & -- \\
$P(>D_{\rm obs})$ & 0.12 & 0.53 & 0.37 & 0.27 \\
\tableline
\multicolumn{5}{c}{Displaced Power Law Cooling}\\
\multicolumn{5}{c}{and Truncated Accretion}\\
$D_{\rm obs}$ & 0.27 & 0.23 & 0.37 & -- \\
$P(>D_{\rm obs})$ & 0.25 & 0.77 & 0.39 & 0.52 \\
\end{tabular}
\end{table}


\begin{references}
\bibitem{B-priors}The Bayes factor depends on the prior ranges for the
parameters (see eqn.~\ref{ockham-app2}).  For the parameters common to
both models, the ranges used were: $[3,50]$~km for $\alpha$; $[0,3]$~s
for offset times.  For the exponential model, the ranges were:
$[0.1,15]$~MeV for $T_0$; $[0.1,30]$~s for $\tau$.  For the cooling
plus accretion model, the ranges were:  $[0.1,15]$~MeV for $T_{c,0}$;
$[0.3,50]$~s for $\tau_c$; $[0.1,10]$~MeV for $T_a$; $[0.03,10]$~s for
$\tau_a$.  In the light of theoretical expectations, these are
generous (and thus conservative) prior ranges for the individual parameters.
\end{references}
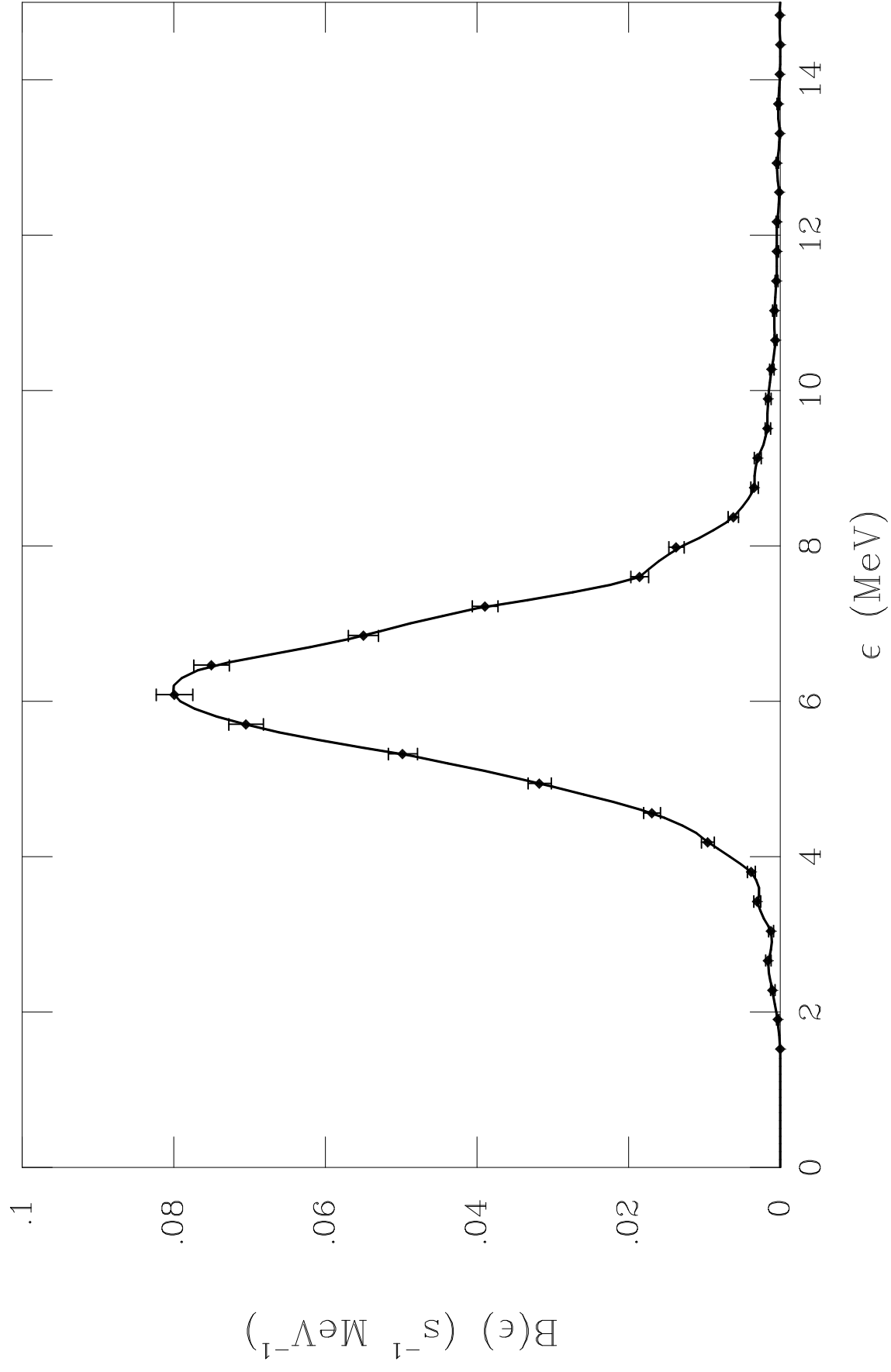
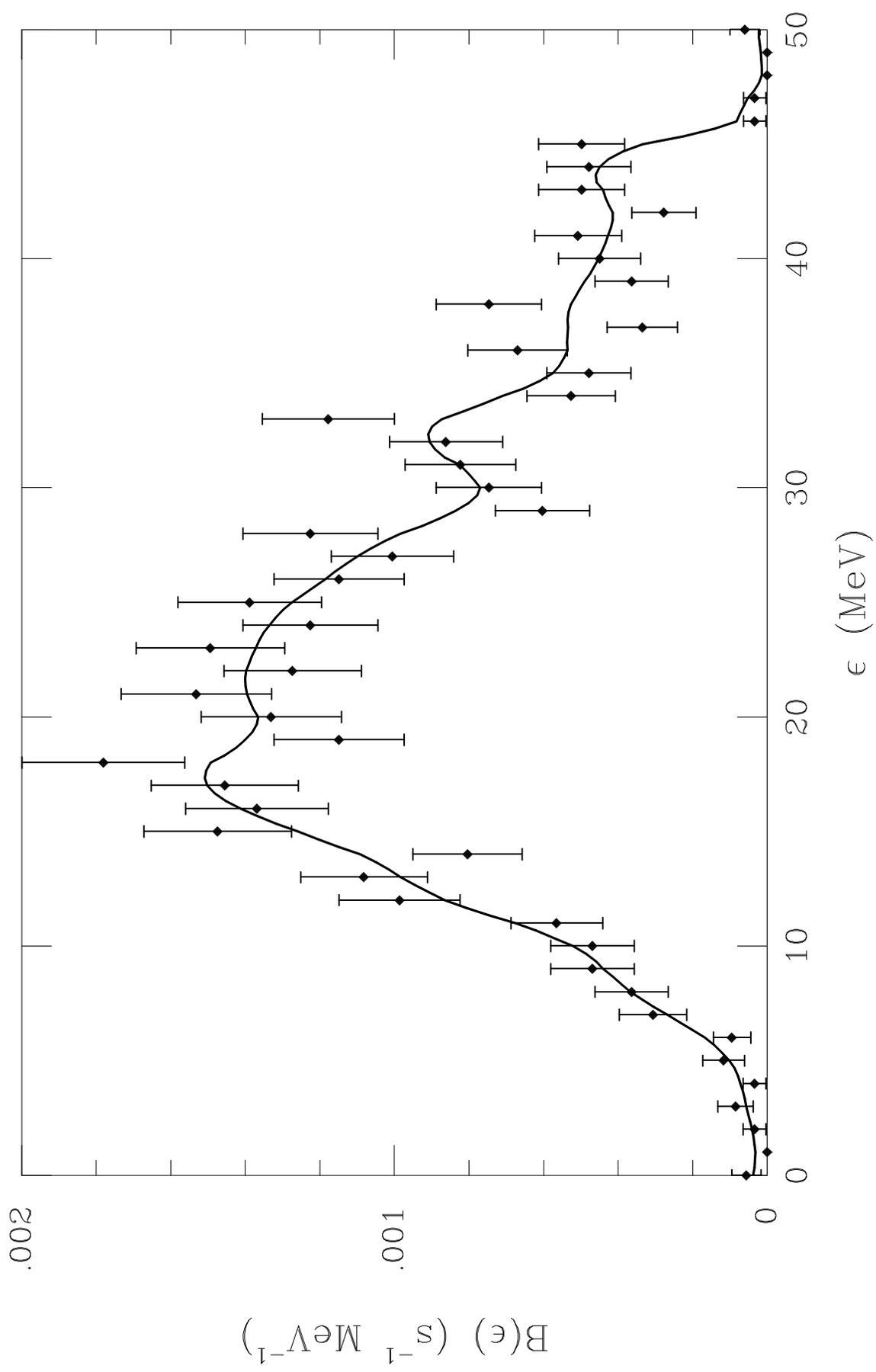
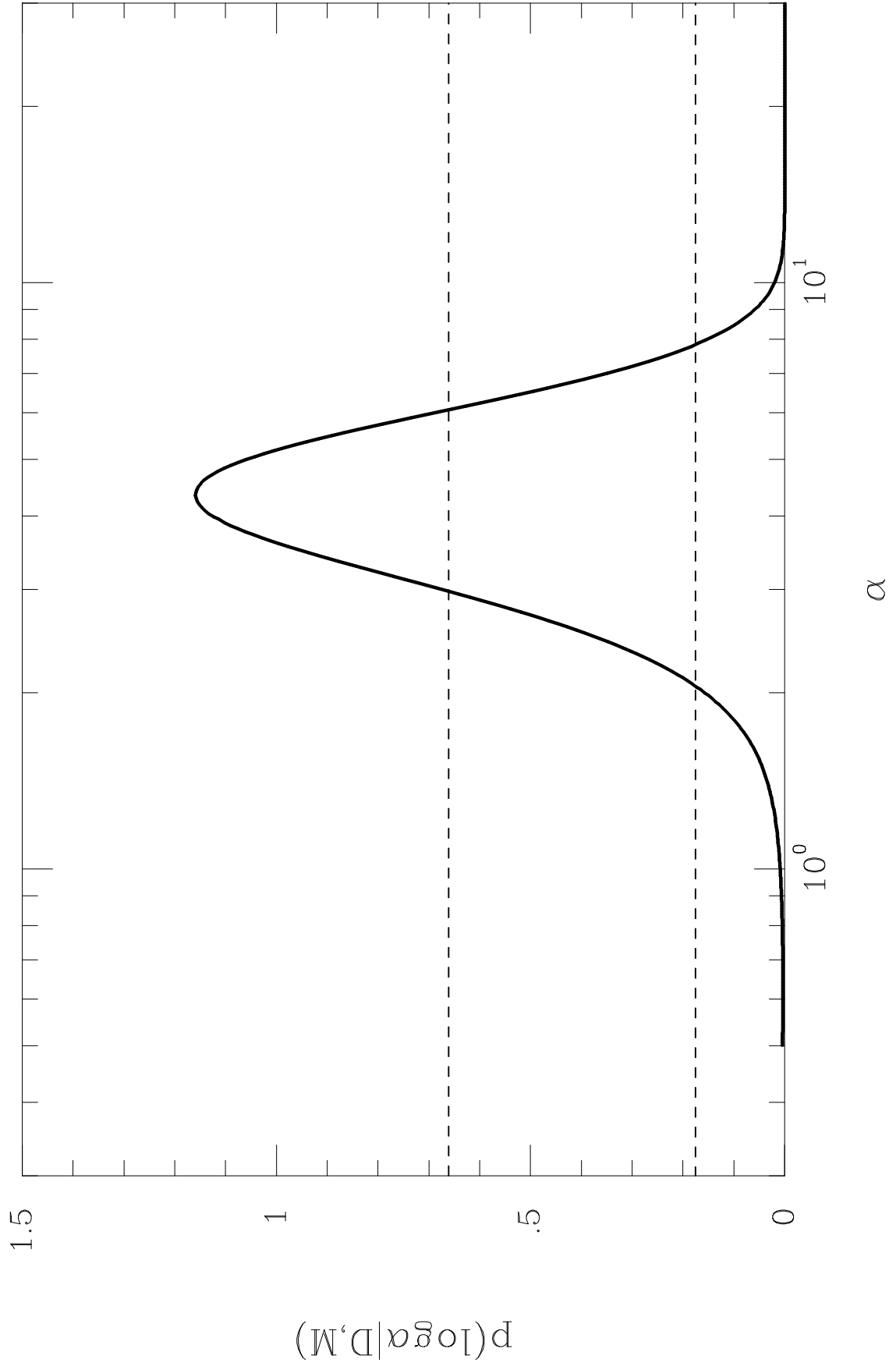
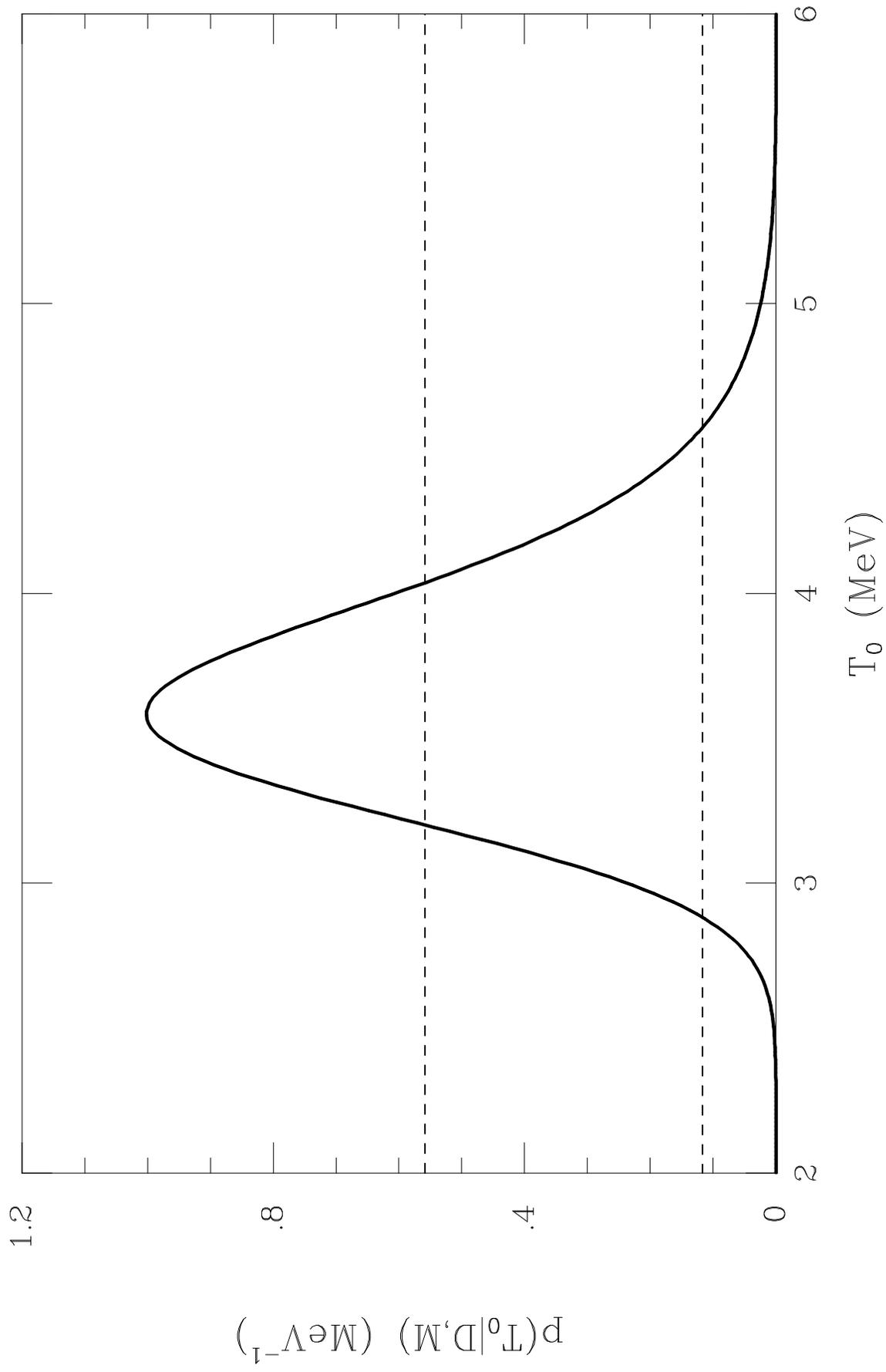
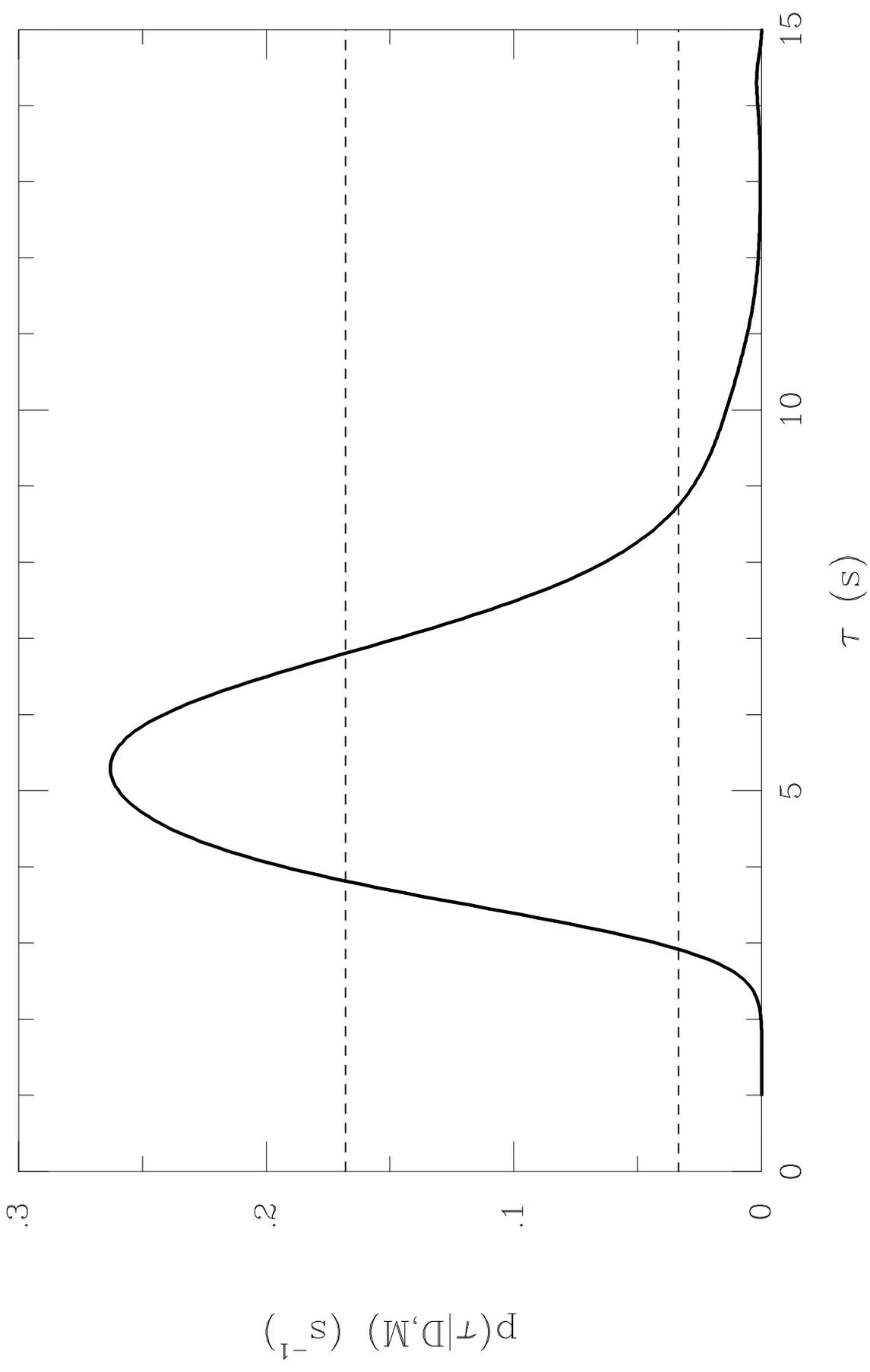
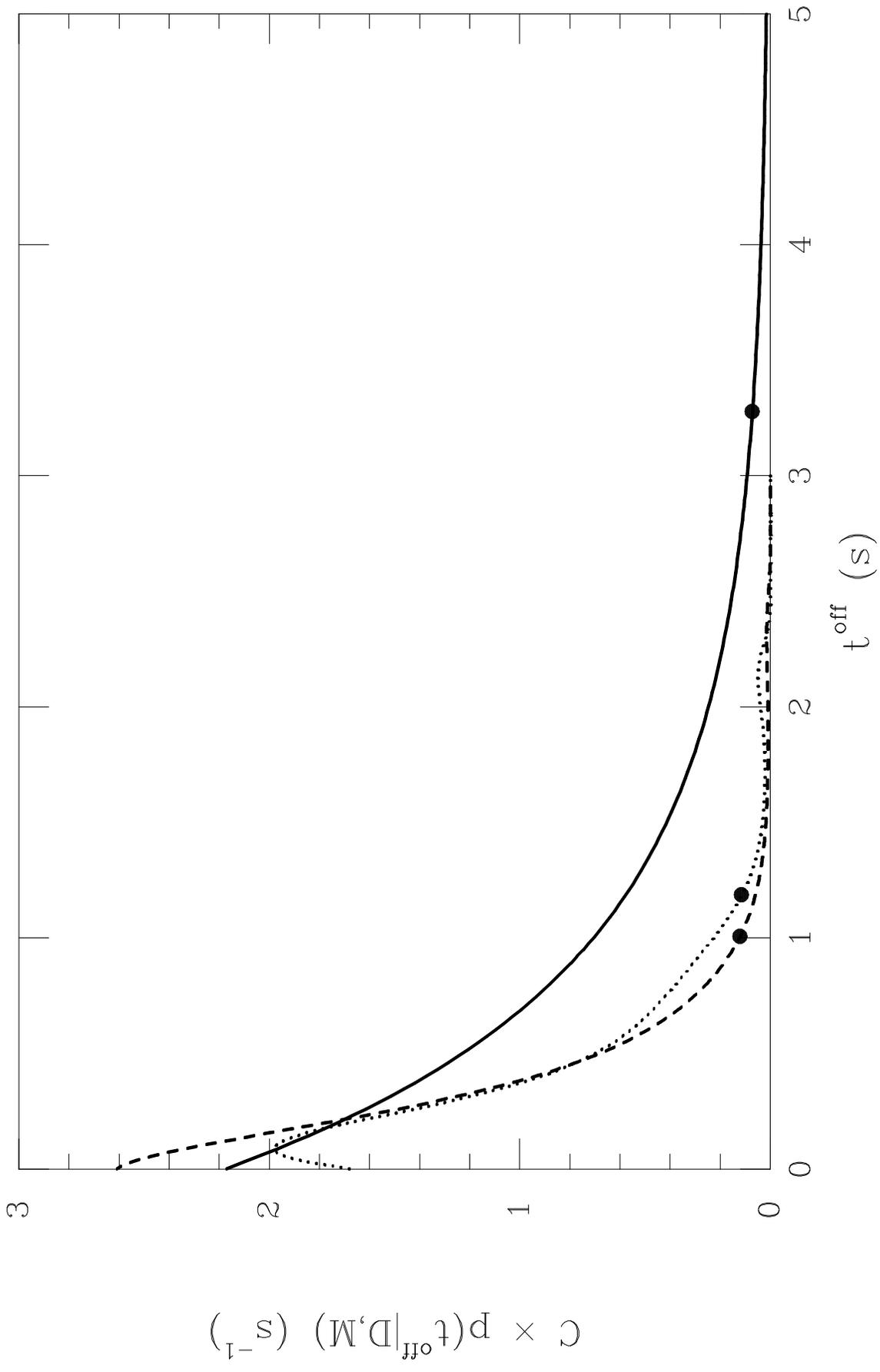
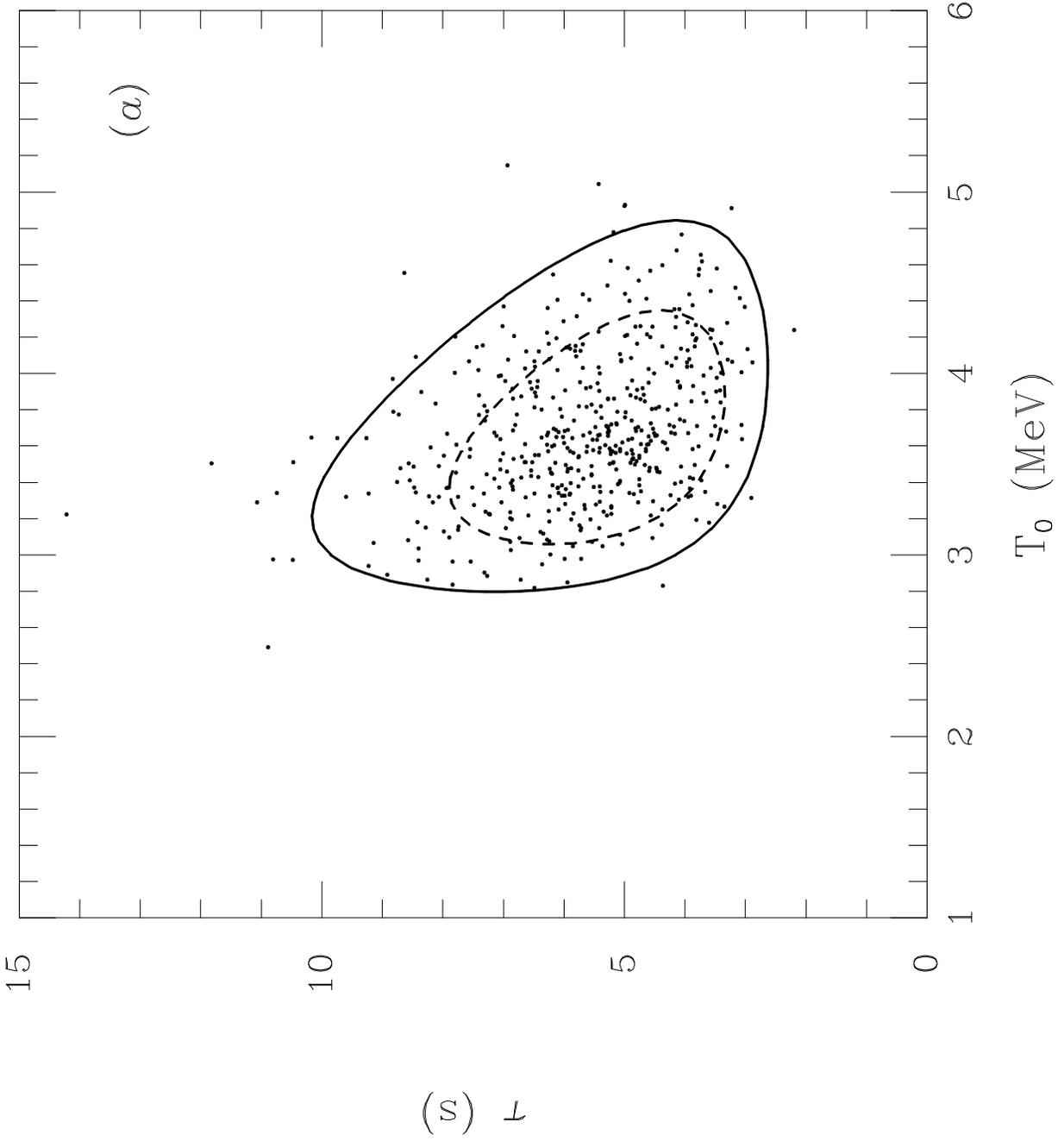
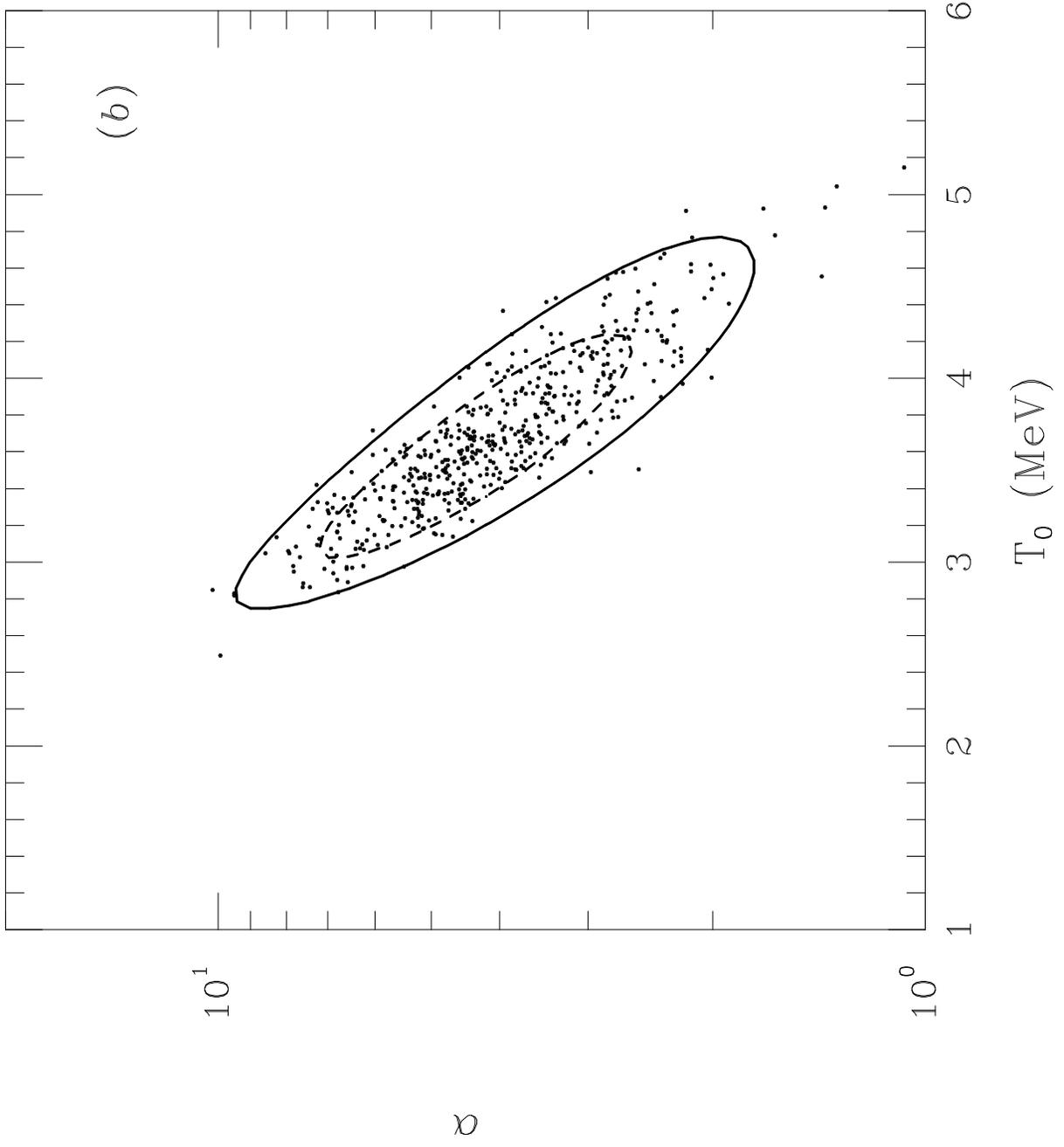
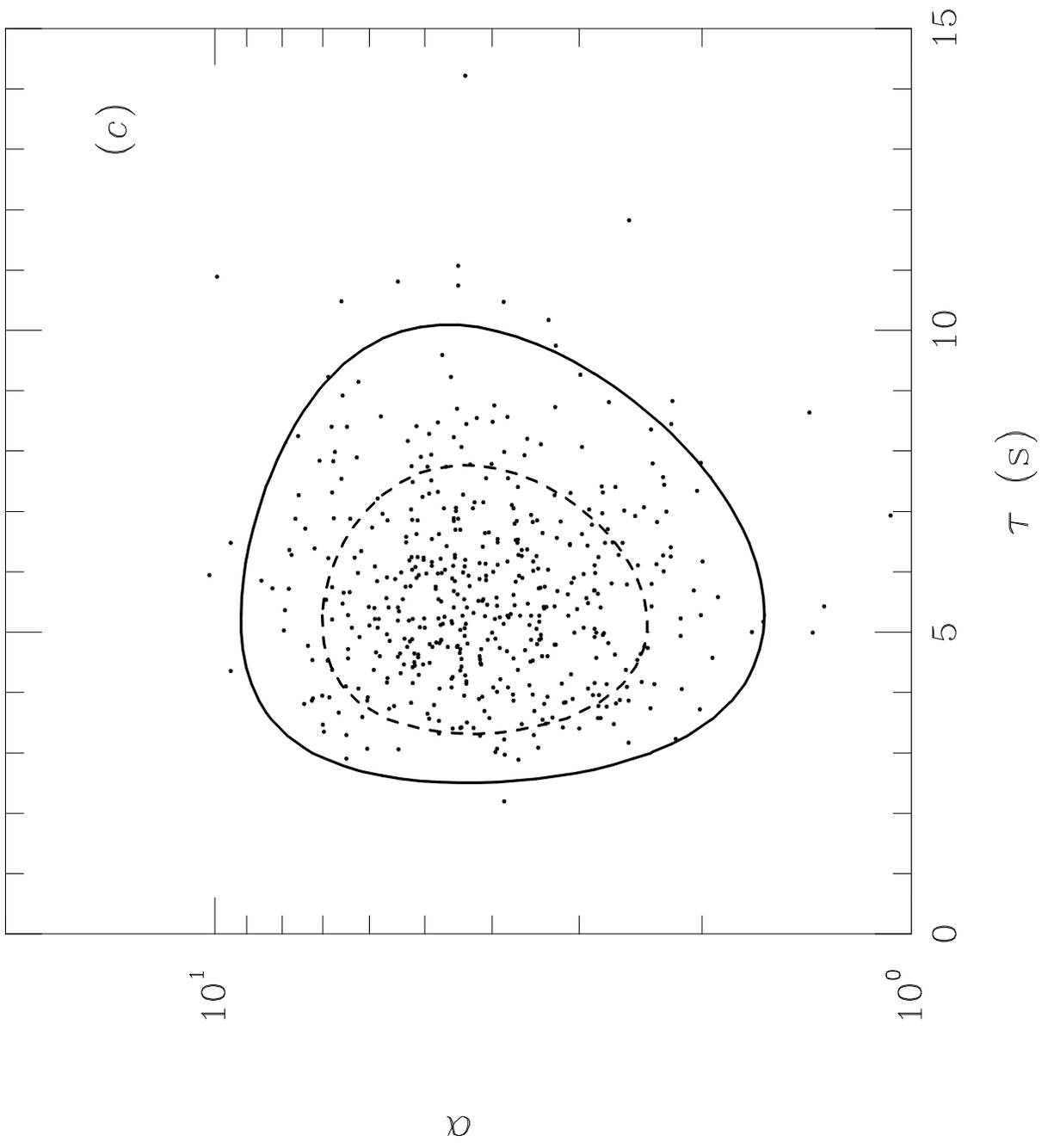
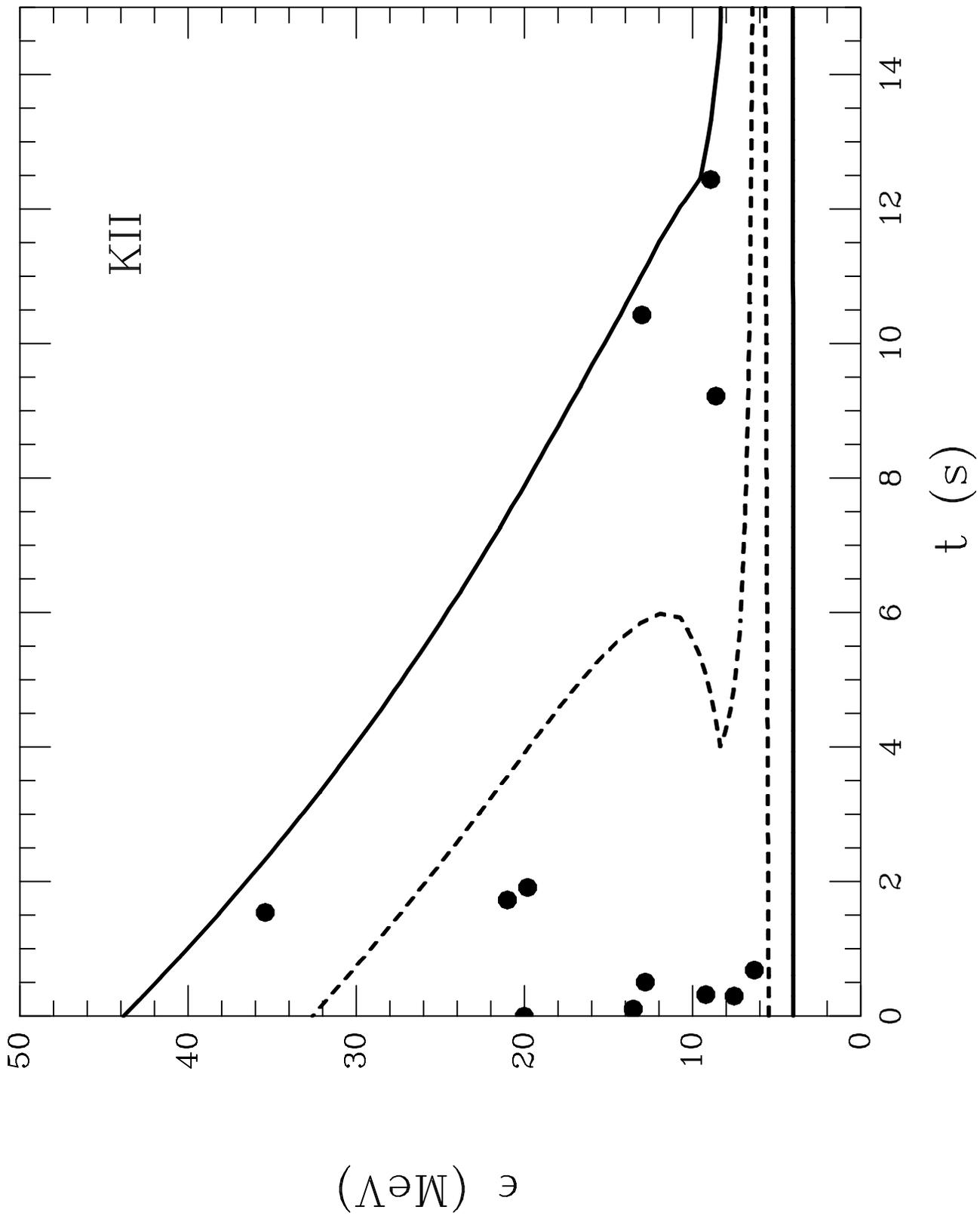
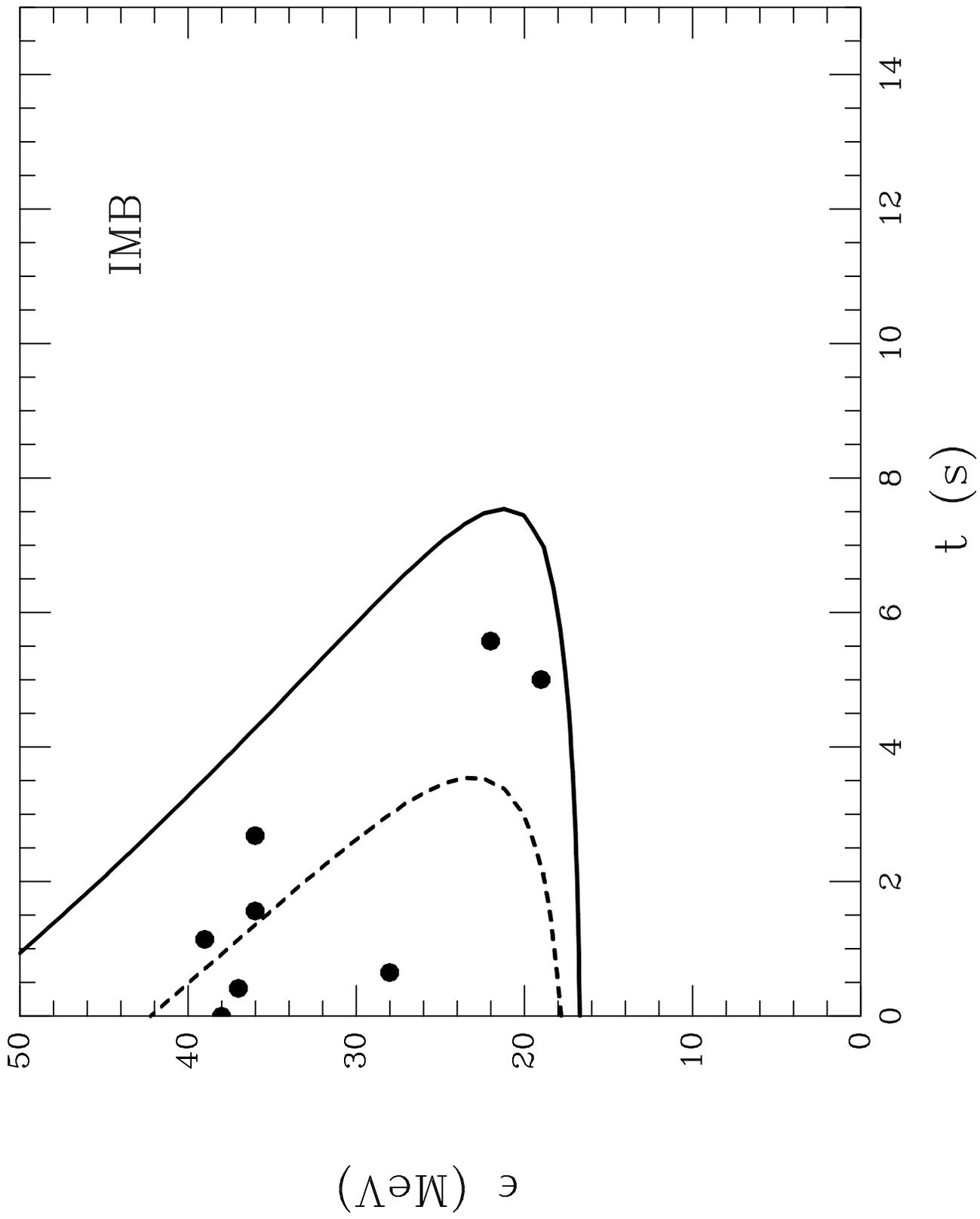
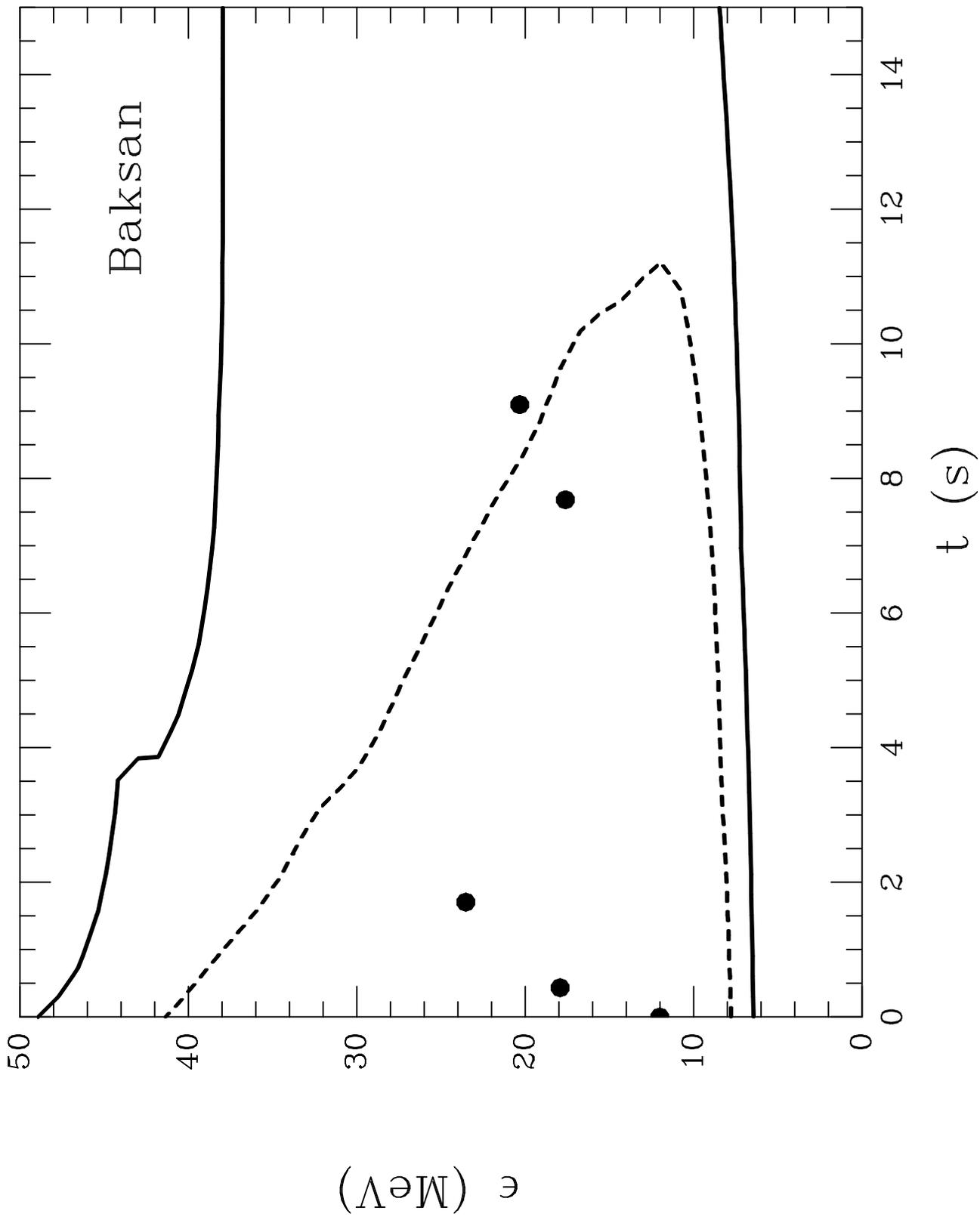
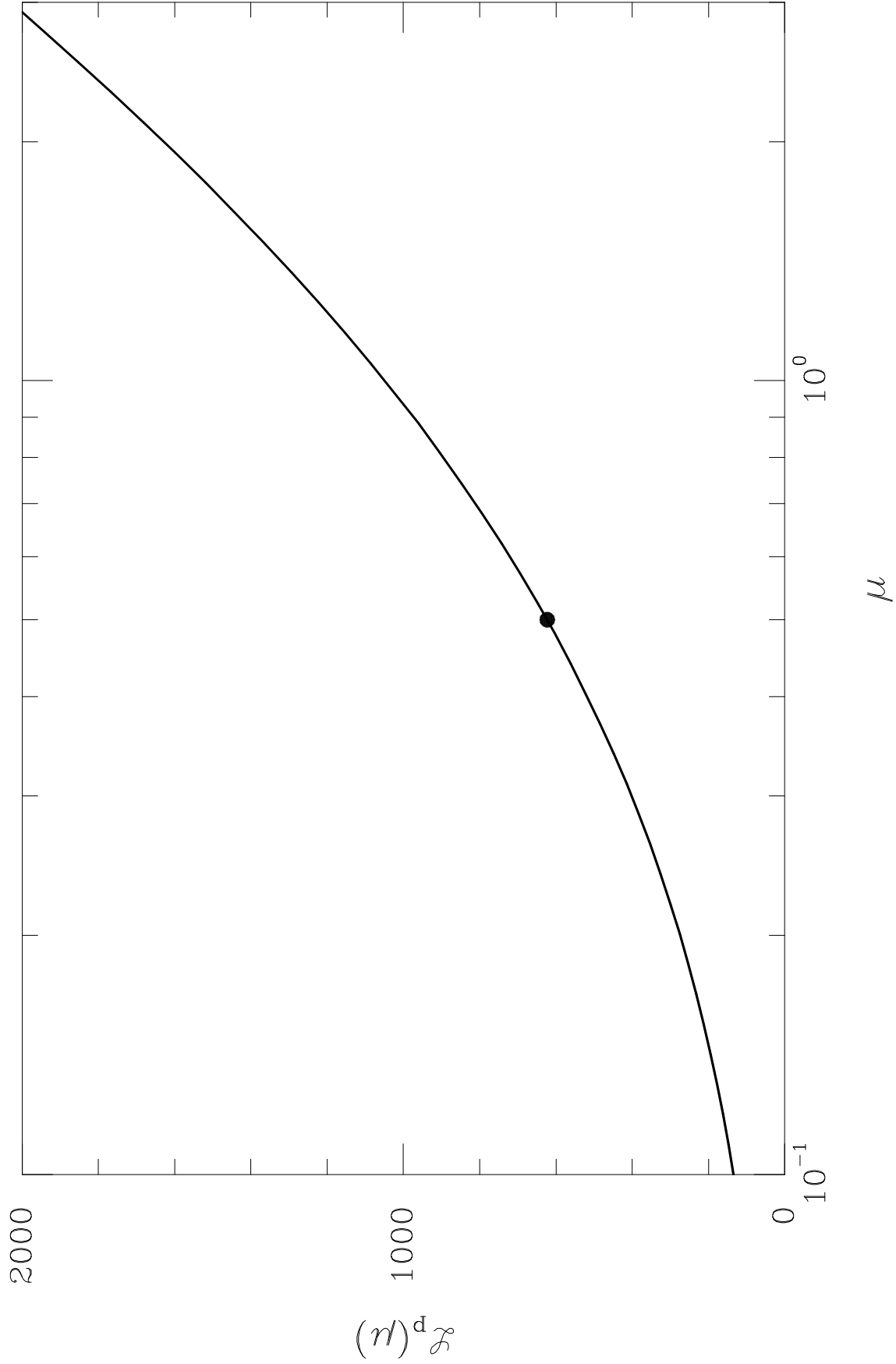
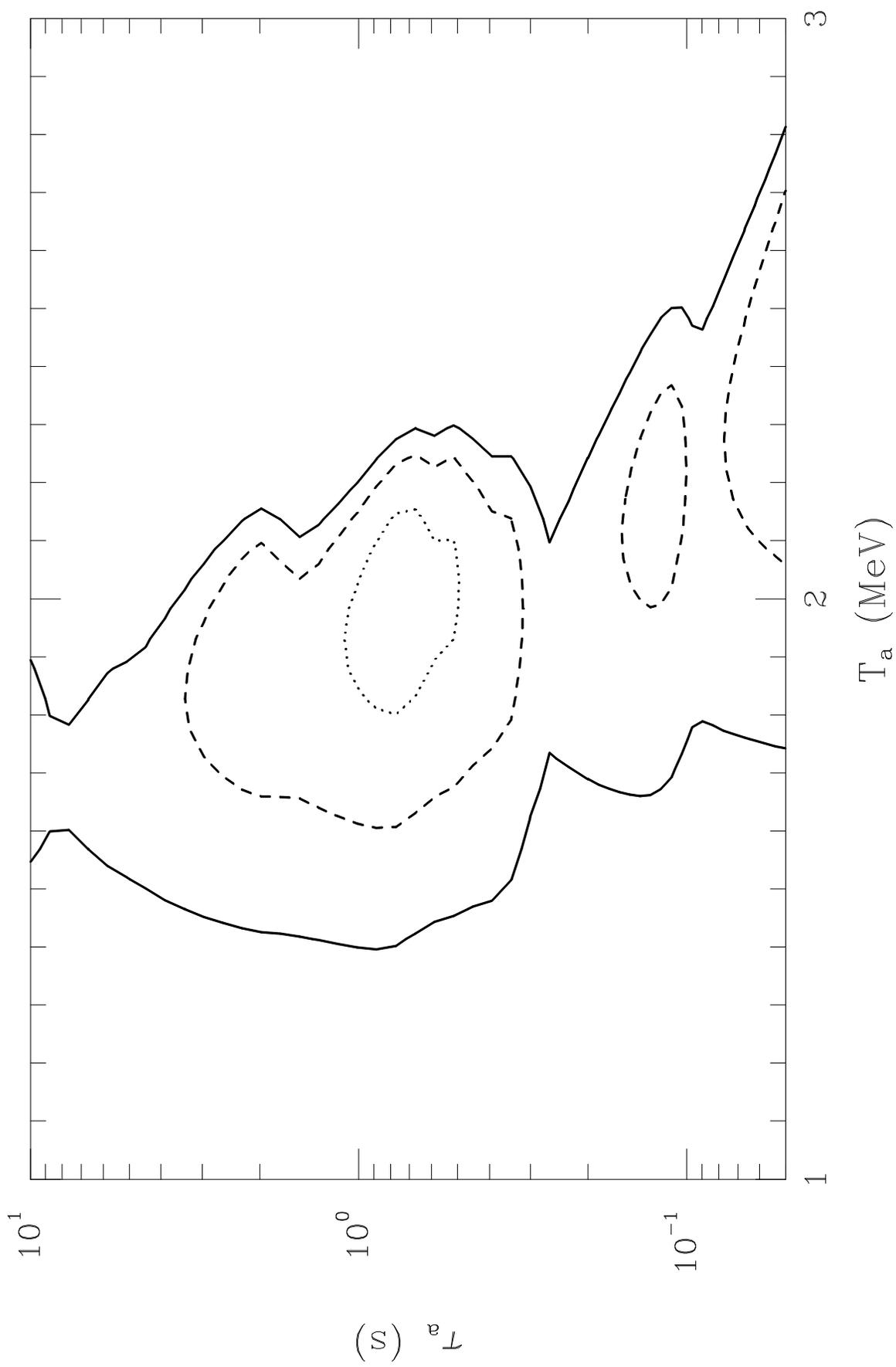
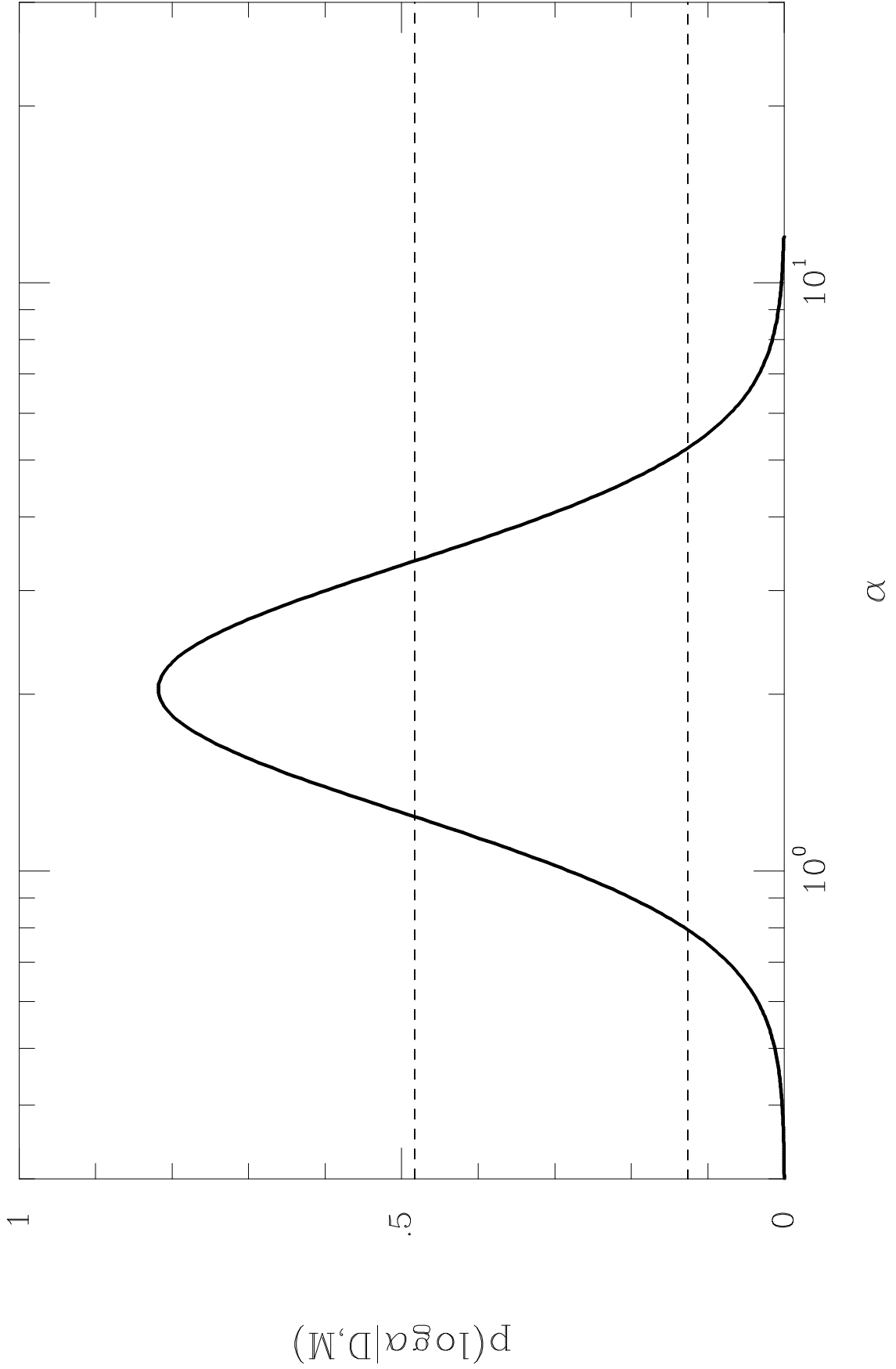
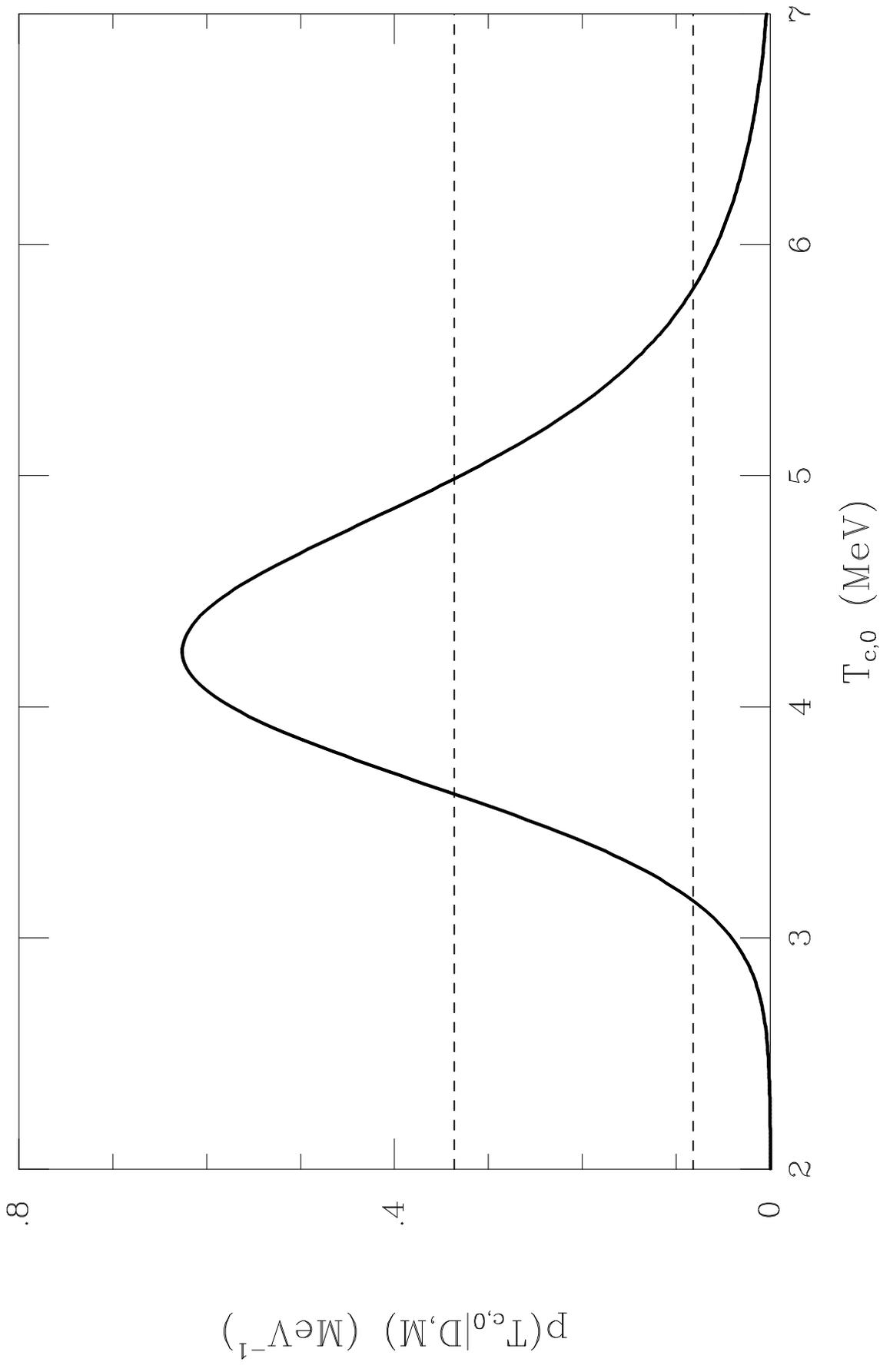
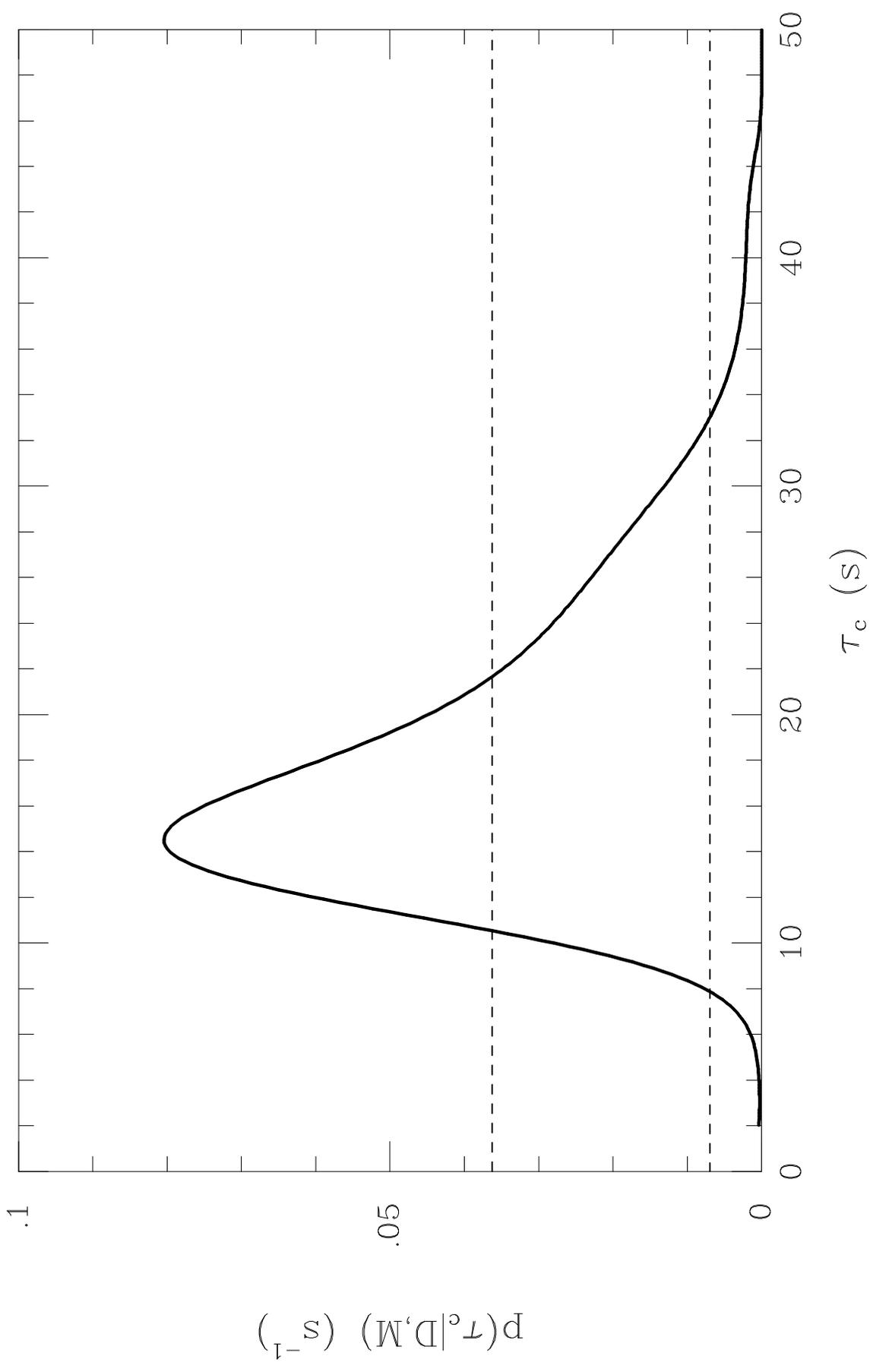
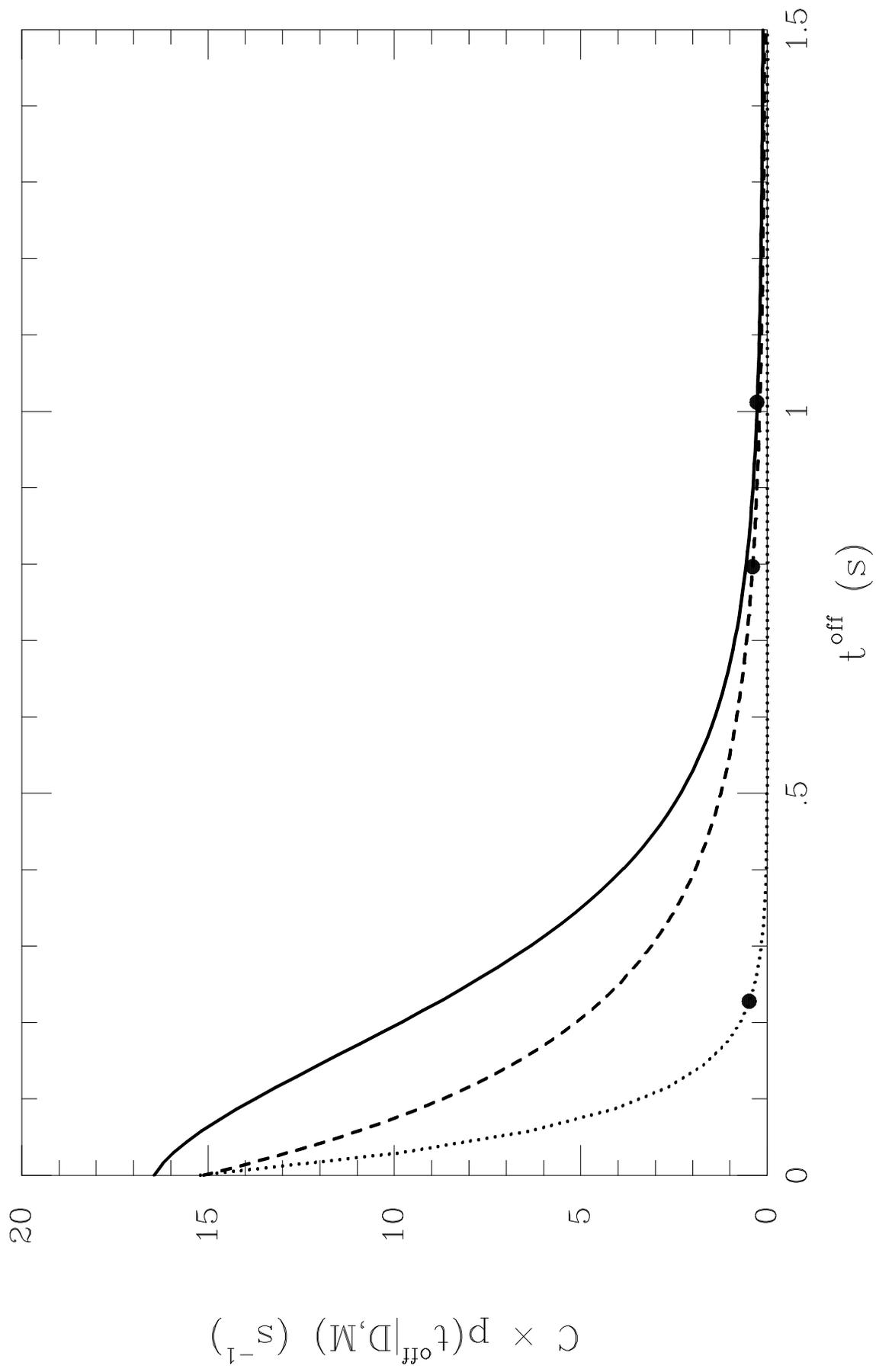
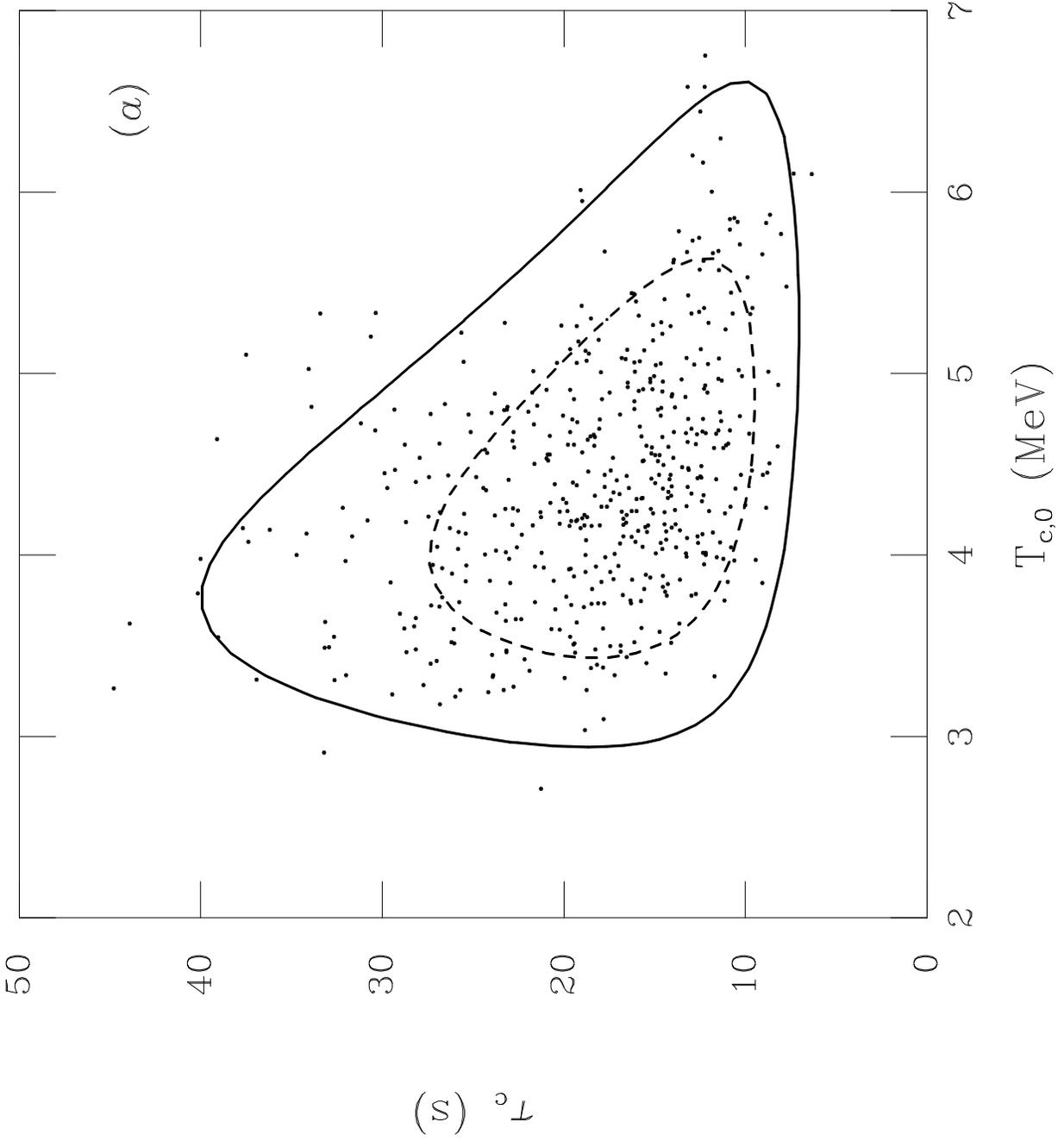
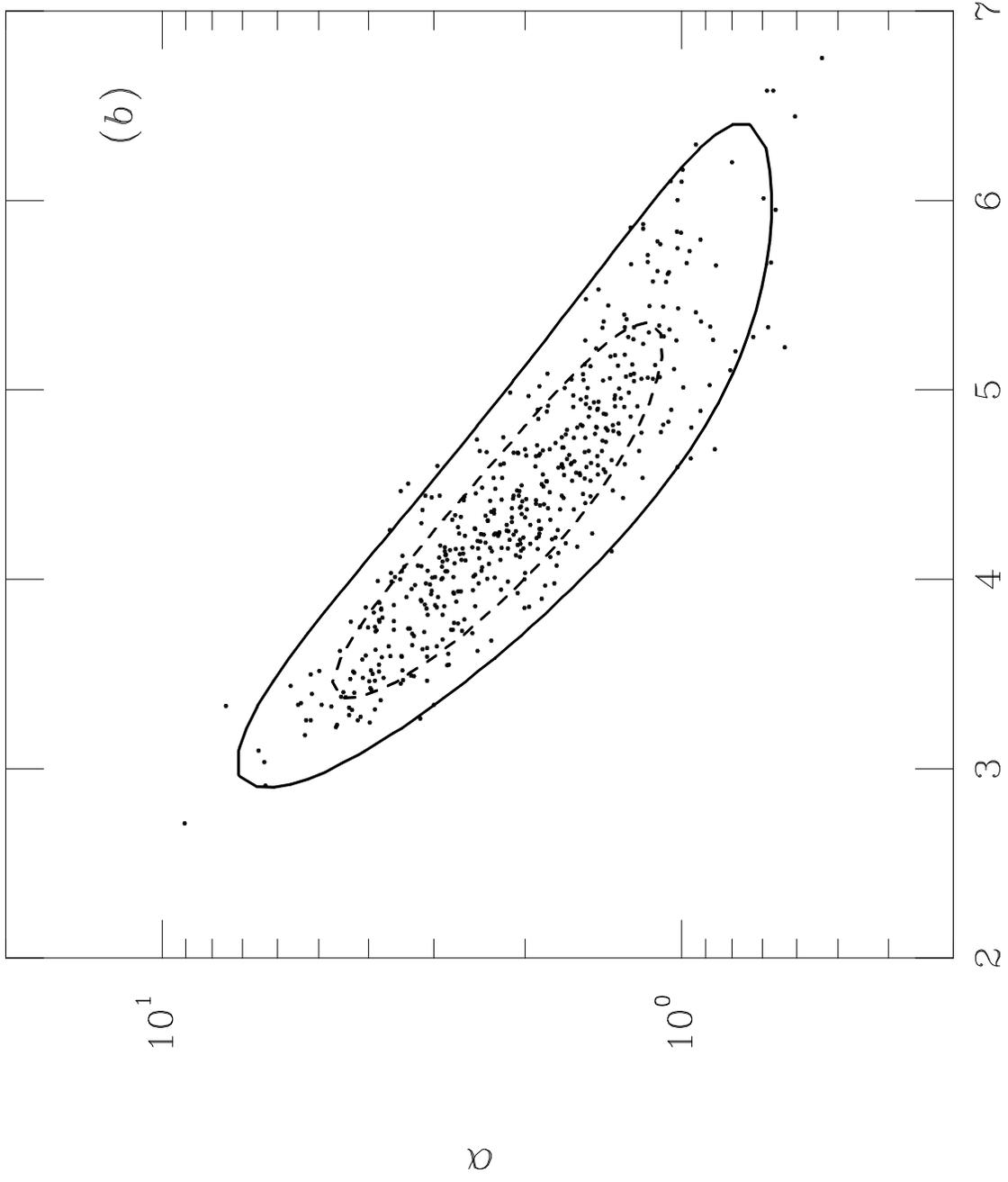
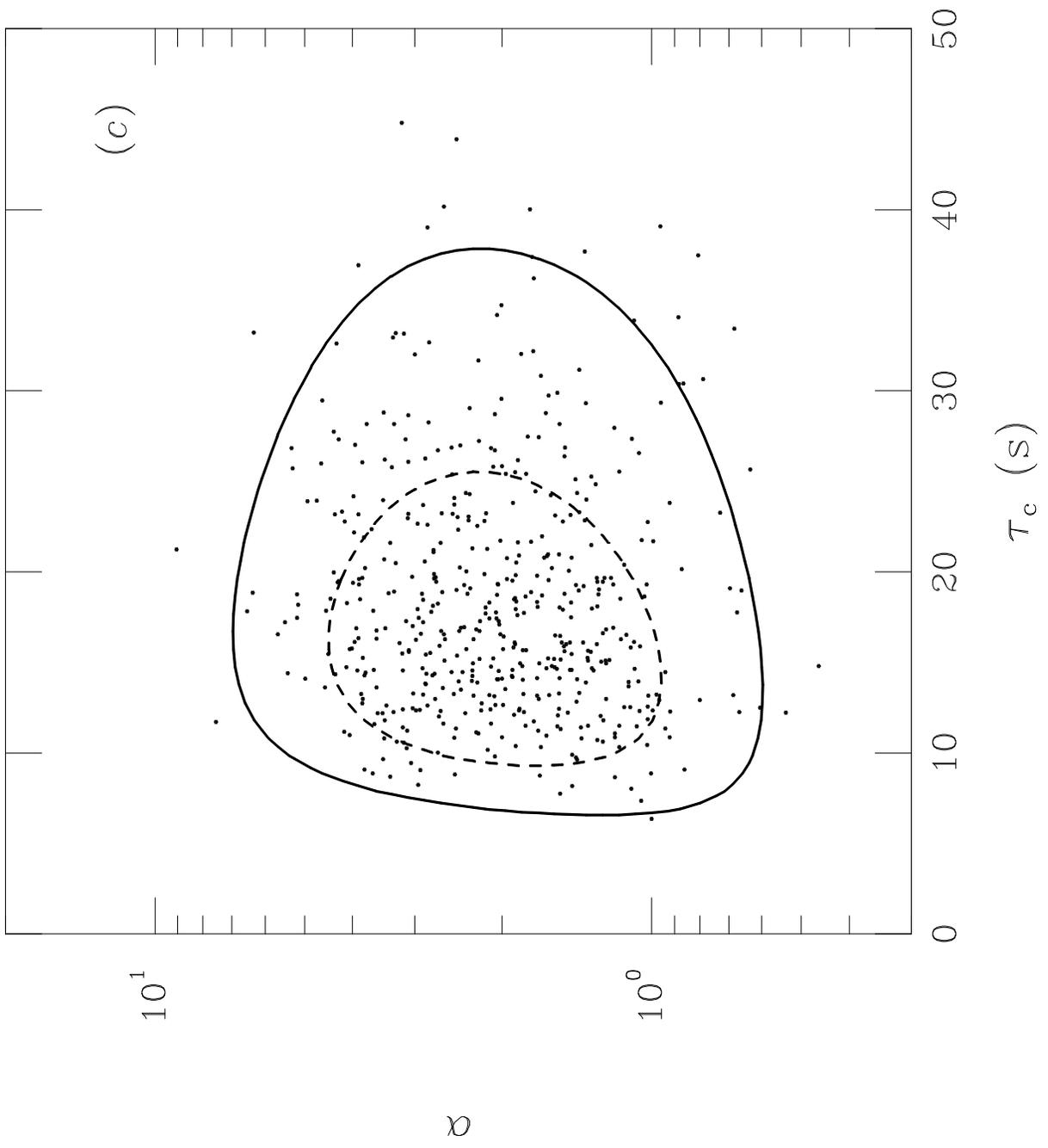
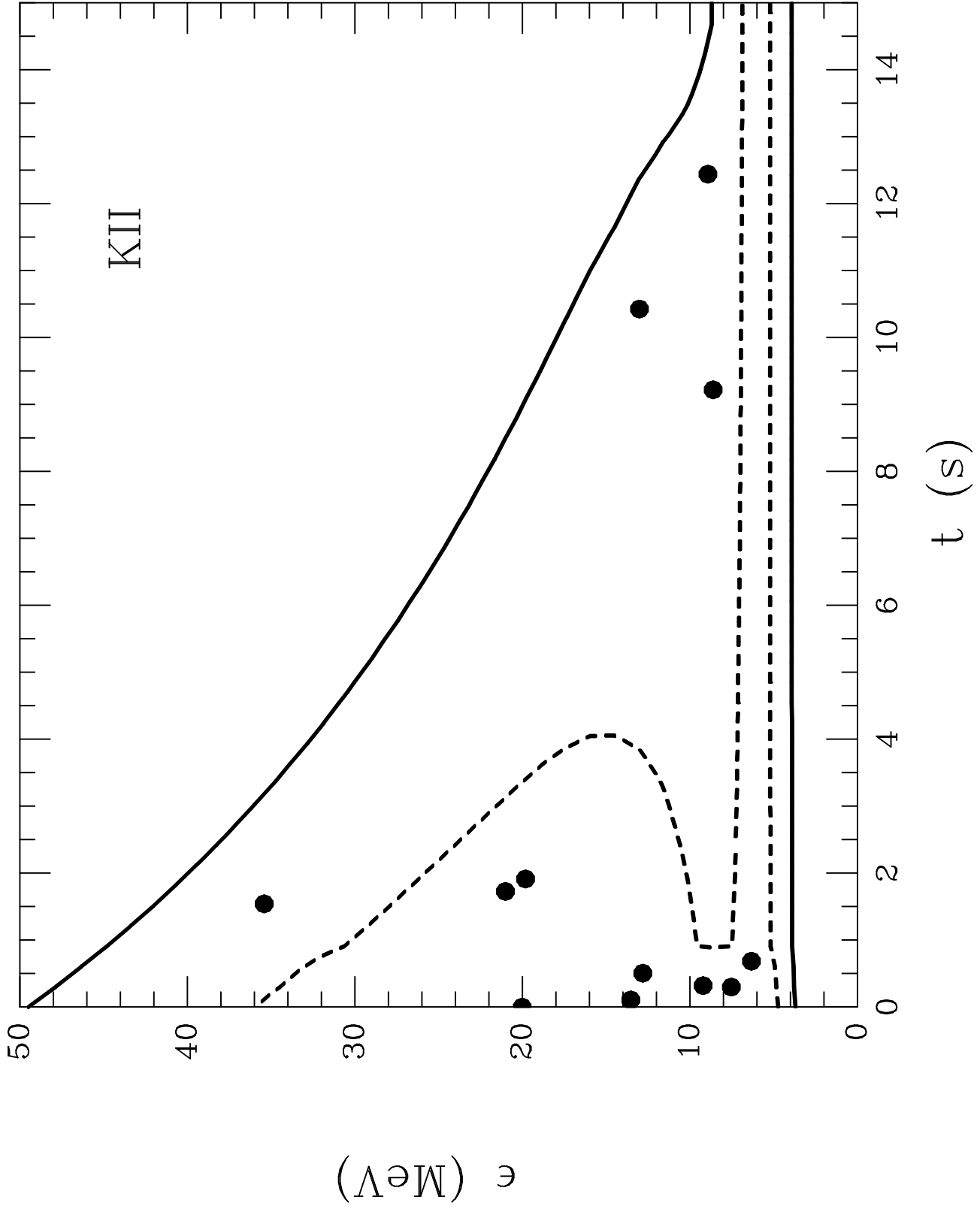
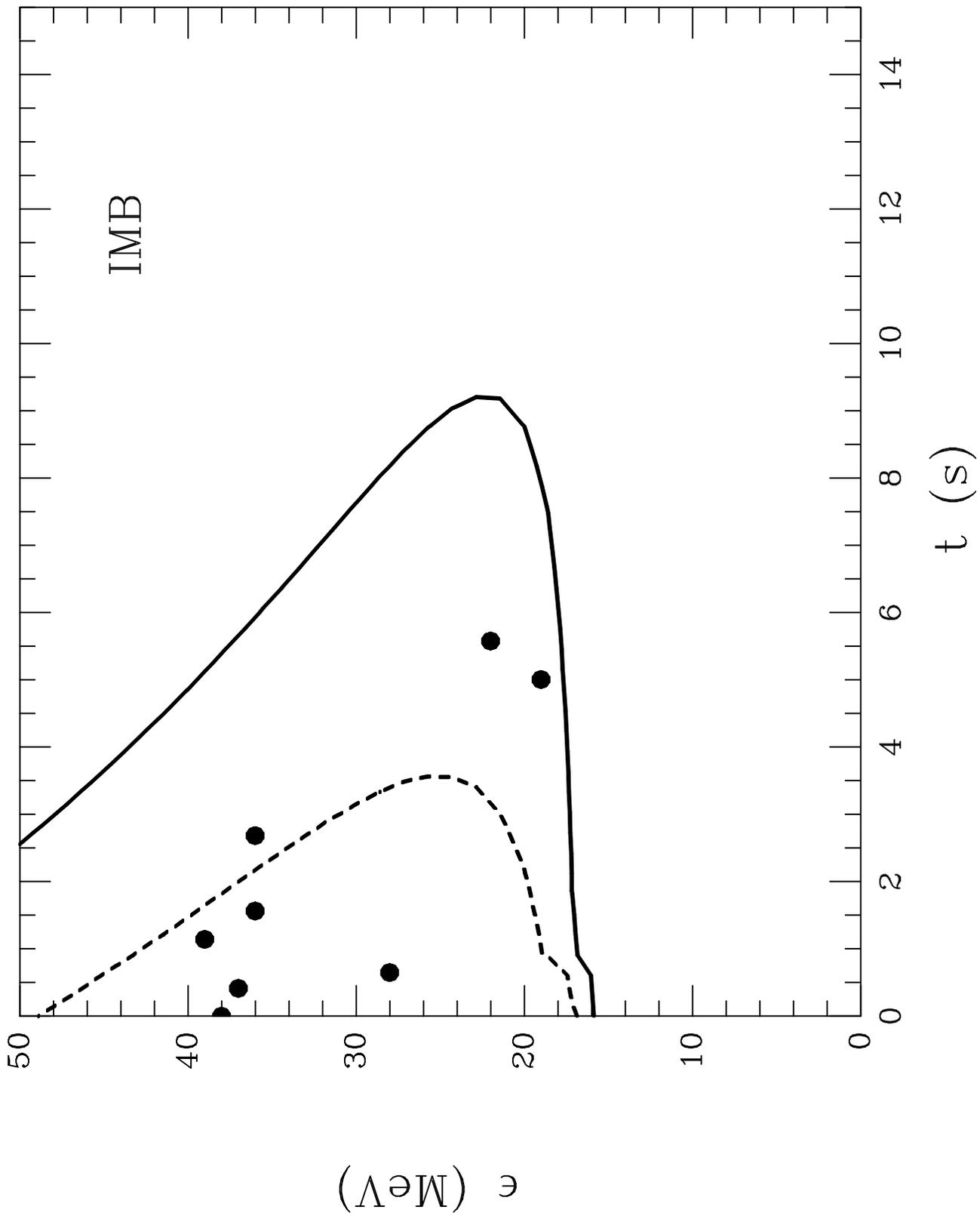
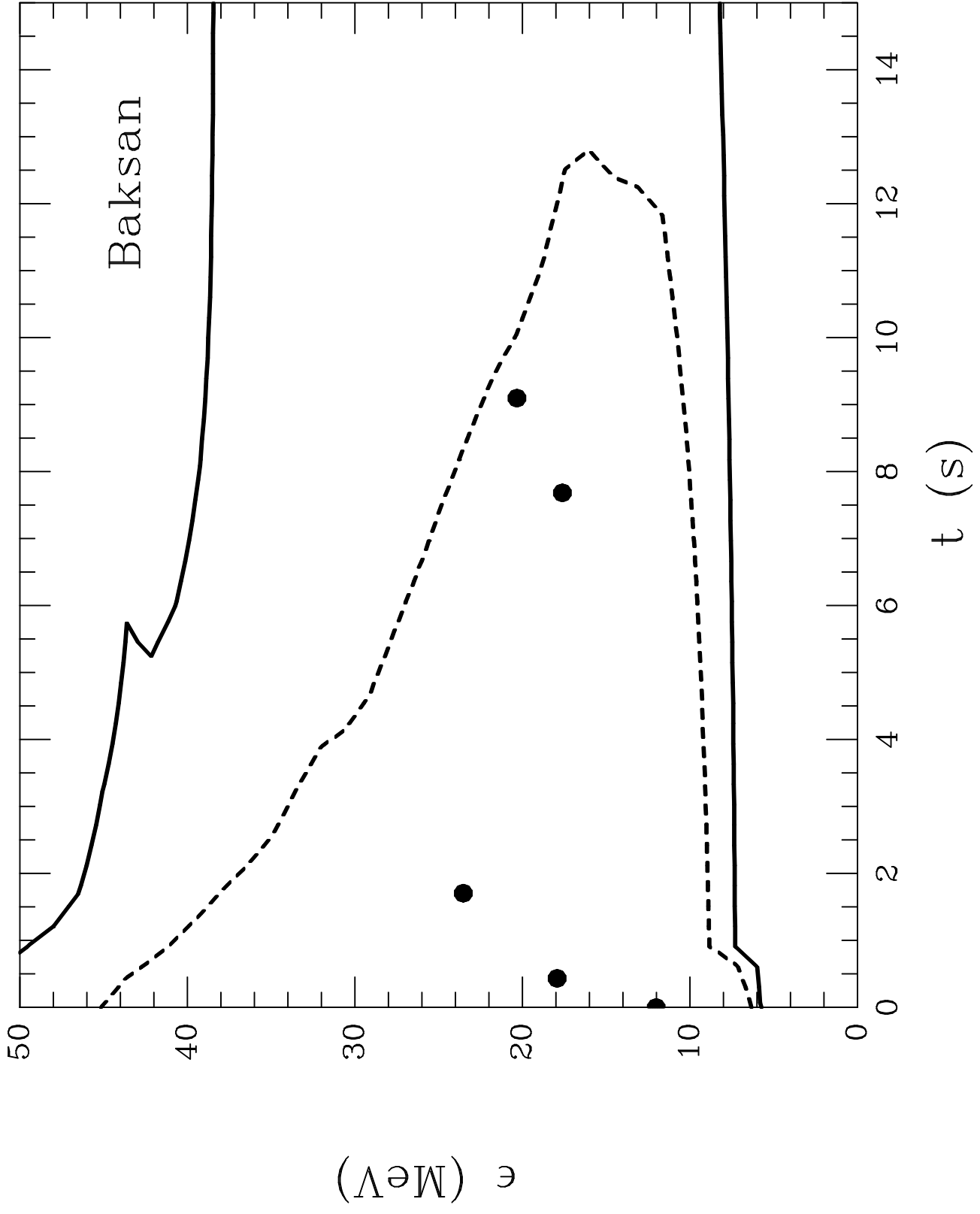
\end{document}